\documentclass[twocolumn,twocolappendix]{aastex63}
 
\usepackage{amsmath}
\usepackage{hyperref}
\usepackage{textcomp}

\graphicspath{{./}}

\submitjournal{AJ}
\accepted{\today}

\shorttitle{SCUBA-2 Calibration}
\shortauthors{Mairs et al.}

\begin{document}

\title{A Decade of SCUBA-2: A Comprehensive Guide to Calibrating\\ \mbox{450 $\mu$m} and \mbox{850 $\mu$m} Continuum Data at the JCMT}

\correspondingauthor{Steve Mairs}
\email{s.mairs@eaobservatory.org}

%%%%%%%%% Author List %%%%%%%%%%%

\author[0000-0002-6956-0730]{Steve Mairs}
\affil{East Asian Observatory,
660 N. A`oh\={o}k\={u} Place, Hilo, HI 96720, USA}

\author[0000-0002-5457-9025]{Jessica T. Dempsey}
\affil{East Asian Observatory,
660 N. A`oh\={o}k\={u} Place, Hilo, HI 96720, USA}

\author[0000-0003-0438-8228]{Graham S. Bell}
\affil{East Asian Observatory,
660 N. A`oh\={o}k\={u} Place, Hilo, HI 96720, USA}

\author[0000-0002-6327-3423]{Harriet Parsons}
\affil{East Asian Observatory,
660 N. A`oh\={o}k\={u} Place, Hilo, HI 96720, USA}

\author[0000-0003-0141-0362]{Malcolm J. Currie}
\affil{East Asian Observatory,
660 N. A`oh\={o}k\={u} Place, Hilo, HI 96720, USA}
\affil{RAL Space, STFC Rutherford Appleton Laboratory, Chilton, Didcot, Oxfordshire OX11 0QX, UK}

\author[0000-0002-8010-8454]{Per Friberg}
\affil{East Asian Observatory,
660 N. A`oh\={o}k\={u} Place, Hilo, HI 96720, USA}

\author[0000-0002-8899-4673]{Xue-Jian Jiang}
\affil{East Asian Observatory,
660 N. A`oh\={o}k\={u} Place, Hilo, HI 96720, USA}

\author[0000-0003-3906-4354]{Alexandra J. Tetarenko}
\affil{East Asian Observatory,
660 N. A`oh\={o}k\={u} Place, Hilo, HI 96720, USA}

\author{Dan Bintley}
\affil{East Asian Observatory,
660 N. A`oh\={o}k\={u} Place, Hilo, HI 96720, USA}

\author{Jamie Cookson}
\affil{East Asian Observatory,
660 N. A`oh\={o}k\={u} Place, Hilo, HI 96720, USA}

\author{Shaoliang Li}
\affil{East Asian Observatory,
660 N. A`oh\={o}k\={u} Place, Hilo, HI 96720, USA}

\author[0000-0002-6529-202X]{Mark G. Rawlings}
\affil{East Asian Observatory,
660 N. A`oh\={o}k\={u} Place, Hilo, HI 96720, USA}

\author[0000-0002-4694-6905]{Jan Wouterloot}
\affil{East Asian Observatory,
660 N. A`oh\={o}k\={u} Place, Hilo, HI 96720, USA}

\author[0000-0001-6524-2447]{David Berry}
\affil{East Asian Observatory,
660 N. A`oh\={o}k\={u} Place, Hilo, HI 96720, USA}

\author[0000-0001-9361-5781]{Sarah Graves}
\affil{East Asian Observatory,
660 N. A`oh\={o}k\={u} Place, Hilo, HI 96720, USA}

\author[0000-0002-7210-6264]{Izumi Mizuno}
\affil{East Asian Observatory,
660 N. A`oh\={o}k\={u} Place, Hilo, HI 96720, USA}

\author{Alexis Ann Acohido}
\affil{East Asian Observatory,
660 N. A`oh\={o}k\={u} Place, Hilo, HI 96720, USA}

\author{Alyssa Clark}
\affil{East Asian Observatory,
660 N. A`oh\={o}k\={u} Place, Hilo, HI 96720, USA}

\author{Jeff Cox}
\affil{Joint Astronomy Centre,
660 N. A`oh\={o}k\={u} Place, Hilo, HI 96720, USA}

\author{Miriam Fuchs}
\affil{East Asian Observatory,
660 N. A`oh\={o}k\={u} Place, Hilo, HI 96720, USA}

\author{James Hoge}
\affil{East Asian Observatory,
660 N. A`oh\={o}k\={u} Place, Hilo, HI 96720, USA}

\author{Johnathon Kemp}
\affil{Joint Astronomy Centre,
660 N. A`oh\={o}k\={u} Place, Hilo, HI 96720, USA}

\author[0000-0003-4102-0385]{E’lisa Lee}
\affil{East Asian Observatory,
660 N. A`oh\={o}k\={u} Place, Hilo, HI 96720, USA}

\author{Callie Matulonis}
\affil{East Asian Observatory,
660 N. A`oh\={o}k\={u} Place, Hilo, HI 96720, USA}

\author{William Montgomerie}
\affil{East Asian Observatory,
660 N. A`oh\={o}k\={u} Place, Hilo, HI 96720, USA}
\affil{SOFIA Science Center, Universities Space Research Association, NASA Ames Research Center, Moffett Field, CA 94035, USA}

\author{Kevin Silva}
\affil{East Asian Observatory,
660 N. A`oh\={o}k\={u} Place, Hilo, HI 96720, USA}

\author{Patrice Smith}
\affil{East Asian Observatory,
660 N. A`oh\={o}k\={u} Place, Hilo, HI 96720, USA}

%%%%%%%%%%%%%%%%%%%%%%%%%%%%%%%%%%%%%%%%
%%%%%%%%%%%%%%%%%%%%%%%%%%%%%%%%%%%%%%%%
%%%%%%%%%%%%%%%%%%%%%%%%%%%%%%%%%%%%%%%%
\begin{abstract}
%%%%%%%%%%%%%%%%%%%%%%%%%%%%%%%%%%%%%%%%
%%%%%%%%%%%%%%%%%%%%%%%%%%%%%%%%%%%%%%%%
%%%%%%%%%%%%%%%%%%%%%%%%%%%%%%%%%%%%%%%%

The Submillimetre Common User Bolometer Array 2 (SCUBA-2) is the James Clerk Maxwell Telescope's continuum imager, operating simultaneously at 450 and \mbox{850 $\mu$m}. SCUBA-2 was commissioned in 2009--2011 and since that time, regular observations of point-like standard sources have been performed whenever the instrument is in use. Expanding the calibrator observation sample by an order of magnitude compared to previous work, in this paper we derive updated opacity relations at each wavelength for a new atmospheric-extinction correction, analyze the Flux-Conversion Factors (FCFs) used to convert instrumental units to physical flux units as a function of date and observation time, present information on the beam profiles for each wavelength, and update secondary-calibrator source fluxes. Between 07:00 and 17:00 UTC, the portion of the night that is most stable to temperature gradients that cause dish deformation, the total-flux uncertainty and the peak-flux uncertainty measured at \mbox{450 $\mu$m} are found to be 14\% and 17\%, respectively. Measured at \mbox{850 $\mu$m}, the total-flux and peak-flux uncertainties are 6\%, and 7\%, respectively. The analysis presented in this work is applicable to all SCUBA-2 projects observed since 2011. 

\end{abstract}

% Keywords need updating:
\keywords{techniques: image processing -- methods: data analysis -- techniques: photometric}

%%%%%%%%%%%%%%%%%%%%%%%%%%%%%%%%%%%%%%%%
%%%%%%%%%%%%%%%%%%%%%%%%%%%%%%%%%%%%%%%%
%%%%%%%%%%%%%%%%%%%%%%%%%%%%%%%%%%%%%%%%
\section{Introduction} 
\label{sec:intro}
%%%%%%%%%%%%%%%%%%%%%%%%%%%%%%%%%%%%%%%%
%%%%%%%%%%%%%%%%%%%%%%%%%%%%%%%%%%%%%%%%
%%%%%%%%%%%%%%%%%%%%%%%%%%%%%%%%%%%%%%%%

Ground-based submillimeter observations are subject to severe 
attenuation of the incoming light due to absorption by 
atmospheric water vapor. There are few sites on Earth that are 
dry enough to allow for significant transmission of submillimeter
light in a series of narrow wavebands; Maunakea, The Atacama Desert, and Antarctica are among them. 
The Submillimetre Common 
User Bolometer Array 2 (SCUBA-2; see \citealt{holland2013} for details), 
installed at the \mbox{15 m} James Clerk Maxwell Telescope (JCMT) in 2009 
capitalises on two such wavebands centred on 450 and 
\mbox{850 $\mu$m}. Situated at an altitude of 4,092 meters near 
the summit of Maunakea, this continuum imager consists of two 
focal planes, each comprising 5,120 Transition Edge Sensors 
(TES) allowing data to be obtained at both wavelengths 
simultaneously. The half-power bandwidths of the bandpass filters are 32 and \mbox{85 $\mu$m} at 450 and \mbox{850 $\mu$m}, respectively.

SCUBA-2 maps the sky more than 100 times faster 
than its predecessor, SCUBA \citep{holland1999}, which was 
decommissioned in 2005. 
Since its installation at the JCMT, SCUBA-2 has proven to
be a workhorse instrument, providing invaluable data on a variety of fields
including studies focused on the thermal dust emission associated with the earliest stages of star formation \citep[e.g.][]{wardthompson2007,herczeg2017,liu2018}, the mass-loss history of evolved stars \cite[e.g.][]{dharmawardena2019}, the properties of active comets \citep[e.g.][]{coulson2020}, the evolution of nearby galaxies \cite[e.g.][]{wilson2009,saintonge2018,li2020},  number counts of distant, submillimeter galaxies \citep[e.g.][]{geach2013,wang2017,simpson2019,shim2020}, black holes \citep[e.g.][]{tetarenko2017}, and transient phenomena \citep[e.g.][]{mairs2018ATel,smith2018ATel,mairs2019}. SCUBA-2 is also the detector used 
in conjunction with the JCMT's linear polarimeter, POL-2 \citep{Pol2Paper2018} 
to obtain Stokes $I$, $Q$, and $U$ parameters, providing information on interstellar magnetic fields \cite[e.g.][]{wardthompson2017}.
 
Despite the unprecedented collecting power and sensitivity of 
SCUBA-2, the atmospheric opacity still poses an issue for calibration as it is 
variable over short timescales and highly dependent on the 
amount of precipitable water vapor (PWV) present along the line
of sight. In order to robustly calibrate the data, this loss of 
light must be well modeled and empirically understood. To this 
end, high-cadence measurements of the PWV are obtained along the
line of sight during SCUBA-2 observations \citep{DempseyPWV2011} 
in order to calculate and correct for the expected 
atmospheric extinction in the images \citep{chapin2013}. 

In addition to the variable atmosphere, changes in the optical 
path of the instrument such 
as thermal-filter changes, secondary-mirror adjustments, or the 
removal of the JCMT's GORE-TEX\texttrademark$\:$ membrane can also affect the 
final data. SCUBA-2 commissioning took place between 2009 October and 2011 September. Instrument performance and 
commissioning details throughout this time are presented in 
\cite{bintley2010,bintley2012} and \cite{dempsey2012}. \cite{dempsey2013} 
(hereafter, D13) present initial 
Flux Conversion Factors (FCFs; used to 
convert instrumental units to physical flux units) 
that were derived over the 
course of one year between 2011 and 2012 by observing 
standard sources, primarily Uranus. They found 
peak-flux calibration uncertainties of $\sim14\%$ and $\sim5\%$ at
450 and \mbox{850 $\mu$m}, respectively. 
After nearly a decade of operation, thousands of calibrator 
observations have been obtained, increasing D13's sample size
 by an order of magnitude. This dataset enables the 
 robust study of  
trends in fluxes as a function of transmission and time of 
observation. Additionally, the influence of 
significant changes to the optical path resulting from hardware 
maintenance and improvement throughout SCUBA-2's history
can be addressed.   

In the global context of astronomy undertaken at $\sim$450 and \mbox{$\sim$850 $\mu$m}, SCUBA-2's flux calibration performance is most naturally compared to other CCD-style bolometer arrays used by ground-based, single-dish telescopes. The first bolometer-array instrument was the Submillimeter High Angular Resolution Camera II (SHARC-II; \citealt{SHARCII}) at the Caltech Submillimetre Observatory, situated approximately 150 meters from the JCMT near the summit of Maunakea. Operating at 350 and \mbox{450 $\mu$m}, the flux-calibration uncertainty is quoted to be $\sim$30\% at both wavelengths \citep[e.g.][]{marsh2006}. SCUBA-2's \mbox{850 $\mu$m} band can be most directly compared to the Atacama Pathfinder Experiment's (APEX) \mbox{870 $\mu$m} bolometer array known as the Large Apex BOlometer CAmera (LABOCA) \citep{LABOCA}. The flux-calibration uncertainty for this instrument has been determined to be $\sim$15\%, \citep[e.g.][]{schuller2009}. The Green Bank Telescope's new Multiplexed Squid TES Array at Ninety Gigahertz-2 (MUSTANG-2; \citealt{MUSTANG2}) operates at $\sim3.3\mathrm{\:mm}$. While more distant in terms of wavelength, this instrument still provides a useful point of comparison. Its quoted flux uncertainty is $\sim10\%$ \citep[e.g.][]{ginsburg2020}, an improvement over the previous MUSTANG array \citep{MUSTANG2009}, which had a systematic flux uncertainty of $\sim20\%$ \citep[e.g.][]{hughes2012}. Most recently, beginning with The Next Generation Balloon-borne Large Aperture Submillimeter Telescope (BLAST-TNG, \citealt{BLASTTNGKIDs}), arrays of Kinetic Inductance Detectors (KIDs) are now being deployed for submillimeter/millimeter observations. At the time of writing, the most direct comparison between a KID array and SCUBA-2 can be drawn between the Institut de Radioastronomie Millim\'{e}trique (IRAM) \mbox{30 m} telescope's ``New IRAM Kids Arrays 2'' (NIKA2, \citealt{NIKA2}). NIKA2 is a dual-band millimetre continuum camera of operating at $\sim$1 and 2 mm wavelengths. The initial rms calibration uncertainty is found to be 6\%, based on 264 scans of point-like sources with flux density $>1\mathrm{\:Jy}$ at 1mm \citep{perotto2020}, the waveband closest to SCUBA-2's longest wavelength. Future comparisons with the Large Millimeter Telescope's (LMT) TolTec \citep{TolTec} KID instrument will also be performed.

In this paper, we present a historical record of the SCUBA-2 FCFs and their uncertainties from 
2011 May 1 (beginning of on-sky commissioning dataset) to 
2021 February 10, updating the
 calibration factors to be used when analyzing both 
archival or new SCUBA-2 data. In Section~\ref{sec:Obs}, we 
present a summary of the observations and data-reduction methods
employed throughout this analysis. In Section~\ref{sec:OR}, we 
discuss the conversion of the monitored opacity of the 
atmosphere at 225 GHz ($\tau_{225}$) to the opacity at the 
wavelengths to which SCUBA-2 is sensitive in order to correct 
for telluric extinction. In Section~\ref{sec:FCFs} we report the
recommended FCF values to use for the calibration of 
SCUBA-2 data obtained during the stable part of the night (07:00--17:00 UTC, 21:00--07:00 HST)
depending on the date the data were observed. In Section~\ref{sec:FCFsnightly},
we present corrections to the 
standard FCFs for observations obtained outside the stable portion of the night, 
when the telescope dish is unstable to contraction and expansion due to changing 
temperature gradients in the support beams with the
setting and rising of the Sun. In Section~\ref{sec:SecCals}, we update
the submillimeter fluxes of the most frequently used secondary calibrators, 
\mbox{CRL 2688}, \mbox{CRL 618}, \mbox{Arp 220}, and \mbox{HL Tau}
based on the new calibration results. 
In Section~\ref{sec:CaseStudy} we apply the new
results to nearly a decade of \mbox{850 $\mu$m} observations of Quasar \mbox{3C 84} and compare
the light curve with data obtained by the Submillimeter Array (SMA) and the 
Atacama Large Millimeter/Submillimeter Array (ALMA) as a case study.
Finally, in Section~\ref{sec:Summary} we provide a summary.

%%%%%%%%%%%%%%%%%%%%%%%%%%%%%%%%%%%%%%%%
%%%%%%%%%%%%%%%%%%%%%%%%%%%%%%%%%%%%%%%%
%%%%%%%%%%%%%%%%%%%%%%%%%%%%%%%%%%%%%%%%
\section{Observations and Data Reduction} 
\label{sec:Obs}
%%%%%%%%%%%%%%%%%%%%%%%%%%%%%%%%%%%%%%%%
%%%%%%%%%%%%%%%%%%%%%%%%%%%%%%%%%%%%%%%%
%%%%%%%%%%%%%%%%%%%%%%%%%%%%%%%%%%%%%%%%

SCUBA-2 obtains continuum data at 450 and \mbox{850 $\mu$m} 
simultaneously with beam sizes of 10.0$^{\prime\prime}$ and 14.4$^{\prime\prime}$, respectively (see D13, \citealt{chapin2013}, and Section \ref{subsec:Beam} for more information). All observations that are used in this work 
were performed using the ``CV Daisy'' scan pattern which yields a 
$\sim12\arcmin$ diameter map with a central 3$\arcmin$ region 
exhibiting a highly uniform 
sensitivity \citep[see][for details]{holland2013}. This scan 
pattern is ideal for recovering compact structures and it is used for both 
flux calibrator and science observations. 

All the data were reduced using the iterative map-making 
software, {\sc{makemap}} \citep{chapin2013}, which 
is part of Starlink's \citep{currie2014} Submillimetre 
User Reduction Facility ({\sc{smurf}}) package 
\citep{jenness2013}. The default recipe for bright, compact 
sources was employed in order to robustly recover 
each calibrator. In order to achieve this, structures larger 
than 3.3$\arcmin$ are spatially filtered and low-frequency noise
is removed. To suppress the ringing introduced by such a high-pass filter, 
the map is constrained to zero beyond 
a radius of 1$\arcmin$ from the source centre 
(significantly outside the power received by the beam's central lobe) on all iterations 
but the last. A pixel size of 1$\arcsec$ 
was chosen for both the 450 and \mbox{850 $\mu$m} final maps. 

The first SCUBA-2 calibrator observations addressed by 
this study were performed in 2011 May during the instrument's 
commissioning phase. D13 analyzed the 
calibrator data taken over one year beginning on 2011 May 1 to derive 
the initial ``Flux Conversion Factors'' (FCFs; see
Section~\ref{sec:FCFs} for details), beam profiles, and secondary-calibrator
brightnesses. Now, we continue the analysis through 
2021 February 10.

Over the course of 10 years, more than 1200 reliable observations of the primary calibrator, Uranus, alone have been obtained. Together with the secondary calibrators, this is a sufficient amount of data to probe changes in the measured calibrator 
fluxes as a function of atmospheric transmission, UTC date, and 
UTC time (hour of night). The UTC-date analysis reveals significant 
changes to the measured fluxes due 
to changes in hardware, while the UTC time highlights 
patterns in dish settling and expansion with the ambient temperature. 
Early in the evening and after the Sun has risen in the 
morning, the JCMT dish experiences significant temperature gradients 
that affect its shape and, consequently, the 
calibration (see Section~\ref{sec:FCFsnightly} for more details). 

Only observations that achieved a
signal-to-noise ratio (SNR) of at least 4 were included 
in the analysis to ensure 
an accurate fit could be made to peak flux of the  
calibrator source. 
The \mbox{450 $\mu$m} data are much more sensitive to wet weather 
than their \mbox{850 $\mu$m} counterparts, so there are 
fewer usable observations at \mbox{450 $\mu$m} in these conditions. 
``Evening'' observations, when dish settling is expected, takes 
place between 03:00 and 07:00 UTC (17:00 and 21:00 
Hawaii-Aleutian Standard Time, HST). ``Night'' observations, 
when observations are expected to be most stable, take place 
between 07:00 and 17:00 UTC (21:00 and 07:00 HST). ``Morning'' 
observations, when dish expansion is expected, are typically 
undertaken during  ``Extended Observing'' time, 17:00 to 22:00 
UTC (07:00 to 12:00, HST). Extended Observing began in October, 
2013. Even before all JCMT observations were carried out remotely 
beginning 2019 November 1 \citep{Parsonsremoteops}, the telescope was always operated remotely 
from sea level during Extended Observing with the ``handover'' 
taking place between 16:00 and 18:00 UTC. 

%%%%%%%%%%%%%%%%%%%%%%%%%%%%%%%%%%%%%%%%
%%%%%%%%%%%%%%%%%%%%%%%%%%%%%%%%%%%%%%%%
%%%%%%%%%%%%%%%%%%%%%%%%%%%%%%%%%%%%%%%%
\section{Extinction Correction: Opacity Relations}
\label{sec:OR}
%%%%%%%%%%%%%%%%%%%%%%%%%%%%%%%%%%%%%%%%
%%%%%%%%%%%%%%%%%%%%%%%%%%%%%%%%%%%%%%%%
%%%%%%%%%%%%%%%%%%%%%%%%%%%%%%%%%%%%%%%%

% Opacity Relation Raw versus Transmission
\begin{figure*}
\plottwo{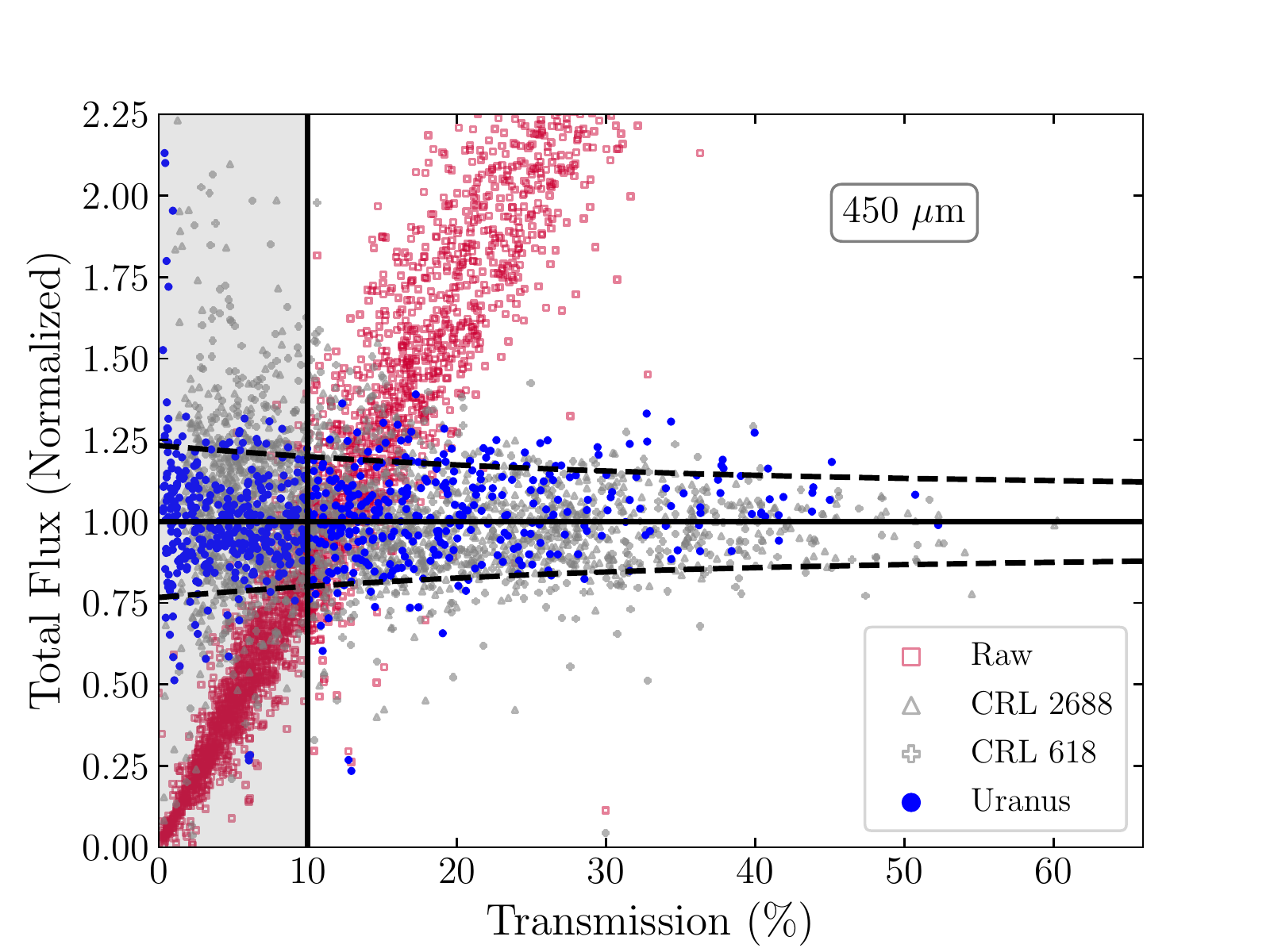}{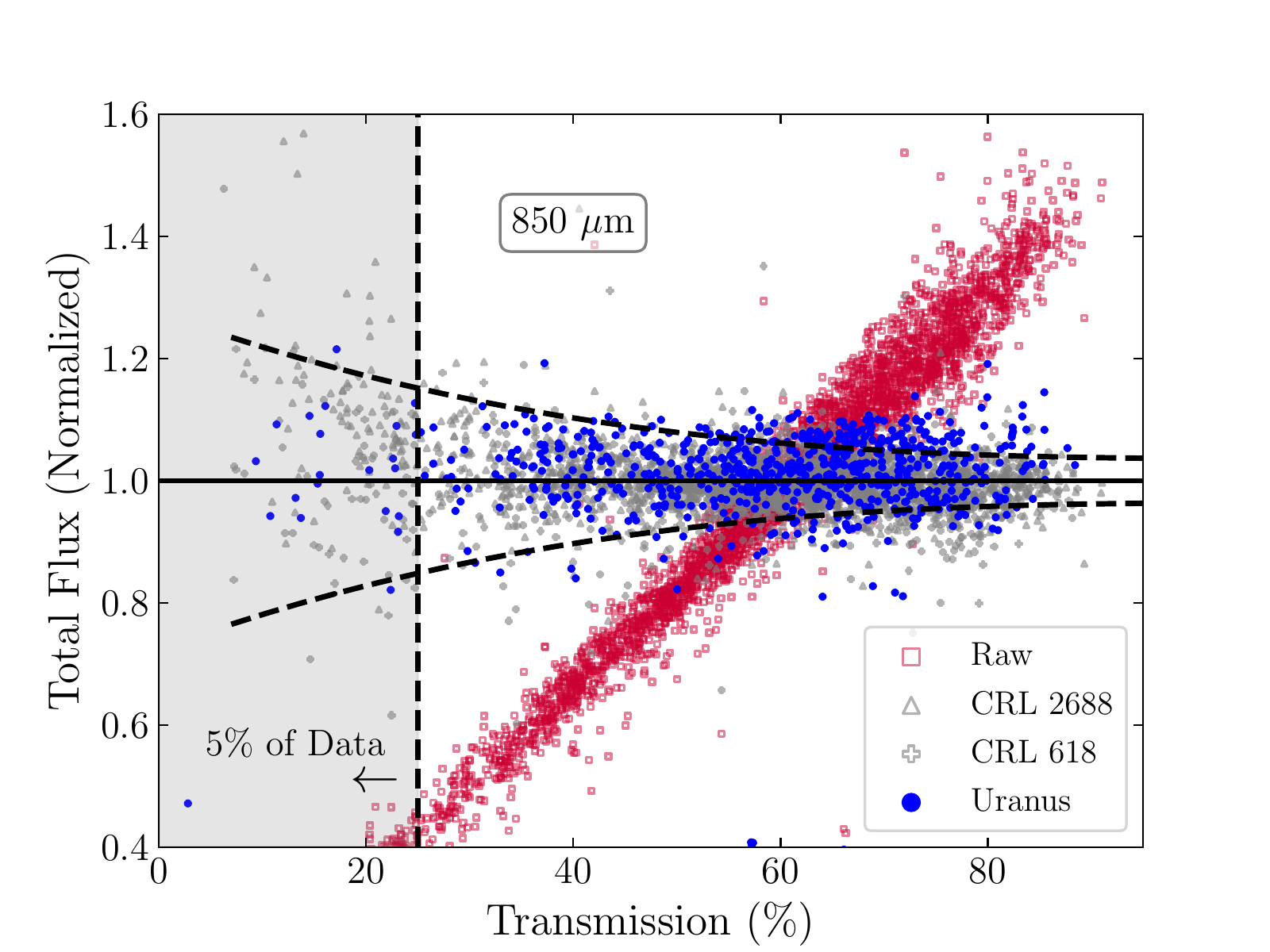}
\caption{Normalized raw (no atmospheric extinction correction applied) total-flux measurements of Uranus, \mbox{CRL 2688}, and \mbox{CRL 618} as a function of transmission are shown as red squares. The atmospheric-extinction corrected peak fluxes are overlaid with the primary calibrator, Uranus, shown in blue (circles) and the secondary calibrators \mbox{CRL 2688} and \mbox{CRL 618} shown in gray (triangles and pluses, respectively). The applied extinction correction uses the opacity relations of the form presented in Equation~\ref{eq:ORForm} with coefficients detailed in Table \ref{tab:OR}. Dashed lines represent the measured median absolute deviation in the data in ten evenly spaced transmission bins. The corrected calibrator fluxes are approximately flat across the range of atmospheric transmissions at which the JCMT operates, correcting for the increased signal attenuation seen in the raw data at higher transmissions. \textit{Left:} \mbox{450 $\mu$m}. The vertical dashed line represents a transmission of 10\%, below which a negligible amount of \mbox{450 $\mu$m}  data can be used due to low signal to noise (grayed region). \textit{Right:} \mbox{850 $\mu$m}. The vertical dashed line represents a transmission of 25\%, marking the 5\% of observed data that begins to have a significantly larger scatter than their higher transmission counterparts (grayed region).}
\label{fig:RawCorFluxOR}
\end{figure*}

The default data-reduction scheme calculates an atmospheric 
extinction correction for 
each observation based on the airmass and the amount of 
precipitable water vapor (PWV) 
along the line of sight (the opacity of the atmosphere 
is used as a proxy for the PWV). The correction assumes an initial 
signal, $I_{0}$, is exponentially attenuated as follows,
\begin{equation}
\label{eq:EXTCOR}
    I_{0} = \frac{I_{\mathrm{m}}}{\mathrm{exp}(-\tau_{\nu,\mathrm{zen}} A)},
\end{equation}
where $I_{\mathrm{m}}$ is the measured signal, $\tau_{\nu,\mathrm{zen}}$ 
is the zenith opacity of the atmosphere at the observed 
frequency $\nu$, and $A$ is the airmass of the source. Equation~\ref{eq:EXTCOR}
assumes a plane-parallel atmosphere and an opacity that is independent of the frequency across the filter profile.
The expression in the denominator:
\begin{equation}
\label{eq:transmissionexpression}
\mathrm{exp}(-\tau_{\nu,\mathrm{zen}} A)
\end{equation}
is defined as the \textit{atmospheric transmission} (hereafter, \textit{transmission}). 
Therefore, the derivation of an accurate correction depends 
on how closely the atmosphere resembles the plane-parallel 
assumption and how well $\tau_{\nu,\mathrm{zen}}$ can be measured.

A water-vapor monitor (WVM) operating at \mbox{183 GHz} is 
mounted in the JCMT receiver cabin, obtaining measurements 
every 1.2 seconds via a pick-off mirror. The WVM measures three
brightness temperatures at frequencies sampling the wings of the
prominent \mbox{183 GHz} water line. Fitting the properties of
the water line allows for a conversion to the amount of PWV 
that is present along the line of sight during an observation \citep[see][for details on the WVM]{Wiedner1998}. 
The measured PWV is converted to its 
zenith value, $\mathrm{PWV}_{\mathrm{zen}}$, by dividing by the 
airmass, so it can be directly related to the zenith opacity at 
\mbox{225 GHz}, $\tau_{225,\mathrm{zen}}$ (D13),
\begin{equation}
\label{eq:PWVTau225}
    \tau_{225,\mathrm{zen}} = 0.04\times \mathrm{PWV}_{\mathrm{zen}} + 0.017.
\end{equation}
The reason for this conversion is historical: it allows direct comparison with 
the fixed-azimuth 
\mbox{225 GHz} tipping radiometer formerly 
situated at the Caltech Submillimeter Observatory (CSO). In 2015, 
however, the CSO 
radiometer was hit by a falling piece of ice, which had the 
effect of shifting the measured opacity to generally lower 
values than measured at the JCMT with no consistent systematic 
trend\footnote{See The Submillimeter Array Memo 164, Radford, S. \url{https://lweb.cfa.harvard.edu/sma/memos/164.pdf}}. In 2015 October, it was moved
from the CSO site (150 meters southeast of the JCMT) to the 
Submillimeter Array site (100 meters northwest of the JCMT). 

In order to perform a reliable extinction correction, a 
conversion must be made from $\tau_{225,\mathrm{zen}}$ to 
$\tau_{666,\mathrm{zen}}$ and $\tau_{345,\mathrm{zen}}$, 
corresponding to the observing
frequencies of SCUBA-2 (\mbox{450 $\mu$m  = 666 GHz} and 
\mbox{850 $\mu$m = 345 GHz}). D13 report 
opacity relations of
\begin{equation}
\label{eq:nom450OR}
    \tau_{666,\mathrm{zen}} = 26.0\times(\tau_{225,\mathrm{zen}} - 0.012)
\end{equation}
\begin{equation}
\label{eq:nom850OR}
    \tau_{345,\mathrm{zen}} = 4.6\times(\tau_{225,\mathrm{zen}} - 0.0043)
\end{equation}
based on one year of data obtained beginning in 2011 May. 

% Opacity Relations - Compared to D13
\begin{figure*}
\plottwo{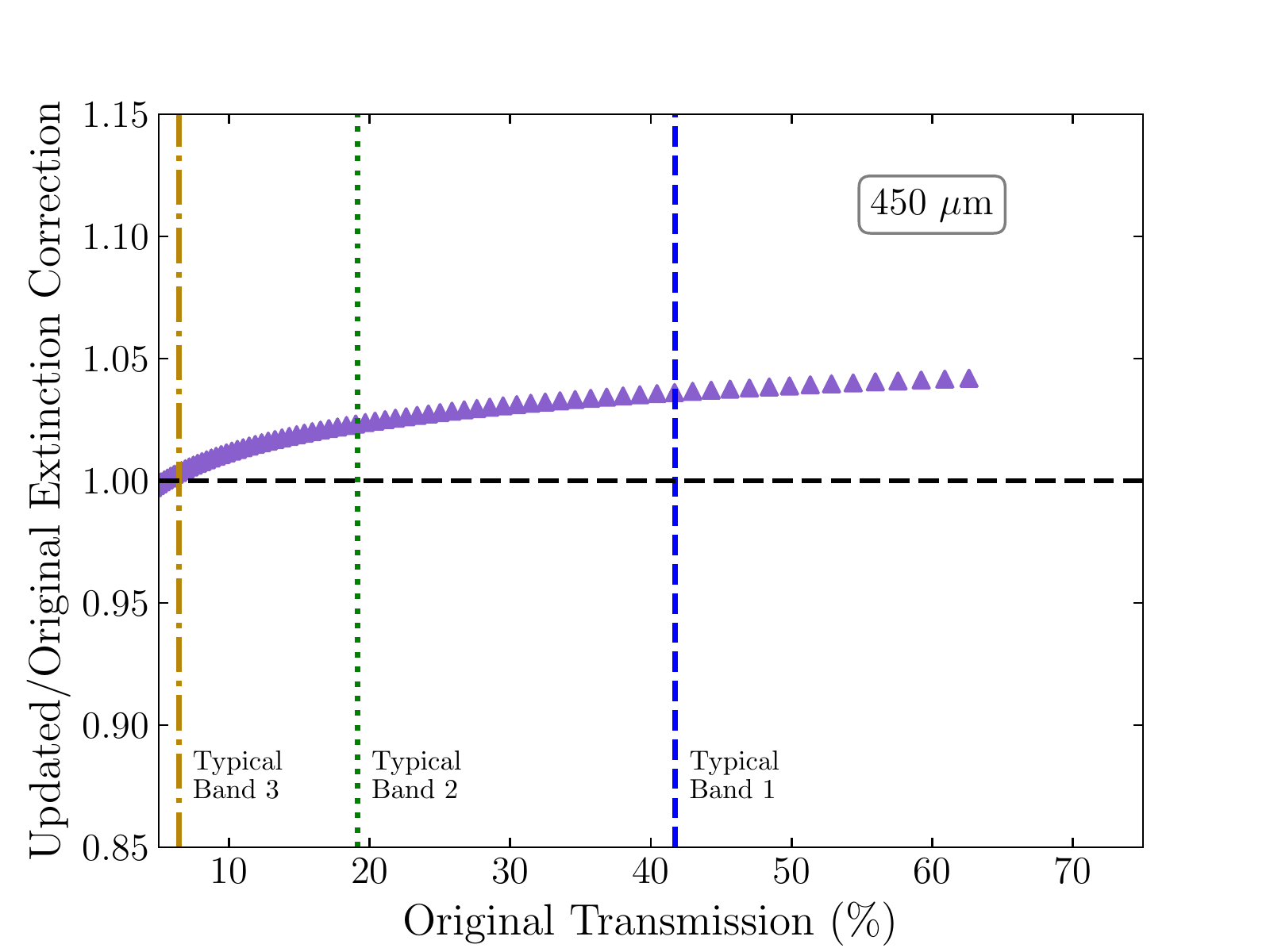}{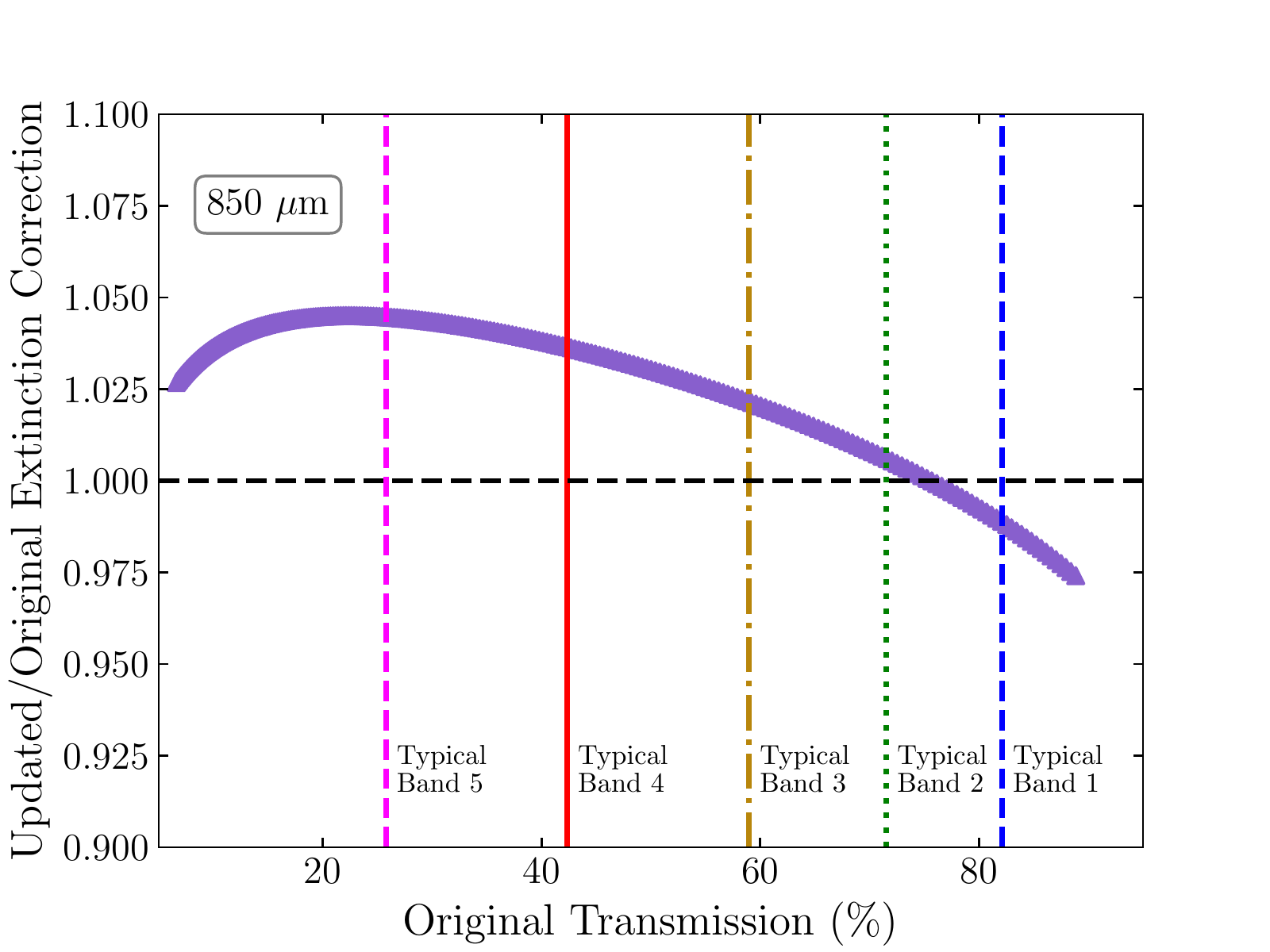}
\caption{The newly derived extinction correction (the reciprocal of Equation~\ref{eq:transmissionexpression} using the opacity relations presented in Equation~\ref{eq:ORForm} with coefficients presented in Table \ref{tab:OR}) divided by the original extinction correction (the reciprocal of Equation~\ref{eq:transmissionexpression} using opacity relations presented in Equations~\ref{eq:nom450OR} and \ref{eq:nom850OR}) as a function of atmospheric transmission.  Vertical lines show the transmission of the typical JCMT weather bands at each wavelength, assuming an airmass of 1.2. \textit{Left:} \mbox{450 $\mu$m}. \textit{Right:} \mbox{850 $\mu$m}. At \mbox{450 $\mu$m}, the original opacity relation is modified by a maximum factor of 5\% in very dry weather while at \mbox{850 $\mu$m} the original opacity relation is modified by a maximum factor of 3--4\% in very dry or very wet conditions. The majority of \mbox{850 $\mu$m} observations, however, require no modification from the original results presented by D13.}
\label{fig:ORCompare}
\end{figure*}

% Opacity Relation Parameter Table
\begin{deluxetable}{ccc}
\tablecaption{Opacity relation coefficients corresponding to Equation~\ref{eq:ORForm}.}
\label{tab:OR}
\tablecolumns{3}
\tablewidth{0pt}
\tablehead{
\colhead{Coefficient} &
\colhead{450 $\mu$m}  &
\colhead{850 $\mu$m}
}
\startdata
$a$ & 23.3 $\pm$ $1.5$ & 3.71 $\pm$ $0.18$ \\
$b$ & -0.018 $\pm$ $0.006$ & -0.040 $\pm$ $0.008$ \\
$c$ & 0.05 $\pm$ $0.04$ & 0.202 $\pm$ $0.044$ \\
\enddata
\end{deluxetable}

To test these opacity relations on the larger 10-year data set,
a more general form of the opacity relation was defined to include a 
higher-order term, expecting a deviation from linearity. It was found that 
a square-root dependence produces a marginally flatter flux response as 
a function of transmission when compared with Equations~\ref{eq:nom450OR}
and \ref{eq:nom850OR}, especially when the atmosphere is highly opaque. 
The form of the updated opacity relation is
\begin{equation}
\label{eq:ORForm}
\tau_{\nu,\mathrm{zen}} = a\times(\tau_{225,\mathrm{zen}} + b + c\times\sqrt{\tau_{225,\mathrm{zen}}}),
\end{equation}
where $a$, $b$, and $c$ are fitted parameters. 
Using the raw, \textit{uncorrected} 
peak flux measurements ($I_\mathrm{m}$, Equation~\ref{eq:EXTCOR})
of the three most-used calibrator sources: 
\mbox{Uranus}, \mbox{CRL 2688}, and \mbox{CRL 618}, 
along with the airmass at the time of 
each observation, the optimal parameters ($a$, $b$, $c$) 
were derived by minimizing the variance in the normalized\footnote{The measured peak fluxes of Uranus 
were normalised to the varying, expected brightness of the 
planet on each observation date to take into account seasonal flux changes.} 
\textit{corrected} peak flux ($I_{0}$, Equation~\ref{eq:EXTCOR}) 
versus transmission curve. The minimization routine used was 
the Broyden-Fletcher-Goldfarb-Shanno (BFGS) algorithm, 
employed by {\sc{scipy}}'s {\sc{optimize.minimize}} 
function in the Python programming language\footnote{\url{https://docs.scipy.org/doc/scipy/reference/generated/scipy.optimize.minimize.html}}. 
The final parameters and associated uncertainties are 
derived by bootstrapping the minimization over 1000 iterations, during which random perturbations are added to the residual between the 
normalised corrected flux and a value of 1.0. The perturbations were 
generated by drawing random values from a normal 
distribution with a standard deviation equivalent to 
the standard deviation of the original set of 
residuals.  
Before the minimum variance was calculated, the flux values 
were averaged over $\tau_{225}$ bins of width 0.01 and airmass bins of 0.1. 
The binning was performed in order to mitigate the effect of 
highly sampled areas of parameter space dominating 
less-sampled areas, causing poor fits for conditions in 
which fewer observations are performed. In this way, the 
flux values used to fit $a$, $b$, and $c$ represent the best knowledge of 
the behavior of each particular region of the parameter space. Times 
during which the JCMT WVM was behaving abnormally, the 
protective GORE-TEX\texttrademark$\:$ membrane\footnote{In order to shield the telescope and sensitive electronics from wind and dirt,  the JCMT is covered by a GORE-TEX\texttrademark$\:$ membrane that is transparent at submillimeter/millimeter wavelengths.} was removed from the telescope, or the 
secondary mirror was malfunctioning were not considered in the analysis (see Section~\ref{sec:FCFs}). 

% Weather PWV Table
\begin{deluxetable}{ccc}
\tablecaption{Weather band to PWV conversion.}
\label{tab:Weather}
\tablecolumns{3}
%\tablenum{1}
\tablewidth{0pt}
\tablehead{
\colhead{Weather Band$^{a}$} &
\colhead{$\tau_{225,zen}^{b}$}  &
\colhead{PWV$^{c}$ (mm)}
}
\startdata
1 (Typical) & 0.04  & 0.58 \\
1/2 Boundary & 0.05 & 0.83 \\
2/3 Boundary & 0.08 & 1.58 \\
3/4 Boundary & 0.12 & 2.58 \\
4/5 Boundary & 0.20 & 4.58 \\
\enddata
\tablecomments{$^{a}$The JCMT classifies 5 different ``weather bands'' according to the amount of precipitable water vapor in the atmosphere.\\$^{b}$The zenith opacity of the atmosphere at \mbox{225 GHz} (see Equation \ref{eq:PWVTau225}).\\$^{c}$Precipitable water vapor.} 
\end{deluxetable}

Throughout the history of SCUBA-2, the opacity relations 
at both 450 and \mbox{850 $\mu$m} remain consistent with 
($a$, $b$, $c$) coefficients corresponding to Equation~\ref{eq:ORForm}
presented in Table~\ref{tab:OR}.

Figure~\ref{fig:RawCorFluxOR} shows the normalised 
raw flux measurements of Uranus, \mbox{CRL 2688}, and \mbox{CRL 618} as a function of transmission 
(red squares) overlaid by the corrected fluxes 
after applying the extinction correction that is 
calculated using these new opacity relations. 
The result is a consistent flux response across the 
full range of transmissions typically used during 
observations. Vertical, dashed lines at 10\% and 25\% for 
450 and \mbox{850 $\mu$m} data, respectively, highlight 
transmissions below which the opacity relations become
significantly more uncertain. In the case of the
\mbox{850 $\mu$m} relation in the very poor transmission
regime, the secondary calibrators show evidence of an 
over-correction (see also Appendix~\ref{appsec:SingleObs}). 
At \mbox{450 $\mu$m},  a transmission of 
$\leq10\%$ occurs when observing at an elevation at or 
below 42$^{\circ}$ ($\mathrm{Airmass} = 1.5$) in typical JCMT 
Weather Band 2 conditions ($\tau_{225,\mathrm{zen}} = 0.065$; 
see Table~\ref{tab:Weather}). 
In the case of \mbox{850 $\mu$m}, a transmission of 25\% occurs 
when observing at an elevation of 
42$^{\circ}$ ($\mathrm{Airmass} = 1.5$) at the border of 
Weather Band 4/5 ($\tau_{225,\mathrm{zen}}$ = 0.2), or when 
observing at an elevation of 
30$^{\circ}$ ($\mathrm{Airmass} = 2.0$) at a 
Weather Band 4 opacity of 
$\tau_{225,\mathrm{zen}}$ = 0.15. In Weather Bands 1, 2, or 3, 
the \mbox{850 $\mu$m} transmission is always higher than 25\% (less than 5\% of SCUBA-2 science data is obtained below this threshold). 

In Figure~\ref{fig:ORCompare}, the extinction correction 
using the newly derived opacity relations (Equation~\ref{eq:ORForm}) is directly 
compared with the 
original extinction correction using the opacity relations derived in 
D13 across the observable transmission range. 
As expected, the \mbox{450 $\mu$m} data have a high enough 
transmission to be analyzed only in Weather Bands 1 and 2, while wetter 
weather quickly renders peak-flux measurements less certain. The new extinction correction produces 
 fluxes that are 2--5\% higher than the original factor yields depending on
 the transmission, with the largest factor affecting the driest weather observations, 
 which were previously under-corrected.
 At \mbox{850 $\mu$m} the peak-flux measurements are robust in all 
 weather bands. The new extinction correction modifies the 
 \mbox{850 $\mu$m} flux by a maximum of 3--4\% at atmospheric 
 transmissions above 25\%. The largest differences to data occur in the extremes of the wettest and driest weather which were previously under and over-corrected, respectively. The majority of \mbox{850 $\mu$m} data has a negligible change ($<3\%$) in corrected flux when using the new opacity relations.

%%%%%%%%%%%%%%%%%%%%%%%%%%%%%%%%%%%%%%%%
%%%%%%%%%%%%%%%%%%%%%%%%%%%%%%%%%%%%%%%%
%%%%%%%%%%%%%%%%%%%%%%%%%%%%%%%%%%%%%%%%
\section{Flux Conversion Factors (FCFs)}
\label{sec:FCFs}
%%%%%%%%%%%%%%%%%%%%%%%%%%%%%%%%%%%%%%%%
%%%%%%%%%%%%%%%%%%%%%%%%%%%%%%%%%%%%%%%%
%%%%%%%%%%%%%%%%%%%%%%%%%%%%%%%%%%%%%%%%

After flat-fielding, raw data obtained by \mbox{SCUBA-2} 
initially have units of picowatts (pW; see \citealt{chapin2013}). After an 
atmospheric-extinction correction is applied (see Section~\ref{sec:OR}), 
a conversion must be made to flux units. 
There are two types of ``Flux-Conversion Factors'' (FCFs) 
that can be applied depending on a project's scientific motivation.
\begin{enumerate}
    \item Peak FCF (FCF$_{\mathrm{peak}}$): Also known as ``Beam FCF'', this factor
    converts every pixel in a map from pW into \mbox{Jy beam$^{-1}$} (with units of \mbox{Jy beam$^{-1}$ pW$^{-1}$}). 
    This is useful when peak fluxes of discrete point-source objects are being 
    measured. The brightest pixel in the source represents the flux in Jy 
    distributed over the full area of the beam. To calculate the Peak FCF, we 
    fit the calibrator source with a two-dimensional Gaussian function using 
    {\sc{cupid}}'s \citep{CUPID2013} {\sc{gaussclumps}} \citep{stutzki1990} program, found in the Starlink software suite \citep{currie2014}, to measure the raw peak brightness, $I_{\mathrm{m,peak}}$ (in units of pW). We then compare the result with the modeled peak flux of the source, $S_{\mathrm{peak}}$, (in units of \mbox{Jy beam$^{-1}$}),
    
    \begin{equation}
        \label{eq:FCFP}
        \mathrm{FCF}_{\mathrm{peak}} = \frac{S_{\mathrm{peak}}}{I_{\mathrm{m,peak}}}.
    \end{equation}

    \item Arcsecond FCF (FCF$_{\mathrm{arcsec}}$): Also known as the ``Aperture FCF'', this factor converts every pixel in a map from pW to  Jy arcsec$^{-2}$ 
    (with units of Jy arcsec$^{-2}$ pW$^{-1}$). This conversion factor is necessary when 
    measuring the total flux of extended emission structures. To calculate the 
    Arcsecond FCF, we first measure the total flux in pW, 
    $I_{\mathrm{m,total}}$, by integrating over a 60$\arcsec$ diameter aperture 
    centered on the calibrator source and subtracting a background level averaged 
    over an annulus with inner radius 90$\arcsec$ and outer radius 120$\arcsec$. We 
    then compare the result with the modeled total flux of the calibrator source, 
    $S_{\mathrm{total}}$ (in units of Jy), dividing by the area of a pixel, 
    $A_{\mathrm{pix}}$, in units of arcsec$^{2}$,
    \begin{equation}
        \label{eq:FCFA}
        \mathrm{FCF}_{\mathrm{arcsec}} = \frac{S_{\mathrm{total}}}{I_{\mathrm{m,total}} \times A_{\mathrm{pix}}}
    \end{equation}
    Note that the pixel area of the calibrator maps is \mbox{1 arcsec$^{2}$}. Additional tests have been performed at different pixel 
    sizes and the resulting FCF$_{\mathrm{arcsec}}$ value remains consistent for maps with pixels of side length between 1--4$\arcsec$.
\end{enumerate}
For a true point source, the measured peak pixel of a source in a map calibrated in 
units of Jy beam$^{-1}$ is equivalent to the integrated total flux of the same source in a map calibrated in units of Jy (correcting for the pixel size in arcseconds). 

In the following section, the FCF distributions and 
the associated uncertainties of the primary calibrator 
source Uranus are analyzed as a function 
of date to determine significant historical impacts on 
SCUBA-2's optical path. The recommended FCFs
to use on data obtained during the stable part of the night 
(07:00--17:00 UTC) are presented in Table~\ref{tab:3EpochFCFs} 
and corrections to these values for data obtained outside 
of these times are given in Section~\ref{sec:FCFsnightly} and Table~\ref{tab:NightlyFCFMods}. 

An additional correction is required when applying a matched-filter\footnote{See Appendix D of the SCUBA-2 Data Reduction Cookbook for detailed information: \url{https://starlink.eao.hawaii.edu/docs/sc21.pdf}} to data. A matched filter is applied when seeking to optimally identify and characterize sources that are the size of the telescope beam, suppressing residual large-scale noise:

\vspace{2mm}

\textit{When a matched filter is applied to the data, the \textbf{Peak FCFs} presented in Table \ref{tab:3EpochFCFs} must be reduced by 2\%}\footnote{See \url{https://www.eaobservatory.org/jcmt/instrumentation/continuum/scuba-2/calibration/} for current SCUBA-2 calibration information.}.

\vspace{2mm}

Additional correction factors to the FCFs presented in the sections below are also required when the POL-2 instrument is in use. This is largely due to the reduced optical throughput caused by the introduction of a spinning wave-plate. 

\vspace{2mm}

\textit{\textbf{All FCFs} presented in Table \ref{tab:3EpochFCFs}
must be multiplied by 1.96 and 1.35 at \mbox{450 and 850 $\mu$m}, 
respectively, when POL-2 is in use.}

%%%%%%%%%%%%%%%%%%%%%%%%%%%%%%%%%%%%%%%%
%%%%%%%%%%%%%%%%%%%%%%%%%%%%%%%%%%%%%%%%
%%%%%%%%%%%%%%%%%%%%%%%%%%%%%%%%%%%%%%%%
\subsection{FCFs as a Function of Date}
\label{subsec:FCFsLongTerm}
%%%%%%%%%%%%%%%%%%%%%%%%%%%%%%%%%%%%%%%%
%%%%%%%%%%%%%%%%%%%%%%%%%%%%%%%%%%%%%%%%
%%%%%%%%%%%%%%%%%%%%%%%%%%%%%%%%%%%%%%%%

% Main FCF Figure. - Step Function
\begin{figure*}
\plotfour{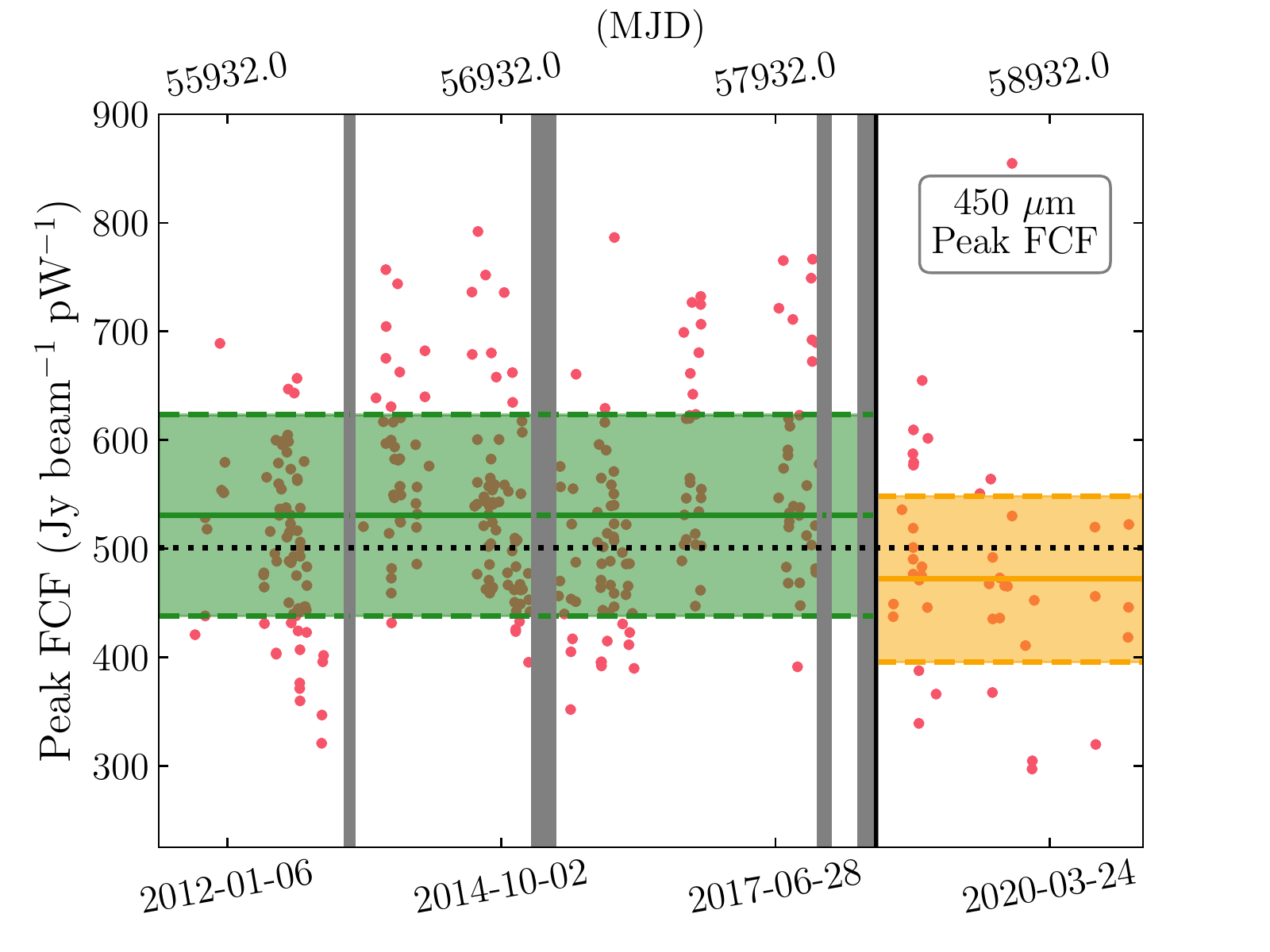}{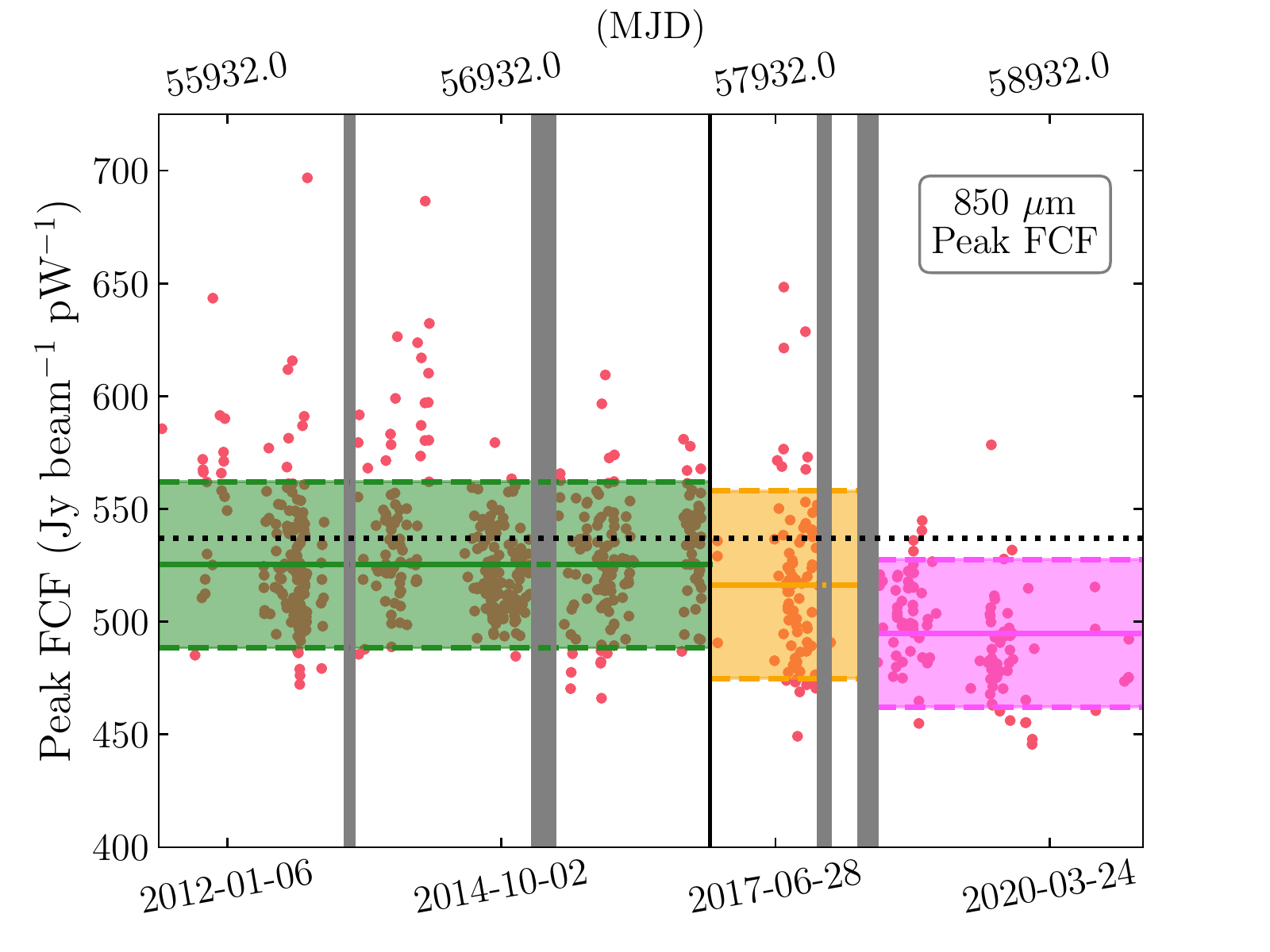}{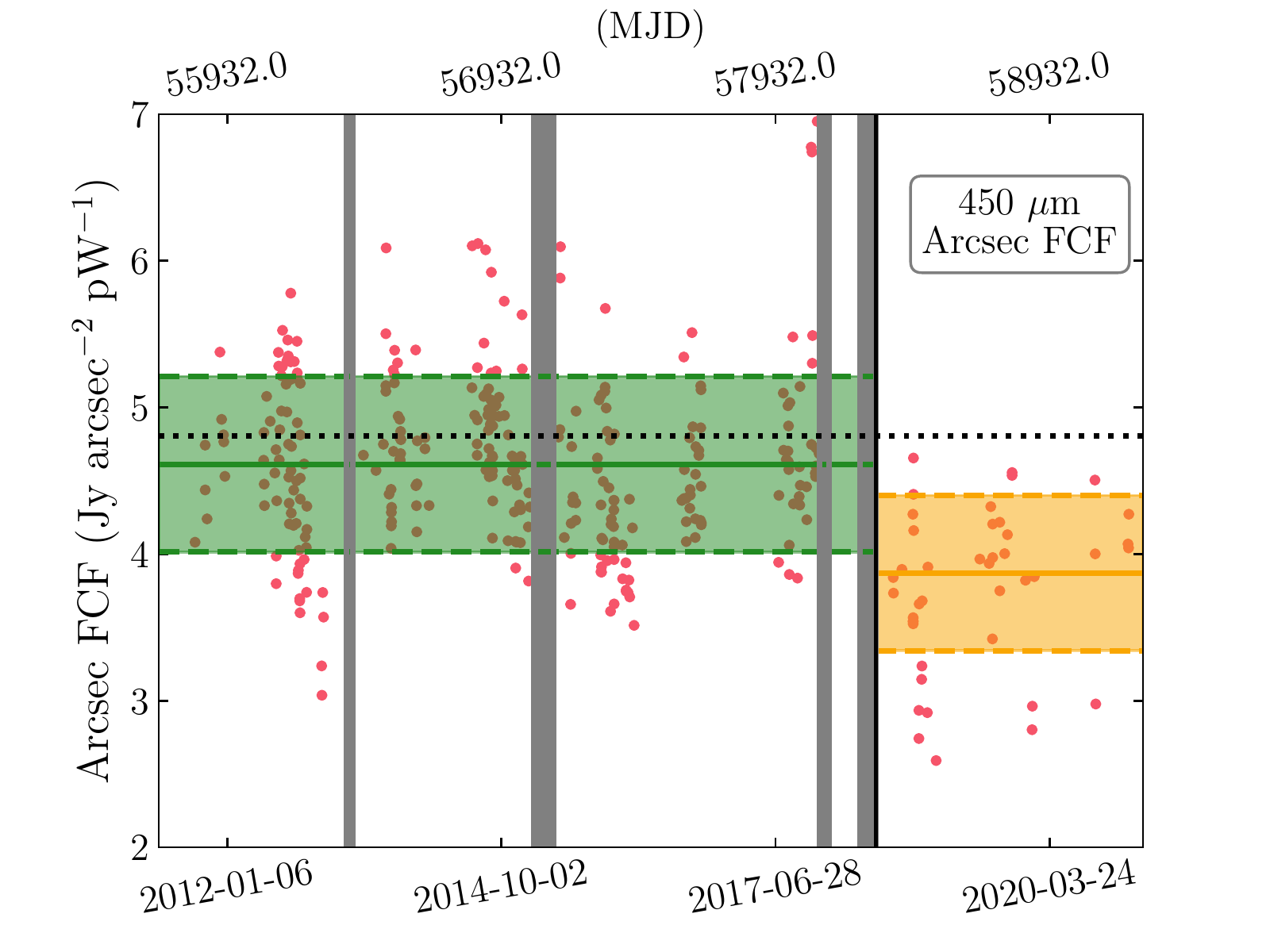}{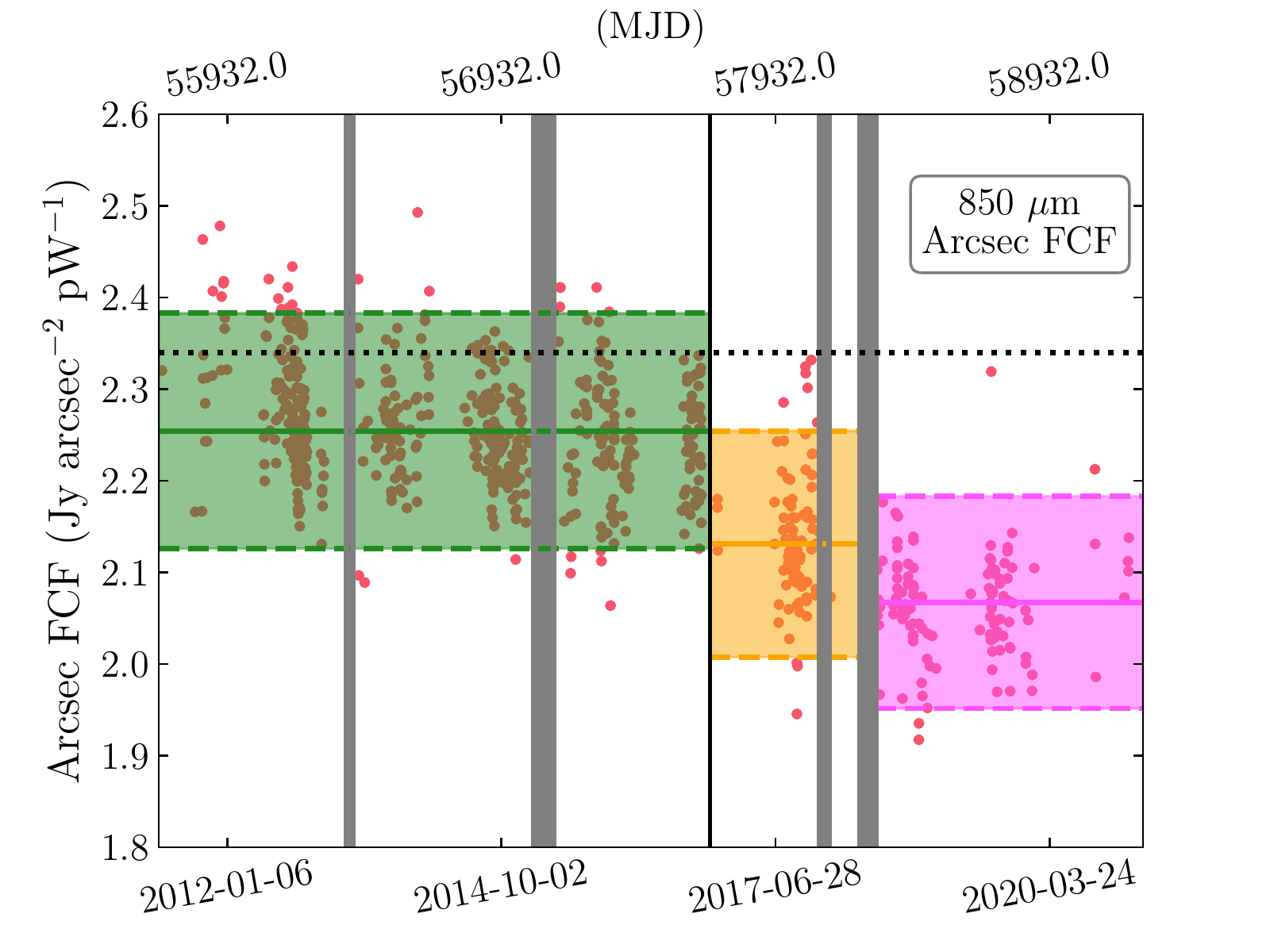}
\caption{Flux conversion factors (FCFs) derived using flux measurements of the primary calibrator Uranus during the stable part of the night (07:00-17:00 UT) as a function of date. The gray shaded regions indicate epochs that are not included in the FCF determinations. The earliest two excluded time intervals indicate when there was no reliable WVM obtaining data from the JCMT. The third interval indicates when the GORE-TEX\texttrademark$\:$ membrane was removed from the telescope for POL-2 commissioning. The latest excluded interval indicates the time when the secondary mirror was malfunctioning (see Section~\ref{subsec:SMUtrouble}). The horizontal, shaded regions indicate the median FCF value over each span of time and the associated median absolute deviation added in quadrature with the 5\% uncertainty in the Uranus flux model. The black (dotted) lines indicate the original FCF value derived by D13, adjusted for the newly derived opacity relation, assuming the most common atmospheric transmissions during observations (see Figure~\ref{fig:ORCompare}). The recommended FCFs to apply to observations obtained in the stable part of the night are summarised in Table~\ref{tab:3EpochFCFs}. \textit{Left:} Peak (\textit{Top}) and Arcsecond (\textit{Bottom}) FCFs derived at \mbox{450 $\mu$m}. The solid, vertical line at the right edge of the latest gray region marks 2018 June 30, when the
secondary-mirror malfunction was fixed. Data wherein the $\mathrm{atmospheric\:transmission}$ are less than $10\%$ are excluded. \textit{Right:} Peak (\textit{Top}) and Arcsecond (\textit{Bottom}) FCFs derived at \mbox{850 $\mu$m}. The solid, vertical line marks 2016 November 19 when the
SCUBA-2 thermal filter stack was updated. Data wherein the $\mathrm{atmospheric\:transmission}$ are less than $25\%$ are excluded.}
\label{fig:FCFmaster}
\end{figure*}

% Table for Uranus Data
\begin{deluxetable*}{ccccccc}
\tablecaption{Uranus-derived FCF Information$^{*}$}
\label{tab:UranusFCF}
\tablecolumns{7}
\tablewidth{0pt}
\tablehead{
\colhead{MJD} &
\colhead{$\tau_{225}^{a}$} &
\colhead{$A^{b}$} &
\colhead{450 $\mu$m Peak FCF$^{c}$} &
\colhead{450 $\mu$m Arcsec FCF$^{d}$} &
\colhead{850 $\mu$m Peak FCF$^{c}$} &
\colhead{850 $\mu$m Arcsec FCF$^{d}$}
\\
\colhead{} &
\colhead{} &
\colhead{} &
\colhead{(Jy pW$^{-1}$ beam$^{-1}$)} &
\colhead{(Jy pW$^{-1}$ arcsec$^{-2}$)} &
\colhead{(Jy pW$^{-1}$ beam$^{-1}$)} &
\colhead{(Jy pW$^{-1}$ arcsec$^{-2}$)}
}
\startdata
... & ... & ... & ... & ... & ... & ... \\
  56088.6892    &   0.066  & 1.085 & 515.80 & 4.91 &   503.46 & 2.25\\
  56102.7249    &   0.054  & 1.066 & 732.73 & 5.51 &   547.90 & 2.35\\
  56103.7297    &   0.054  & 1.077 & 528.94 & 4.92 &   506.83 & 2.27\\
  56105.5995    &   0.041  & 1.221 & 495.46 & 4.55 &   515.29 & 2.25\\
  56105.7215    &   0.043  & 1.073 & 535.79 & 4.88 &   510.93 & 2.27\\
  56109.7449    &   0.048  & 1.153 & 645.76 & 4.53 &   551.56 & 2.30\\
  56109.6842    &   0.050  & 1.048 & 599.89 & 4.71 &   543.25 & 2.30\\
  56110.5168    &   0.043  & 1.859 & 403.92 & 3.80 &   531.33 & 2.27\\
  56110.6040    &   0.046  & 1.151 & 402.80 & 3.99 &   494.46 & 2.22\\
  56112.5498    &   0.046  & 1.408 & 488.38 & 4.36 &   539.04 & 2.31\\
 ... & ... & ... & ... & ... & ... & ... \\
\enddata
\tablecomments{$^{a}$The opacity of the atmosphere at 225 GHz at the time of the observation.\\$^{b}$The airmass at the time of the observation.\\$^{c}$Peak fluxes were measured by performing a Gaussian fit to the source while the data was still in units of picowatts. These measurements were then compared to the expected (model) peak flux of Uranus at the time of observation.\\$^{d}$Total fluxes were measured using aperture photometry while the data was still in units of picowatts. The total flux was calculated within a 1 arcminute diameter aperture centered on the source. The background level was determined using an annulus with inner diameter 1.5 arcminutes and outer diameter 2 arcminutes. These measurements were then compared to the expected (model) total flux of Uranus at the time of observation.\\$^{*}$ The full table is available in the online version.}
\end{deluxetable*}

Figure~\ref{fig:FCFmaster} and Table~\ref{tab:UranusFCF} show the derived FCFs as a function of date using the primary calibrator Uranus\footnote{The full table is available in the online version.}. 
Uranus's flux model is well-known (to within 5\%; \citealt{orton2014}) and it 
produces flat, stable light 
curves when corrected by the opacity relations described above in
Section~\ref{sec:OR}. All observations
included in the figure were obtained in the most stable part of the
night, 07:00--17:00 UTC, in order to avoid effects introduced by 
the changing shape of the dish as it cools in the evening and warms
in the morning (see Section~\ref{sec:FCFsnightly}).
Gray shaded regions, from earliest to latest (left to right in the figure), 
show two intervals during which there was no reliable WVM to indicate the amount of 
water-vapor extinction along the line of sight, one interval during which the 
GORE-TEX\texttrademark$\:$ membrane was removed for POL-2 commissioning,
and one interval during which the Secondary Mirror Unit (SMU) 
was malfunctioning (see Section~\ref{subsec:SMUtrouble}). 

Large outliers in the FCF distributions
are biased to higher values and correspond to focus 
issues or pixel-to-pixel noise that causes the 
Gaussian fit to underestimate the true peak brightness 
of the source. Derived FCFs for each calibrator 
were deemed robust provided a peak flux could be reliably fit using a Gaussian profile.
Transmission lower limits of 10\% and 25\% have also been applied to the 
450 and \mbox{850 $\mu$m} data, respectively (see Section~\ref{sec:OR}).

Table~\ref{tab:3EpochFCFs} presents the recommended FCFs for observations obtained during 
the stable part of the night (07:00-17:00 UTC) as a function of the observation date. The 
FCF value before and after each significant shift in the flux (Figure~\ref{fig:FCFmaster}, see text below for details) 
is represented by the median value of Uranus's FCF distribution. The typical uncertainty, likewise, is represented by the median absolute deviation (MAD) of the FCF distribution added in quadrature with the uncertainty in Uranus's 
flux model ($5\%$). To verify the results, the secondary calibrator source
that is observed most often, 
\mbox{CRL 2688}, was analysed in the same manner and 
there is no significant difference between the derived FCFs 
(see Section~\ref{sec:SecCals} for further information on the secondary calibrators).

The Peak FCF distributions have uncertainties of $17\%$ and $7\%$ at 450 and 
\mbox{850 $\mu$m}, respectively. Both values are slightly higher than the
previously derived $14\%$ at \mbox{450 $\mu$m}, and $5\%$ at \mbox{850 $\mu$m} (D13).  
The Arcsecond FCFs have uncertainties of $14\%$ 
and $6\%$ at 450 and \mbox{850 $\mu$m}, respectively; these are also both slightly larger than the previously derived $11\%$ and $3\%$ derived by D13. 
The smaller uncertainties at \mbox{850 $\mu$m} 
when compared with the \mbox{450 $\mu$m} 
regime is due to the higher transmission and stability of 
the submillimeter atmosphere. A full analysis of the inherent scatter
in the FCF distributions is beyond the scope of this paper, though a
discussion is offered in Appendix~\ref{appsec:SingleObs}.

% Main (Stable) FCF Table
\begin{deluxetable*}{ccc}
\tablecaption{Recommended FCFs for observations obtained between 07:00-17:00 (UTC).}
\label{tab:3EpochFCFs}
\tablecolumns{3}
%\tablenum{1}
\tablewidth{0pt}
\tablehead{
\colhead{Wavelength, Date Range} &
\colhead{FCF$_{\mathrm{peak}}$}  &
\colhead{FCF$_{\mathrm{arcsec}}$}
}
\startdata
450 $\mu$m, Pre 2018 Jun 30            & 531 $\pm$ 93 & 4.61 $\pm$ 0.60 \\
450 $\mu$m, Post 2018 Jun 30           & 472 $\pm$ 76 & 3.87 $\pm$ 0.53 \\
850 $\mu$m, Pre 2016 Nov 19            & 525 $\pm$ 37 & 2.25 $\pm$ 0.13 \\
850 $\mu$m, 2016 Nov 19 to 2018 Jun 30 & 516 $\pm$ 42 & 2.13 $\pm$ 0.12 \\
850 $\mu$m, Post 2018 Jun 30           & 495 $\pm$ 32 & 2.07 $\pm$ 0.12 \\
\enddata

\tablecomments{These FCFs assume the opacity relations presented in Equation~\ref{eq:ORForm} and Table \ref{tab:OR} were applied during the extinction correction. The same extinction correction must be used for a direct comparison with the FCF values presented by D13. The atmospheric transmission lower limits included in the FCF determination are 10\% and 25\% for 450 and \mbox{850 $\mu$m}, respectively. The primary calibrator, Uranus was used to derive these FCFs.} 
\end{deluxetable*}

At \mbox{450 $\mu$m}, there is a single downward shift in FCF values. 
The shift occurs on 2018 June 30 (solid black vertical line on the right edge
of the latest gray region in left panels of
Figure~\ref{fig:FCFmaster}) 
after heavy SMU maintenance was performed wherein:
\begin{enumerate}
    \item The mirror was cleaned;
    \item The accuracy of the $Y$-Axis Linear displacement Variable Transducer (LDVT) was significantly improved;
    \item The chopper actuators that hold the SMU in the zero position were balanced;
    \item The drive motor assemblies for each axis were disassembled, 
    inspected, and improved both mechanically and electrically;
    \item The mirror was re-centred in its position above the primary; and
    \item The E-W chopper electronics were debugged after an intermittent issue
    with source aspect ratios became evident throughout the month of 2018 June
    (see Section~\ref{subsec:SMUtrouble} for more details).
\end{enumerate}
The combination of these adjustments significantly improved the flux 
concentration. This decrease in beam dilution translates to an increase in flux
and a reduction in FCF values. The average \mbox{450 $\mu$m} Peak FCFs are reduced by 
$11\%$ (top left panel of Figure~\ref{fig:FCFmaster}).  
While the drop appears much more significant for the Arcsecond 
FCF values (bottom left panel of Figure~\ref{fig:FCFmaster}), this is because the scatter in the light curves
is significantly lower when compared with the Peak FCF values. Overall, the 
Arcsecond FCF trends at both wavelengths have
less scatter and fewer outliers than their Peak FCF counterparts because they rely on 
aperture photometry and are, therefore, less affected by pixel-to-pixel noise and not
affected by Gaussian-fitting uncertainties. Based on the \mbox{450 $\mu$m} Arcsecond
FCF data, the downward shift in values is $16\%$ (in agreement 
with the Peak FCF results within the derived uncertainties).

At 850 $\mu$m, the FCF values experienced two downward shifts. The first was on 
2016 November 19 after new thermal-filter stacks were installed for each wavelength. The original 
\mbox{60 K}, \mbox{37 cm$^{-1}$} Low-pass Edge (LPE) filter was replaced with a \mbox{55 cm$^{-1}$} 
LPE filter with an improved transmission profile 
in the \mbox{850 $\mu$m} observing band \citep{cookson2018}. The main reason for  the filter change was to lower the detector loading by reducing the temperature of the filters and not, necessarily, to improve the transmission.  However, there is a 
 $2\%$ transmission improvement when measured using the \mbox{850 $\mu$m} Peak FCF distributions and a $5\%$ improvement when 
measured using the Arcsecond FCFs. No significant shift was expected or seen at \mbox{450 $\mu$m} after the 2016 November thermal-filter installation, as there was no improvement in the transmission profile of the new filter stack at this waveband. The second shift at \mbox{850 $\mu$m} occurs 
after the SMU maintenance described, above. Due to the large and more-stable beam 
profile when compared with 450 $\mu$m (see Section~\ref{subsec:Beam}), the magnitude of this 
downward shift is $4\%$ when measured using the Peak FCFs and 
$3\%$ when measured using the Arcsecond FCFs (in agreement to within the derived uncertainties). 

%%%%%%%%%%%%%%%%%%%%%%%%%%%%%%%%%%%%%%%%
%%%%%%%%%%%%%%%%%%%%%%%%%%%%%%%%%%%%%%%%
%%%%%%%%%%%%%%%%%%%%%%%%%%%%%%%%%%%%%%%%
\subsubsection{Secondary-mirror Malfunction}
\label{subsec:SMUtrouble}
%%%%%%%%%%%%%%%%%%%%%%%%%%%%%%%%%%%%%%%%
%%%%%%%%%%%%%%%%%%%%%%%%%%%%%%%%%%%%%%%%
%%%%%%%%%%%%%%%%%%%%%%%%%%%%%%%%%%%%%%%%

% Aspect Ratios showing SMU trouble
\begin{figure*}
\plotfour{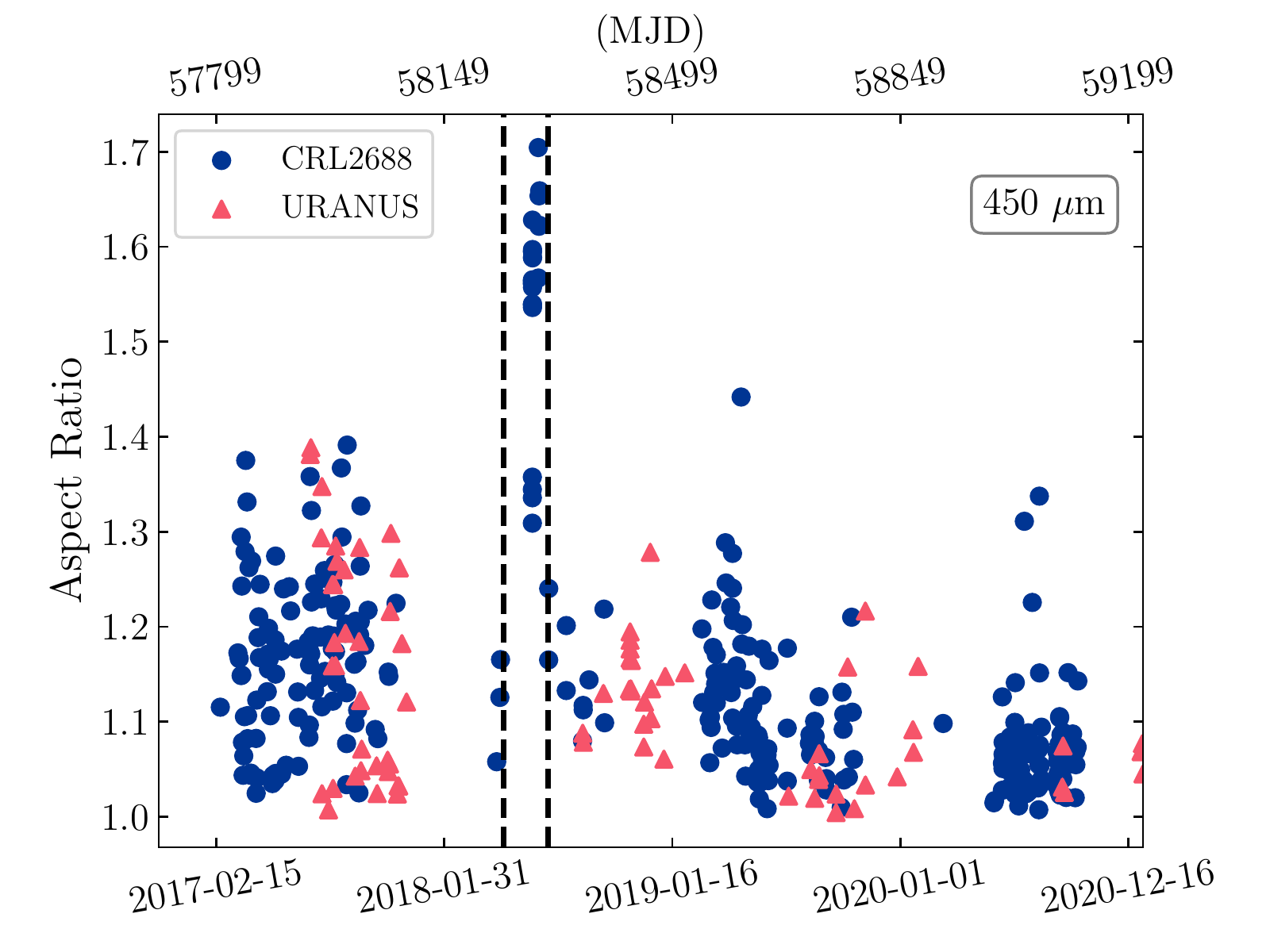}{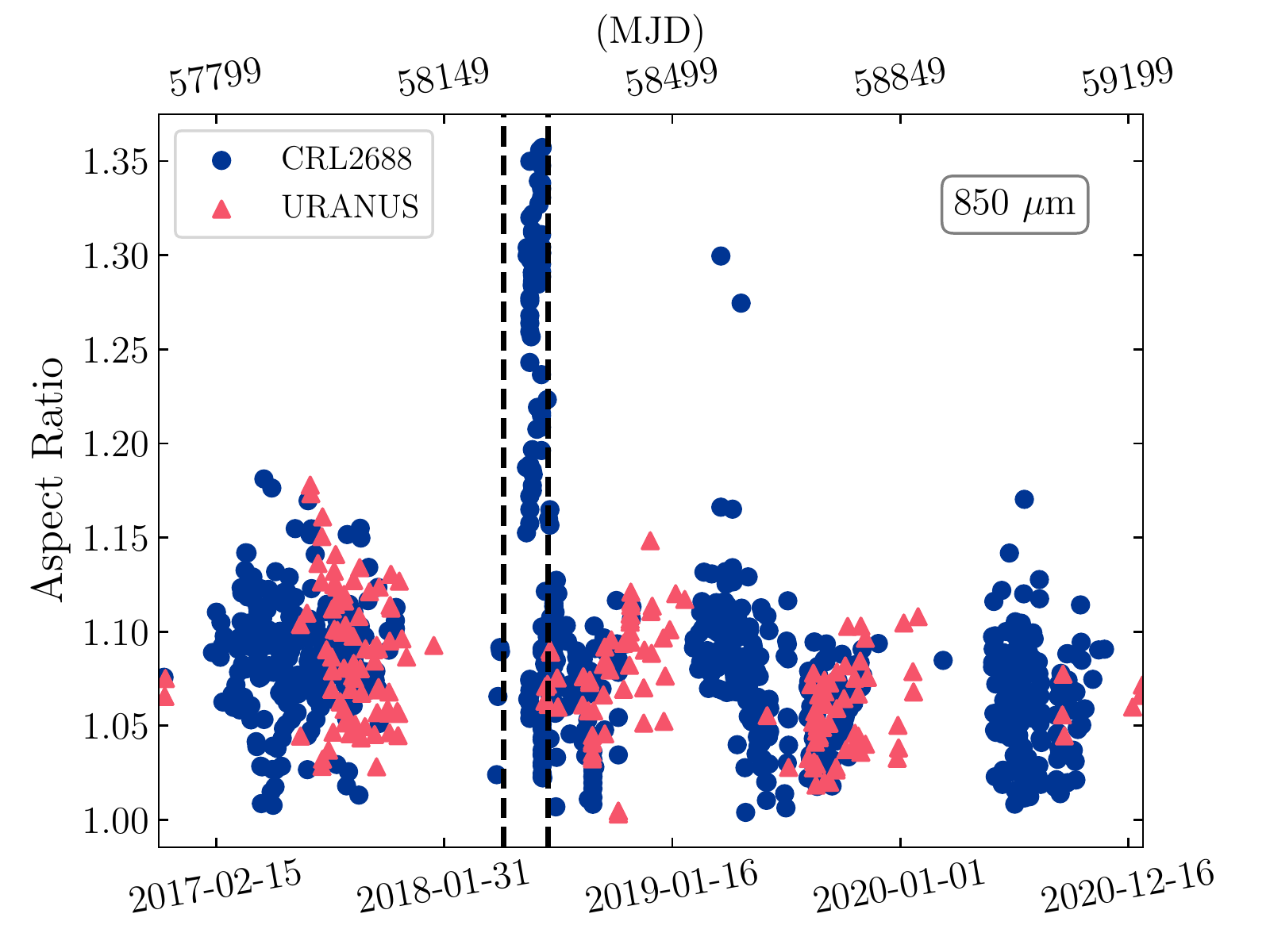}{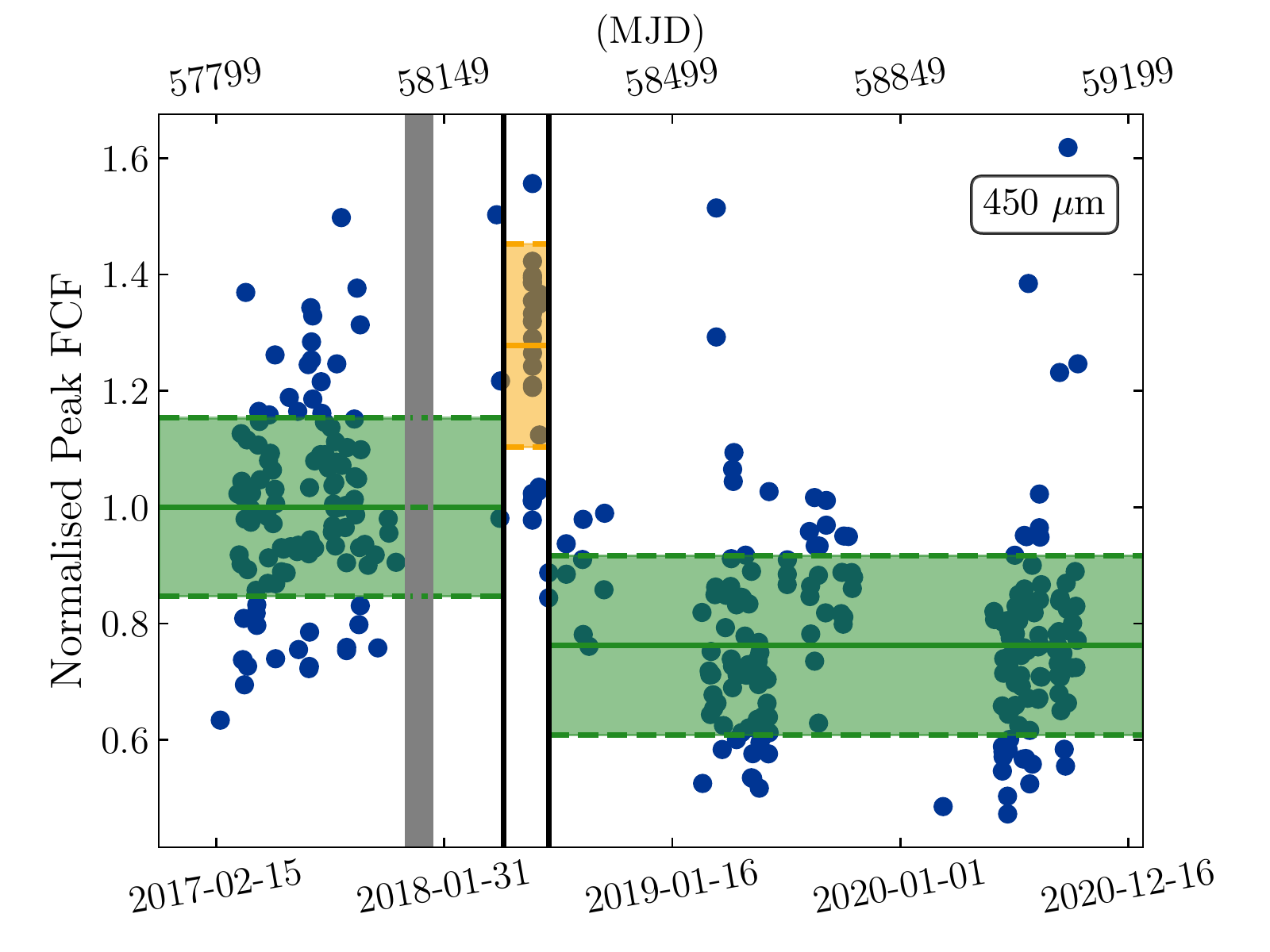}{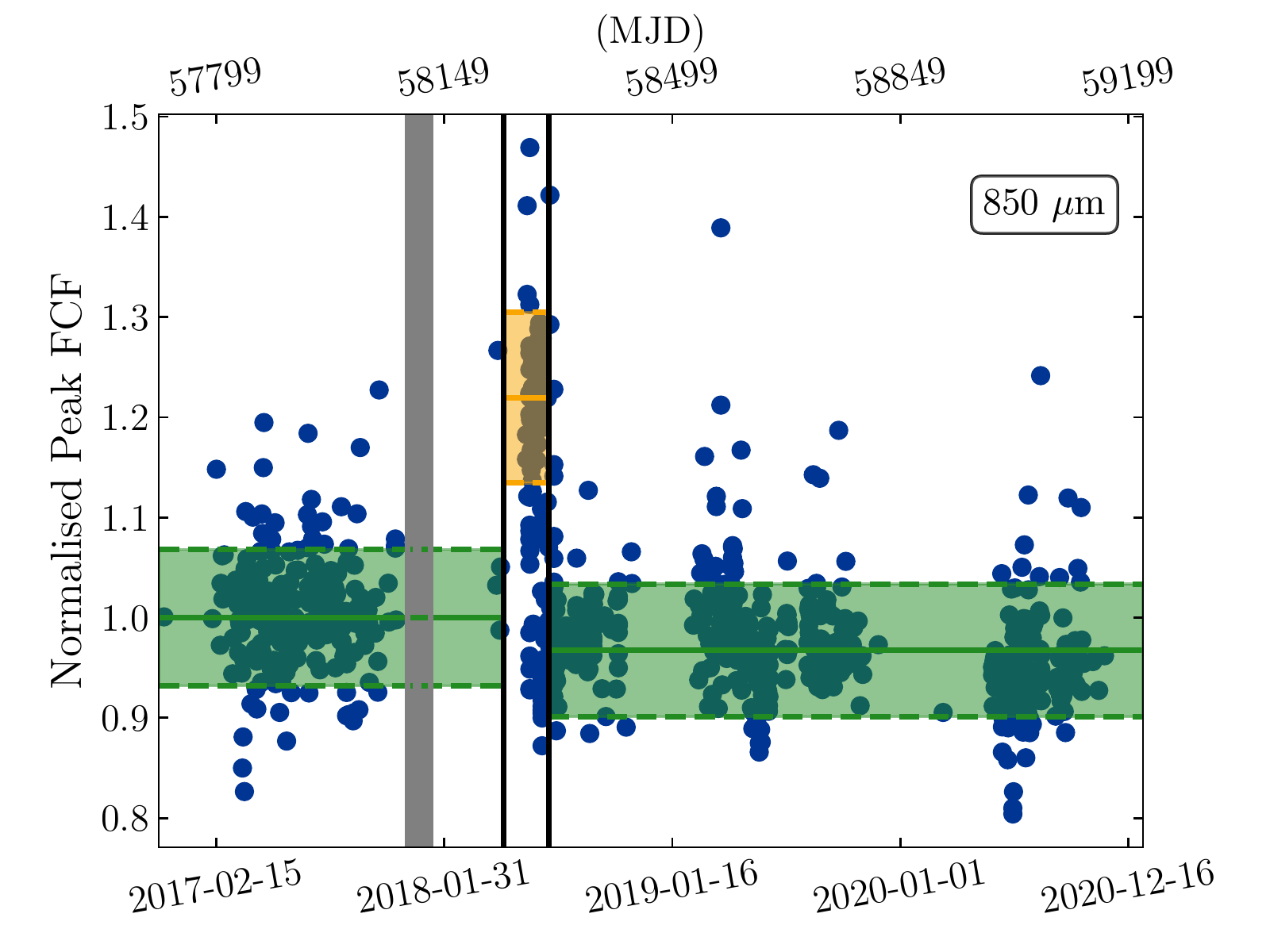}
\caption{\textit{Top:} 450 (\textit{left}) and \mbox{850 $\mu$m} (\textit{right}) aspect ratios of \mbox{CRL 2688} (circles) and Uranus (triangles) observations as a function of date derived by fitting the calibrators with a 2D Gaussian function. The vertical, dashed lines indicate the time interval of the secondary-mirror malfunction, correlating with large aspect ratios. \textit{Bottom:} Normalized 450 (\textit{left}) and \mbox{850 $\mu$m} (\textit{right}) Peak FCFs derived using flux measurements of the secondary calibrator \mbox{CRL 2688} throughout the stable part of the night (0700--1700 UTC) as a function of date. The gray shaded region indicates when the GORE-TEX\texttrademark$\:$ membrane was removed from the telescope for POL-2 commissioning. The two solid, vertical lines mark the time interval in which the secondary-mirror was malfunctioning. The horizontal, shaded regions indicate the median FCF value over each span of time and the associated median absolute deviation. Point source observations observed during June 2018 should be carefully inspected before use. }
\label{fig:SMUTrouble}
\end{figure*}

Throughout 2018 May, the SMU underwent heavy maintenance wherein it 
was removed from the 
telescope (see Section~\ref{subsec:FCFsLongTerm}). For several 
weeks after, until 2018 June 30, there was an intermittent issue with the 
East-West (E-W) chopper that resulted in the elongation of compact sources, 
mainly affecting the Peak-FCF 
measurements during this time. While ``chopping'' is not performed during SCUBA-2 observing, the electronics that control the chopper are responsible for keeping the observation at a fixed angle throughout the observations. Therefore, any oscillation in the electronics affect the SMU position and the FCFs. The top panels of Figure~\ref{fig:SMUTrouble} show 
the aspect ratio of Uranus and \mbox{CRL 2688} plotted as a function of UTC date. The time interval
of the intermittent problem can be clearly seen as a spike in the aspect ratios\footnote{Errant aspect ratios can also occur due to low-signal-to-noise observations when the 2D Gaussian fit is less certain.}. 
Engineering/commissioning 
telescope time was specifically used to observe 
the secondary calibrator source \mbox{CRL 2688} in the time following the SMU maintenance, 
which is why 
the number of observations is relatively high.

The bottom panels of Figure~\ref{fig:SMUTrouble} show the FCFs derived during the 
stable part of the night using \mbox{CRL 2688} before, during, and after the SMU malfunction. Note 
that throughout the duration of the intermittent problem, several FCFs indicated the system was 
performing within normal parameters. Once the E-W chopper was fixed 
(2018 June 30), the SMU maintenance of 2018 was complete and, as noted previously, there was an improvement in flux concentration within the beam. 
There have been no 
further issues with the data as a result of the SMU since 2018 June 30.
While the main science programs being carried 
out during 2018 May were checked to 
ensure that the quality of the 
produced maps was not adversely affected, 
science data (especially point sources) 
obtained during this period should be
carefully examined before use.

%%%%%%%%%%%%%%%%%%%%%%%%%%%%%%%%%%%%%%%%
%%%%%%%%%%%%%%%%%%%%%%%%%%%%%%%%%%%%%%%%
%%%%%%%%%%%%%%%%%%%%%%%%%%%%%%%%%%%%%%%%
\section{FCF Nightly Patterns}
\label{sec:FCFsnightly}
%%%%%%%%%%%%%%%%%%%%%%%%%%%%%%%%%%%%%%%%
%%%%%%%%%%%%%%%%%%%%%%%%%%%%%%%%%%%%%%%%
%%%%%%%%%%%%%%%%%%%%%%%%%%%%%%%%%%%%%%%%

\begin{figure*}
\plotfour{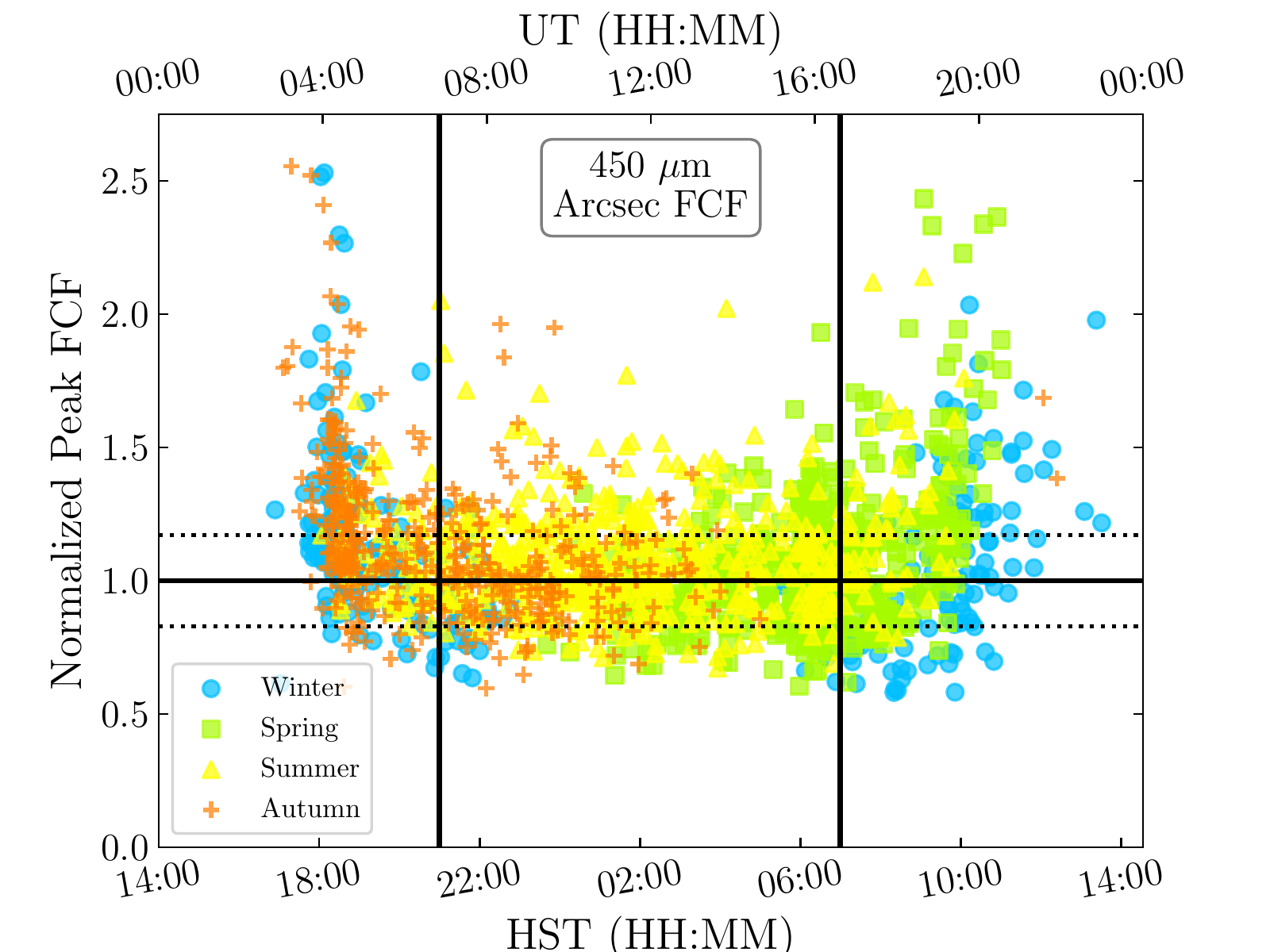}{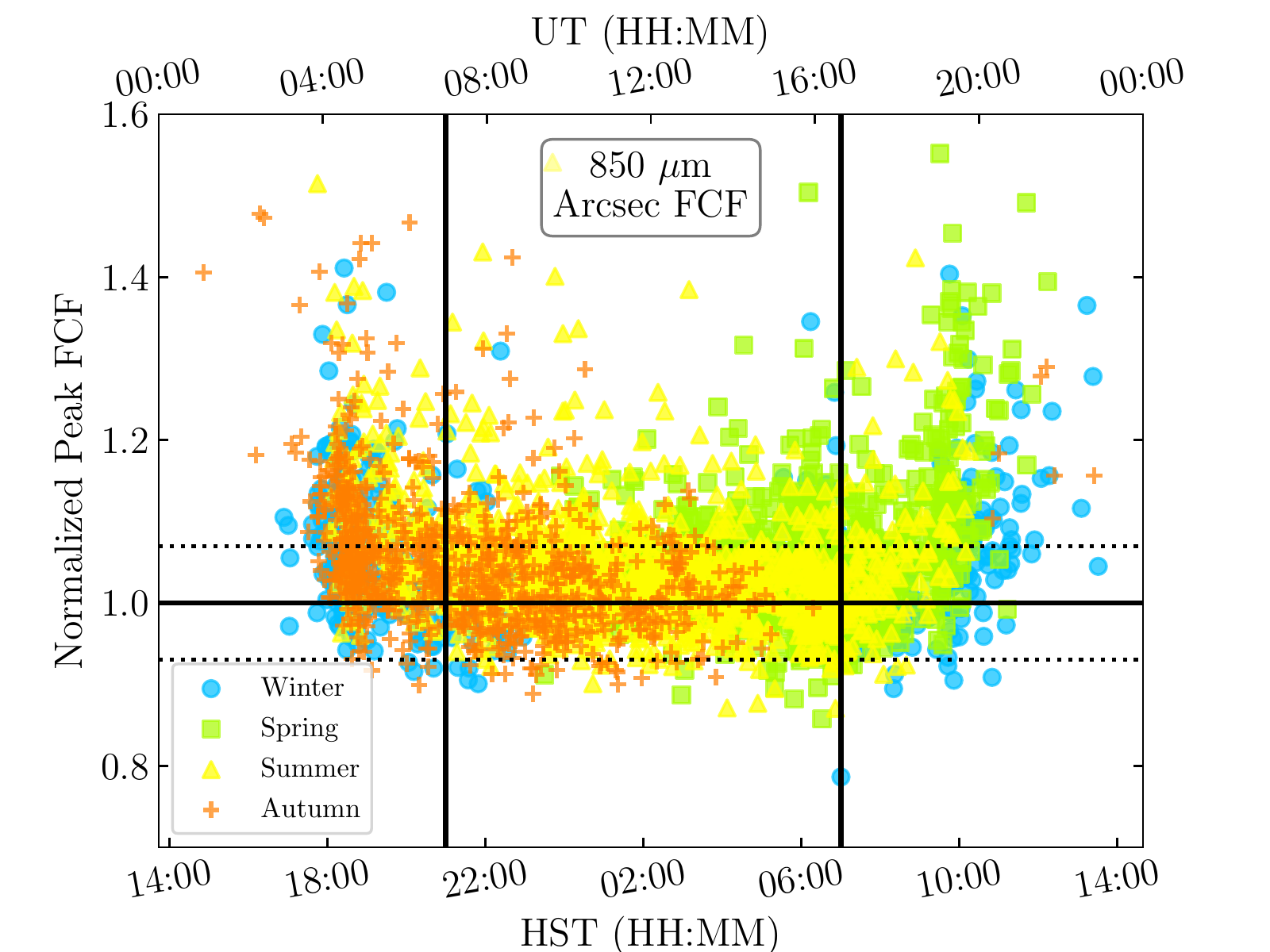}{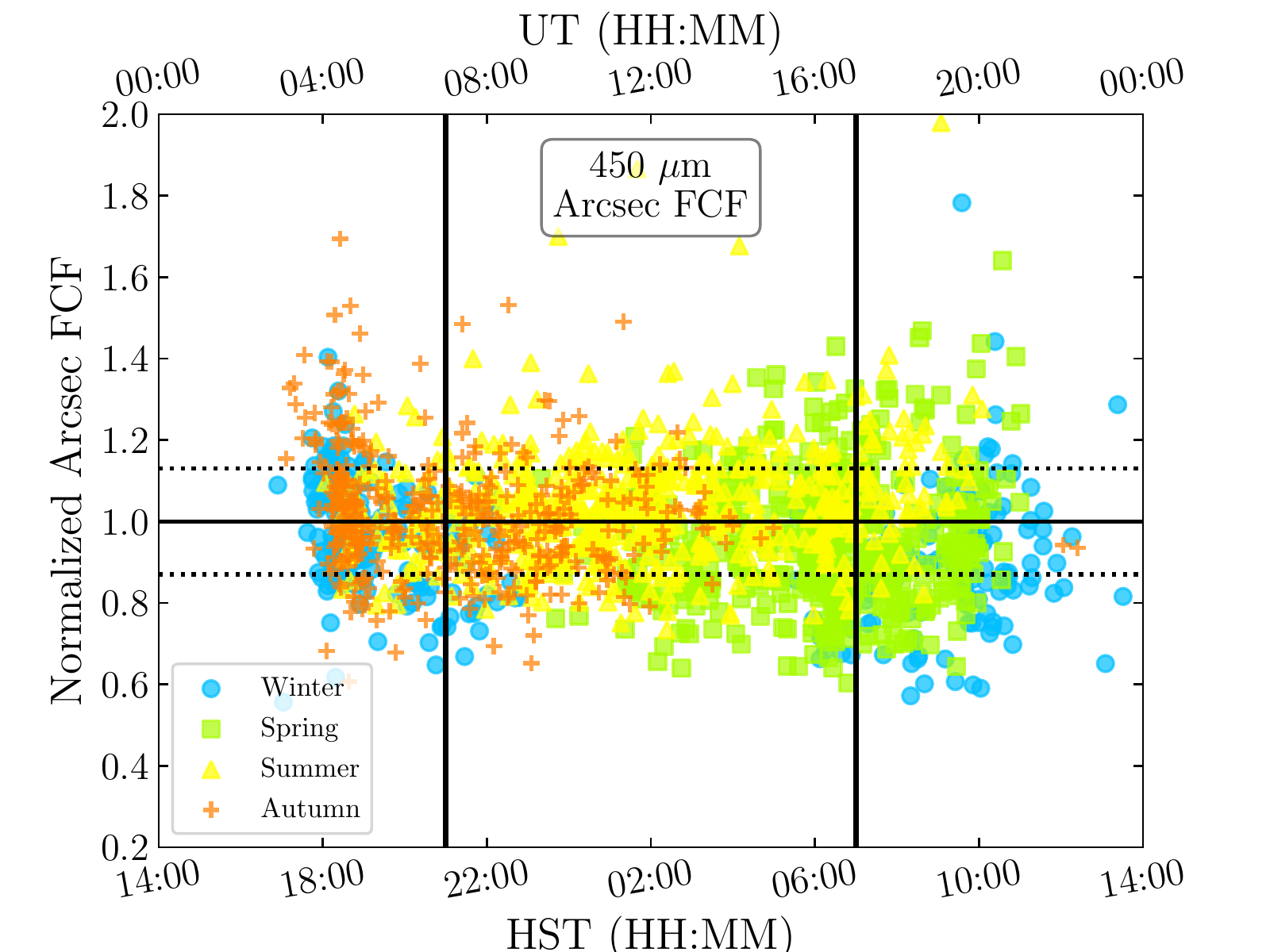}{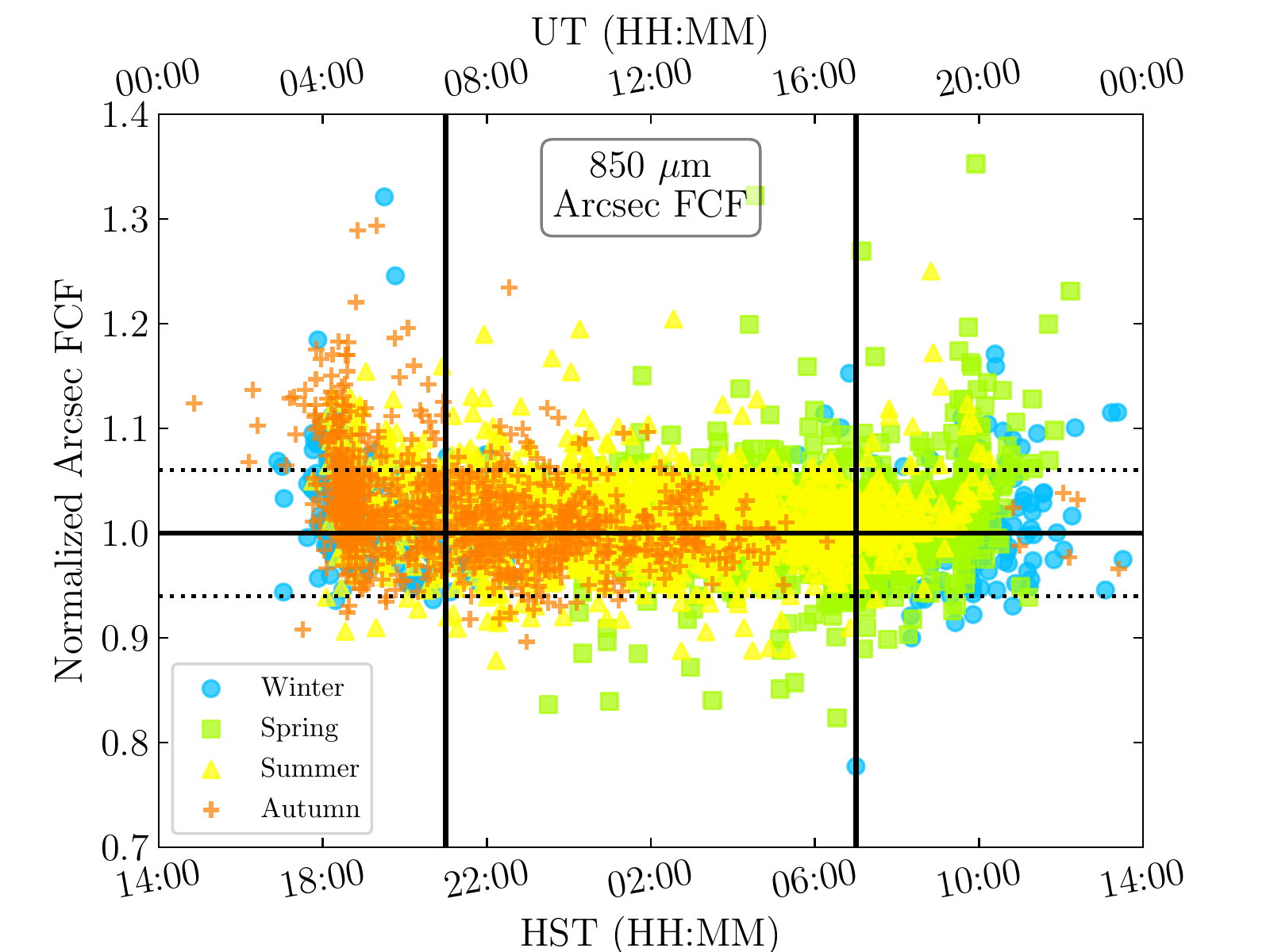}
\caption{Normalized Peak (\textit{top}) and Arcsecond (\textit{bottom}) FCFs at 450 (\textit{left}) and \mbox{850 $\mu$m} (\textit{right}) as a function of observation time. All FCFs are derived using the primary calibrator Uranus and secondary calibrator \mbox{CRL 2688}. The results presented in
Table~\ref{tab:3EpochFCFs} were used for the normalization. No data are included for the periods in which there were no reliable JCMT WVM data, the GORE-TEX\texttrademark$\:$ membrane was removed, or the secondary mirror was malfunctioning (see Figure~\ref{fig:FCFmaster}). The vertical lines mark the beginning and end of the stable observation period from 07:00--17:00 (UTC). The horizontal (dotted) lines show the FCF uncertainties derived for the stable observation period around a value of 1.0 (horizontal, solid line). Data are colored according to season: blue circles represent Winter, green squares represent Spring, yellow triangles represent Summer, and orange crosses represent Autumn. There are no significant trends with the time of year. The early-evening and late-morning Peak FCF observations deviate significantly from the stable observations, while the Arcsecond FCFs remain largely unchanged with a small increase in the uncertainty in the evening and morning.}
\label{fig:NightlyFCF}
\end{figure*}

It is well known that the temperature decrease and increase in the 
early evening and late morning, respectively, causes slight structural 
changes to the dish that can affect the beam shape, focus and, ultimately, the calibration. In Figure~\ref{fig:NightlyFCF}, we present the derived, normalized FCFs as a 
function of time of night. All Uranus and \mbox{CRL 2688} observations taken when the atmospheric transmission was greater than $10\%$ at \mbox{450 $\mu$m} 
and $25\%$ at \mbox{850 $\mu$m} since 2011 May are included in the figure.
The same trends are observed both before and after the key dates when the FCFs shifted (Section~\ref{subsec:FCFsLongTerm}). 

% "U-shape" nightly FCF trends

% Hourly Degradation Plot
\begin{figure*}
\plotsix{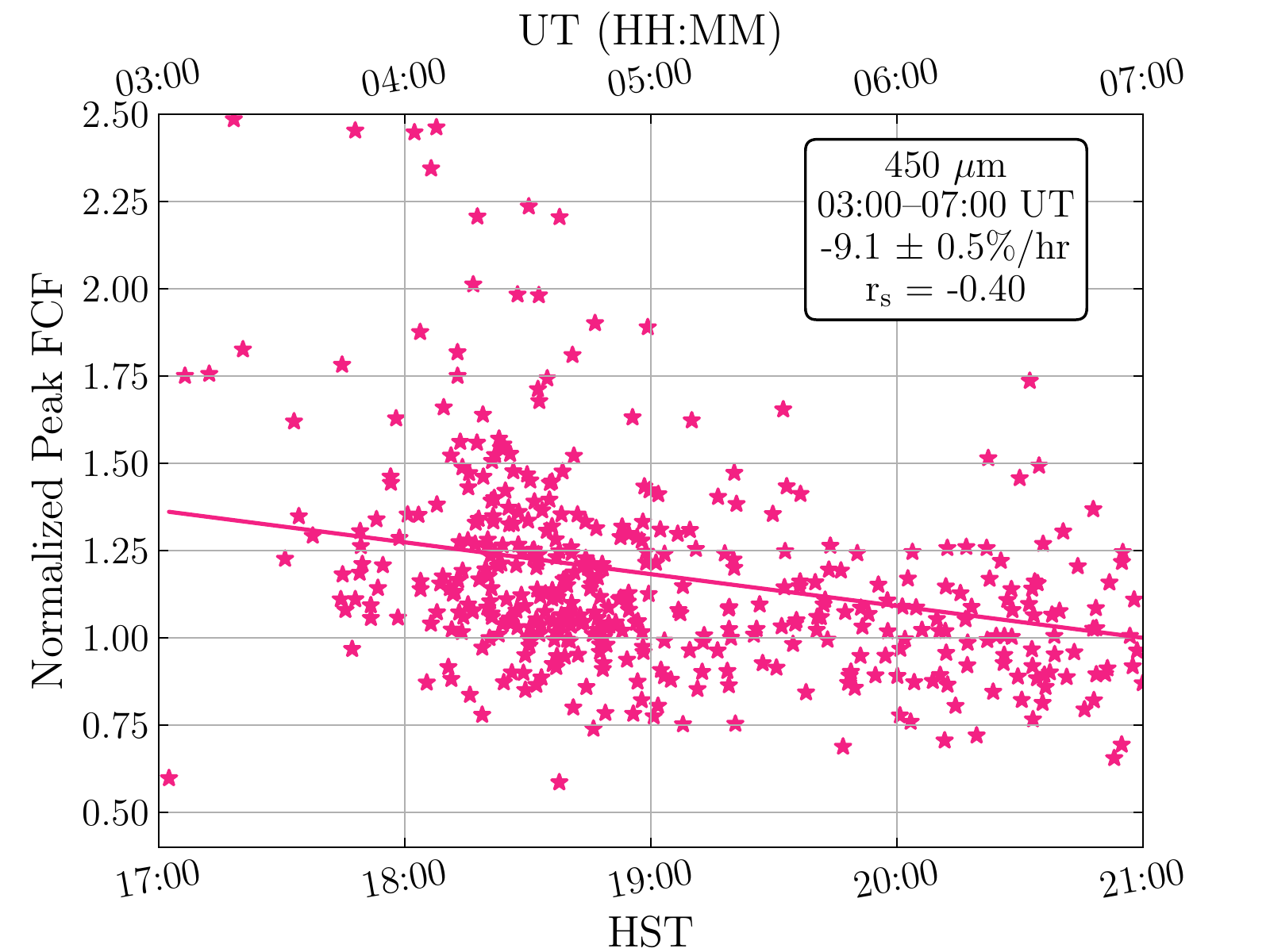}{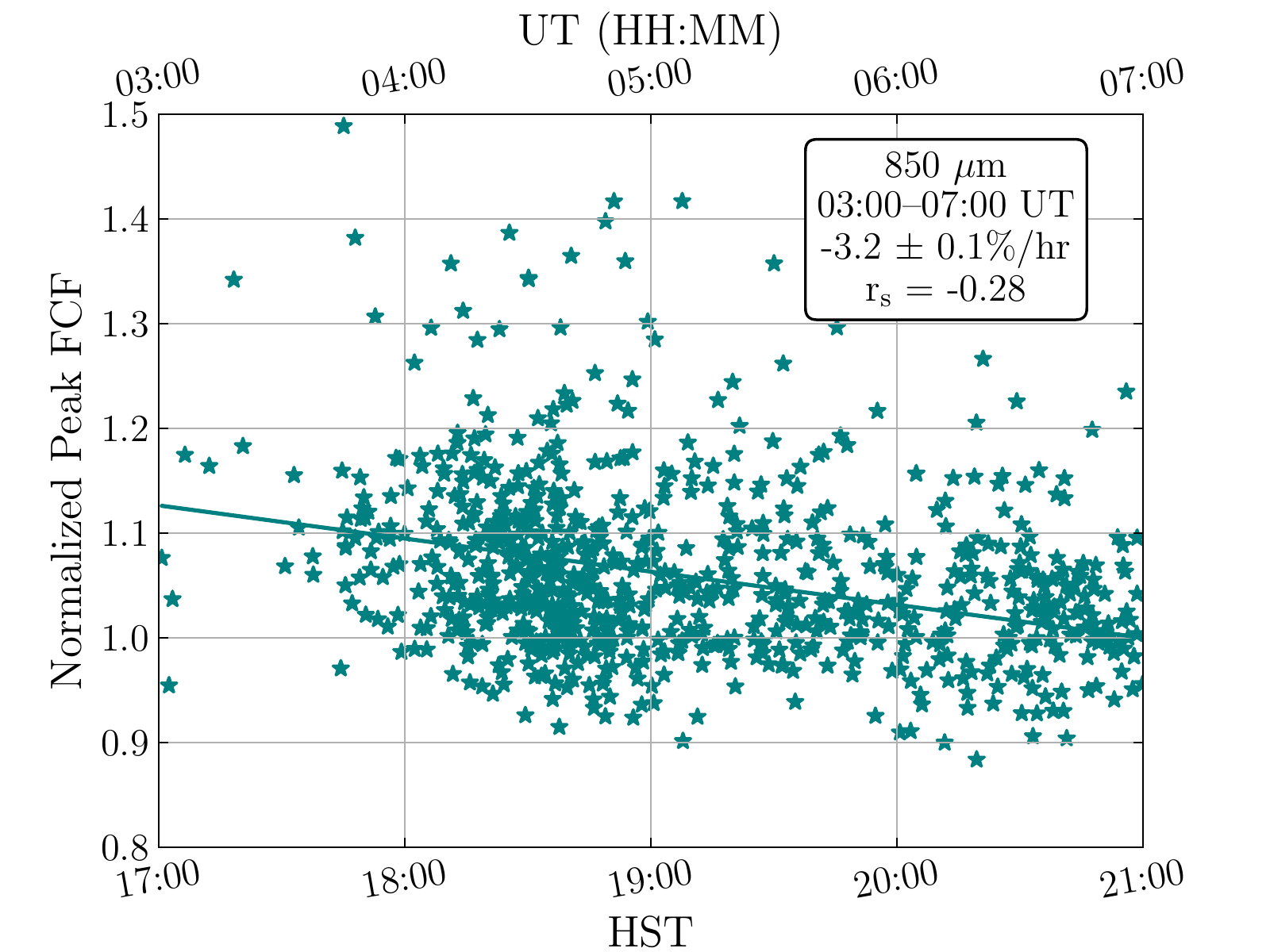}{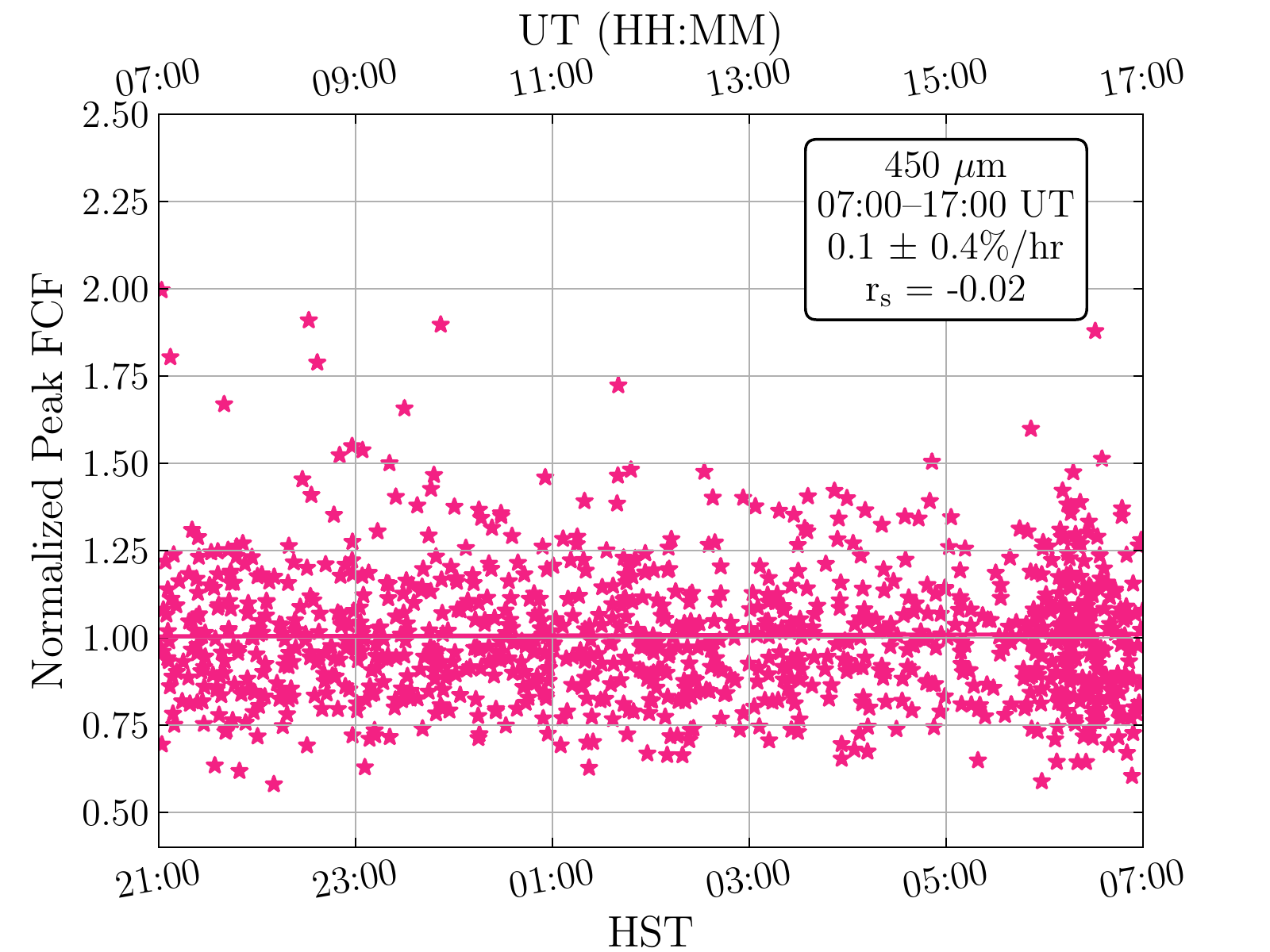}{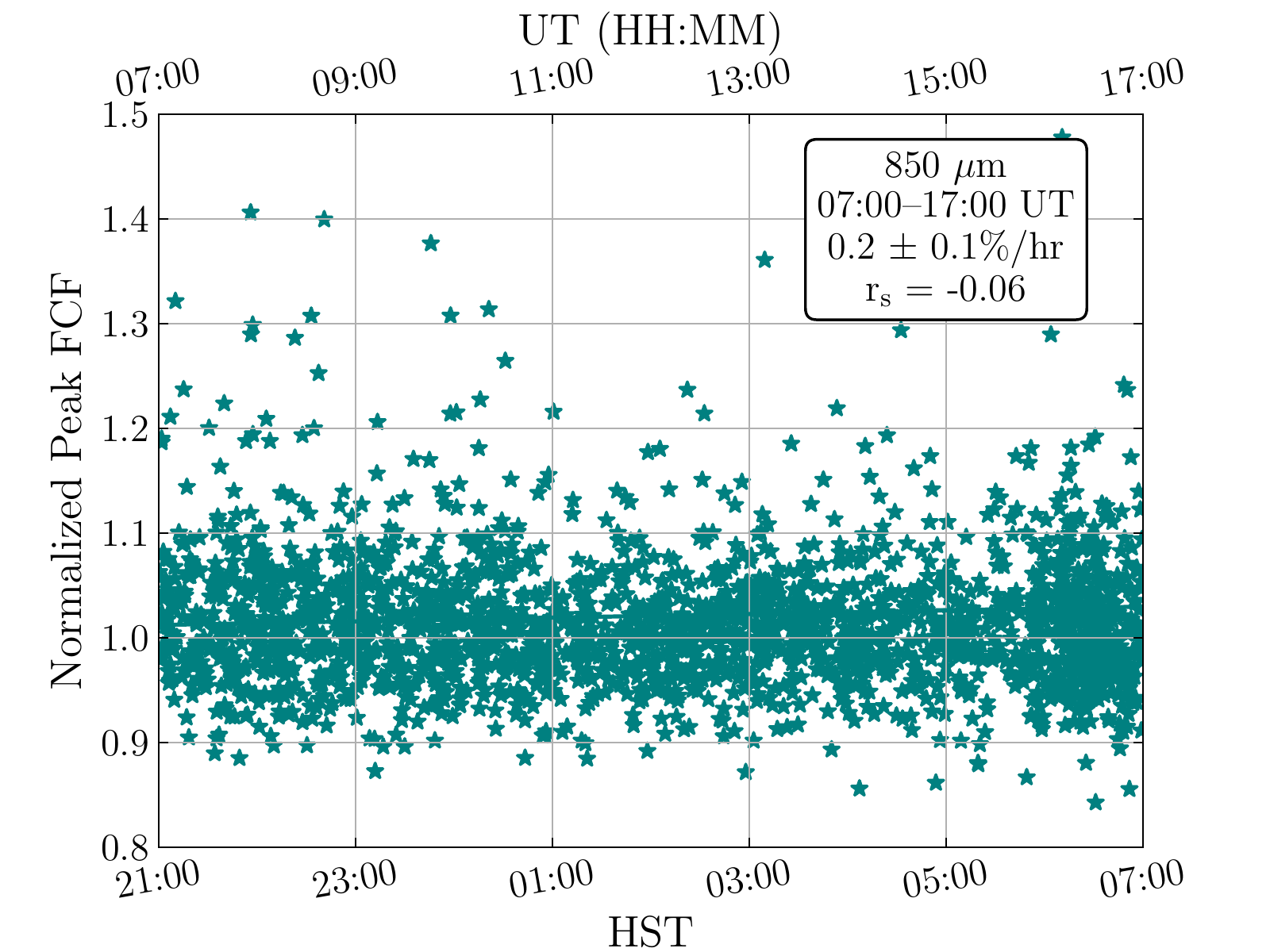}{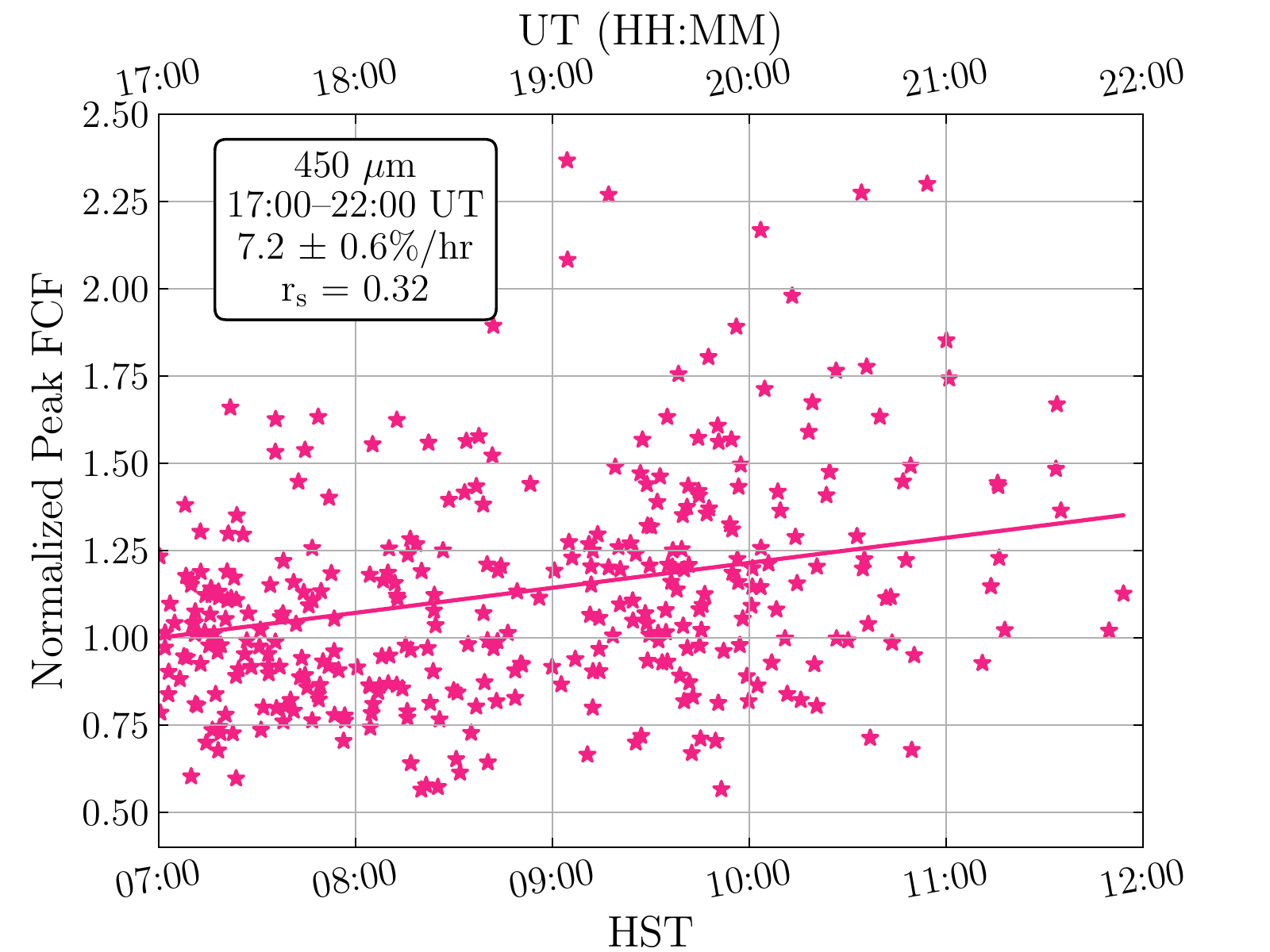}{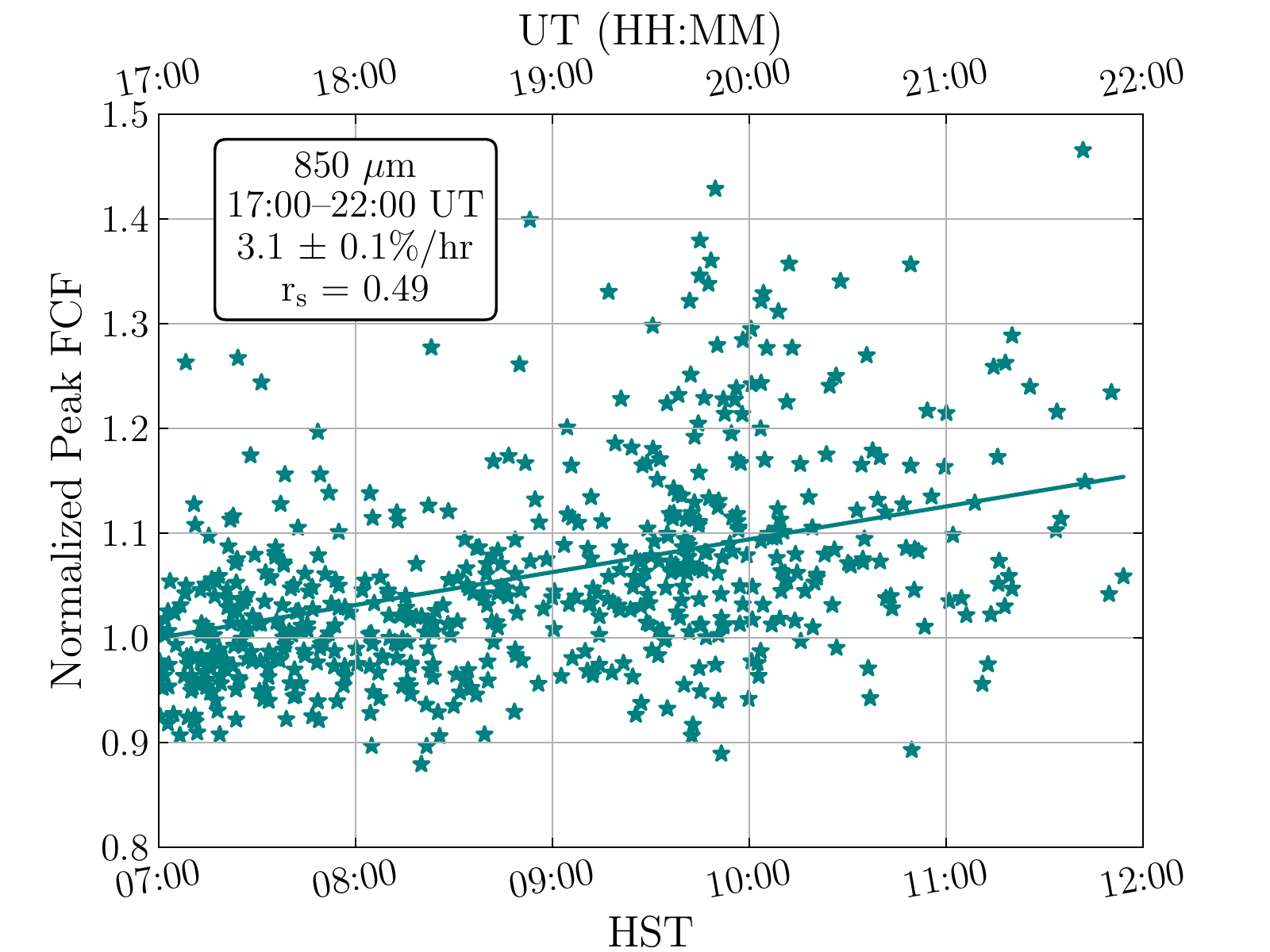}
\caption{Normalized Peak FCFs at 450 (\textit{left}) and \mbox{850 $\mu$m} (\textit{right}) derived using the primary calibrator Uranus and the secondary calibrator \mbox{CRL 2688} as a function of observation time. Linear, bootstrapped least squares fits to the data and the associated Spearman Rank Correlation (r$_{\mathrm{s}}$) are included in each panel. These results and their uncertainties are presented in Table~\ref{tab:NightlyFCFMods} and can be used to modify the FCFs presented in Table~\ref{tab:3EpochFCFs} for observations taken outside of the stable time of night (0700--1700 UTC). \textit{Top:} Evening observations (03:00-07:00 UTC). \textit{Middle:} Stable observations (07:00-17:00 UTC). \textit{Bottom:} Morning observations (17:00-22:00 UTC).}
\label{fig:EveningStableMorning}
\end{figure*}

In the top panels of Figure~\ref{fig:NightlyFCF}, there is an 
obvious U-shape in the nightly Peak FCF data that
corresponds to dish instability. The vertical lines in each figure are 
drawn at 07:00 UTC (21:00 HST) and 17:00 UTC (07:00 HST), marking
when the dish is stable to dynamic deformations. 
The \mbox{450 $\mu$m} data show a more dramatic trend in FCFs as its beam 
profile is more sensitive than its \mbox{850 $\mu$m} counterpart to atmospheric and structural 
effects. Without accounting for the obvious increase in
Peak FCFs in the evening and morning observing regimes, the uncertainties  for a typical calibrator source at \mbox{450 $\mu$m}
varies from $20\%$ in the evening to $17\%$ during the night, and it can reach up to 
$25\%$ in the morning. At \mbox{850 $\mu$m}, 
the Peak FCF uncertainties in each source distribution are $8\%$ in 
the evening, $7\%$ during the night, and up to $10\%$ in the morning.  
 In the day time, uncertainties in the observations are dominated by the ambient temperature increase and air stability decrease (seeing) when compared to night observations. In the afternoon, the opening of the dome typically causes a faster temperature decrease and hence more thermal deformations. As expected, the Arcsecond FCFs
(bottom panels of Figure~\ref{fig:NightlyFCF}) 
show a much weaker nightly trend as the large-aperture 
photometry used to calculate these values is more stable to beam deformation than 
the Gaussian fits used to derive the Peak FCFs. 
With a 60$\arcsec$ diameter aperture used for the photometry, 
only the secondary (``error'') beam is affected (see Section~\ref{subsec:Beam}), 
which has a small effect on
the data at all transmission bands of interest. The \mbox{450 $\mu$m} 
Arcsecond FCF uncertainties range from $17\%$ in the
evening and morning to $14\%$ in the stable part of the night, while the \mbox{850 $\mu$m} Arcsecond FCF 
uncertainties range from $6\%$ in the stable part of the night to $7\%$ in the evening and morning.

Figure~\ref{fig:EveningStableMorning} shows the Peak FCF trends in 
detail for evening, night, and morning observations of Uranus and \mbox{CRL 2688}. In the 
evening, between 03:00 and 
07:00 (UTC), a first-order correction to the instability in the Peak FCFs caused by the settling dish can be derived by bootstrap-fitting (1,000 iterations) a linear function of the relative Peak FCF values over time. Such a correction indicates the Peak FCFs decrease at rates of $9.1 \pm 0.5\%\,\mathrm{hr}^{-1}$ and
$3.2 \pm 0.1 \%\,\mathrm{hr}^{-1}$ at 450 and \mbox{850 $\mu$m}, respectively. This is due to the 
peak flux increasing as the beam profile becomes more stable and the flux more 
concentrated. During the night, 07:00 to 17:00 (UTC), the Peak FCFs remain stable. 
In the morning, after 17:00 (UTC), the dish expansion caused 
by the increasing 
temperature once again causes the beam profiles to degrade 
and the peak flux to 
decrease (Peak FCFs increase) at a 
rate of $7.2 \pm 0.6\%\,\mathrm{hr}^{-1}$ and 
$3.1 \pm 0.1\%\,\mathrm{hr}^{-1}$ at 450 and \mbox{850 $\mu$m}, 
respectively. These results along with their associated Spearman Rank Correlation values (r$_{\mathrm{s}}$) are summarised in Table~\ref{tab:NightlyFCFMods}. 
The Arcsecond FCFs do not vary significantly enough over the evening, night, or morning to warrant additional FCF correction factors during typical observing times.

% Evening/Morning FCF corrections Table 
\begin{deluxetable*}{cccc}
\tablecaption{Modifications to FCFs presented in Table~\ref{tab:3EpochFCFs} for observation times outside 07:00-17:00 (UTC).}
\label{tab:NightlyFCFMods}
\tablecolumns{4}
%\tablenum{1}
\tablewidth{0pt}
\tablehead{
\colhead{Wavelength} &
\colhead{Time Range (UTC)} &
\colhead{FCF$_{\mathrm{peak}}^{a}$ Correction$^{b}$ ($\%\mathrm{\:hr}^{-1}$)} &
\colhead{Spearman Rank Correlation (r$_{\mathrm{s}}$)}
}
\startdata
450 $\mu$m & 03:00--07:00 & 9.1 $\pm$ 0.5 &  -0.40 \\
450 $\mu$m & 17:00--22:00 & 7.2 $\pm$ 0.6 &  0.32 \\
\hline
850 $\mu$m & 03:00--07:00 & 3.2 $\pm$ 0.1 &  -0.28 \\
850 $\mu$m & 17:00--22:00 & 3.1 $\pm$ 0.1 &  0.49 \\
\enddata
\tablecomments{$^{a}$Arcsecond FCFs do not require an evening or morning modification to the stable FCFs presented in Table~\ref{tab:3EpochFCFs} but the uncertainty in the FCFs increase slightly during these times (see text).\\$^{b}$In the evening the Peak FCFs presented in Table~\ref{tab:3EpochFCFs} must be increased by the factor shown multiplied by the number of hours until 07:00 UTC. In the morning, the Peak FCFs presented in Table~\ref{tab:3EpochFCFs} must be increased by the factor shown multiplied by the number of hours since 17:00 UTC. Data obtained during the stable part of the night does not require a modification.} 
\end{deluxetable*}

\subsection{Variations in the Beam}
\label{subsec:Beam}

% Beam FWHM and Relative Volume plot
\begin{figure*}
\plotfour{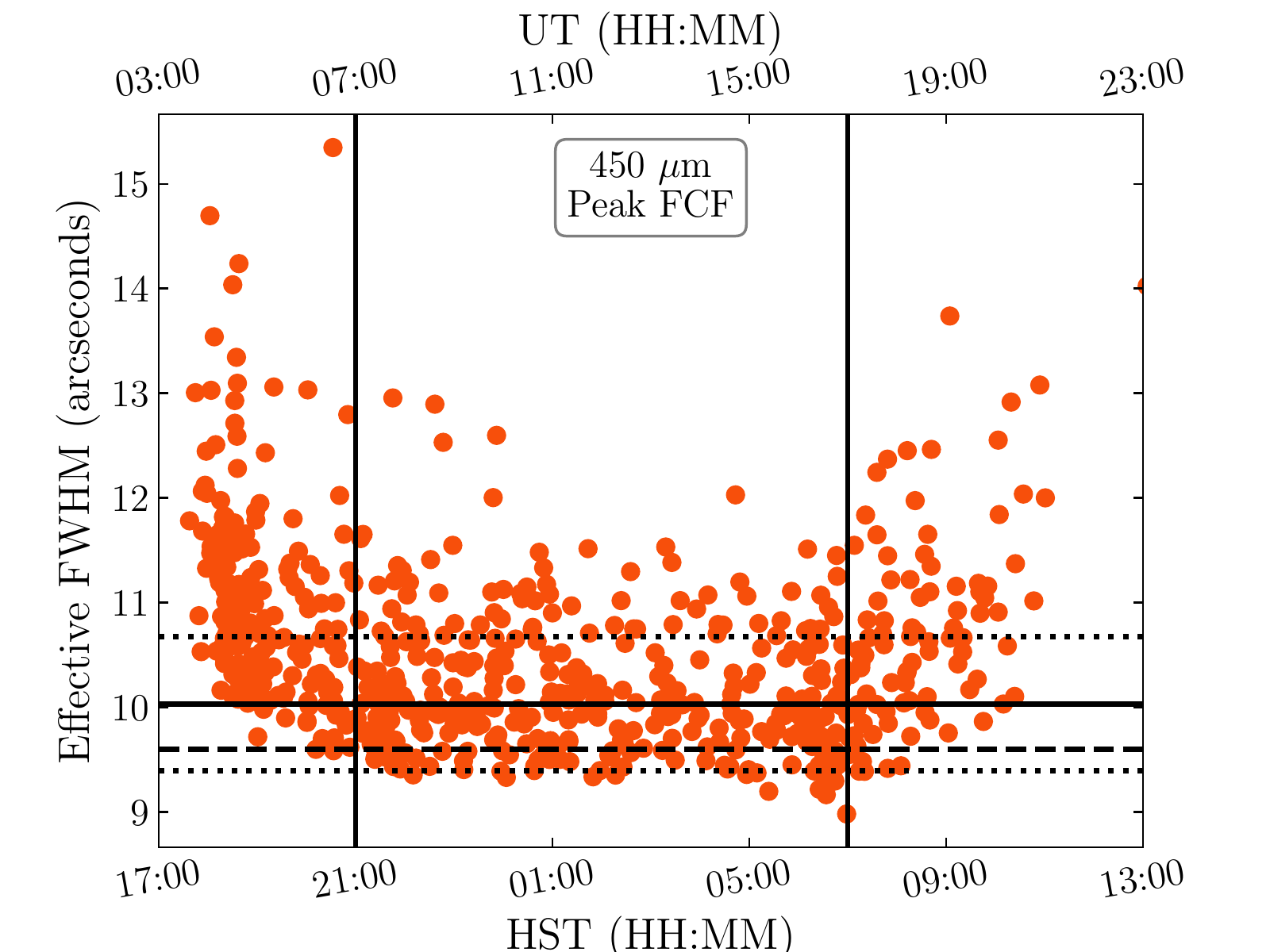}{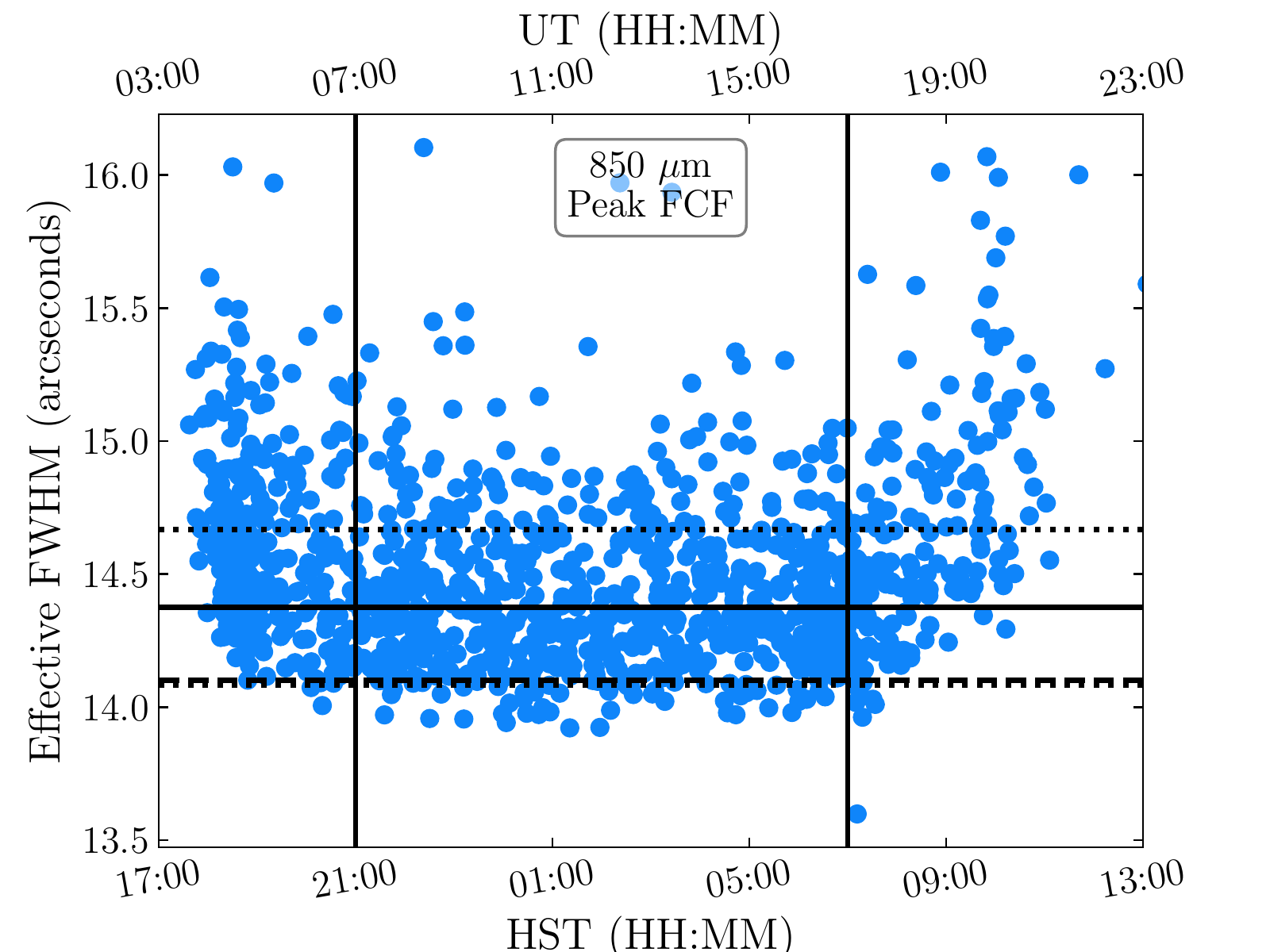}{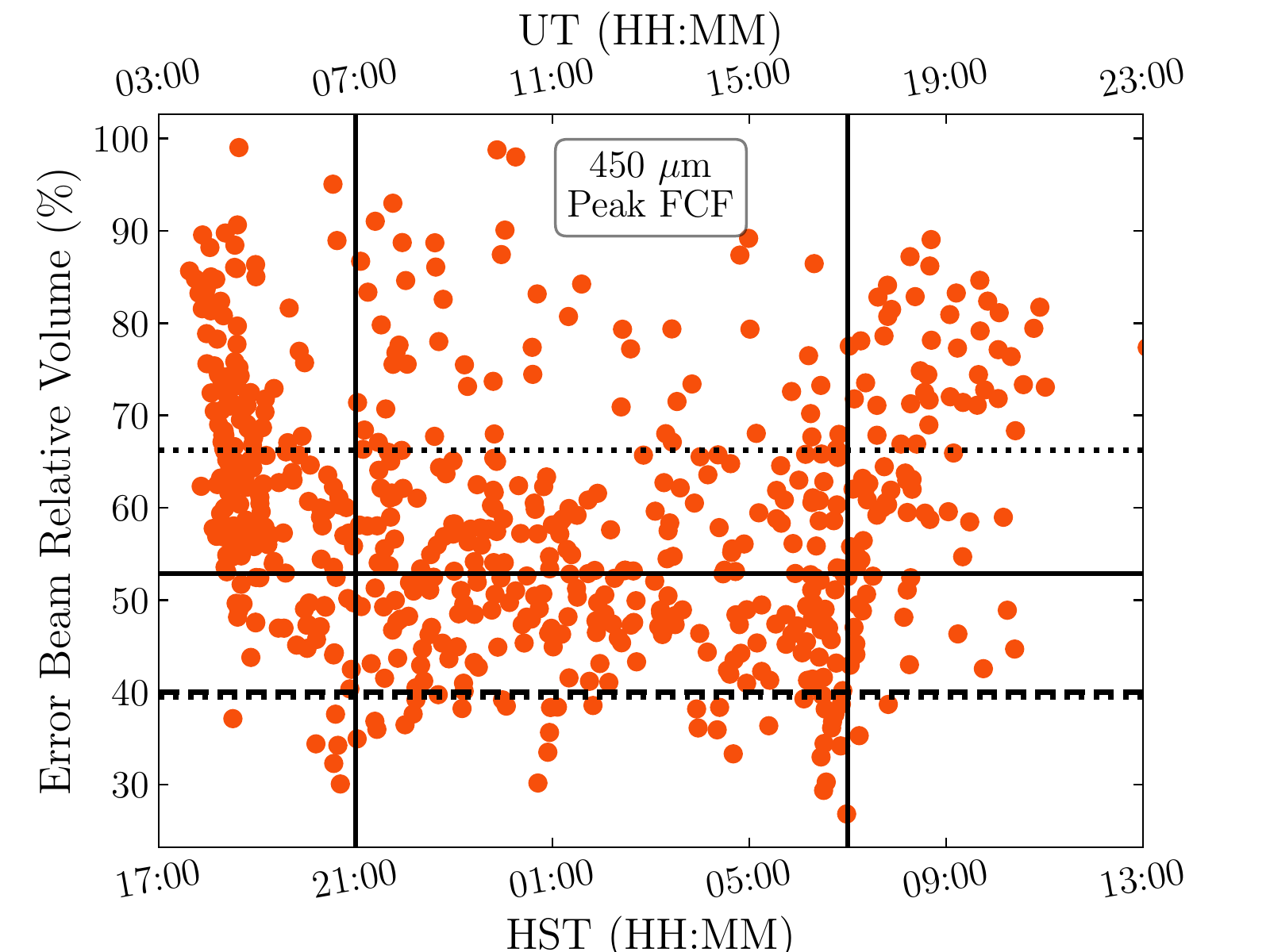}{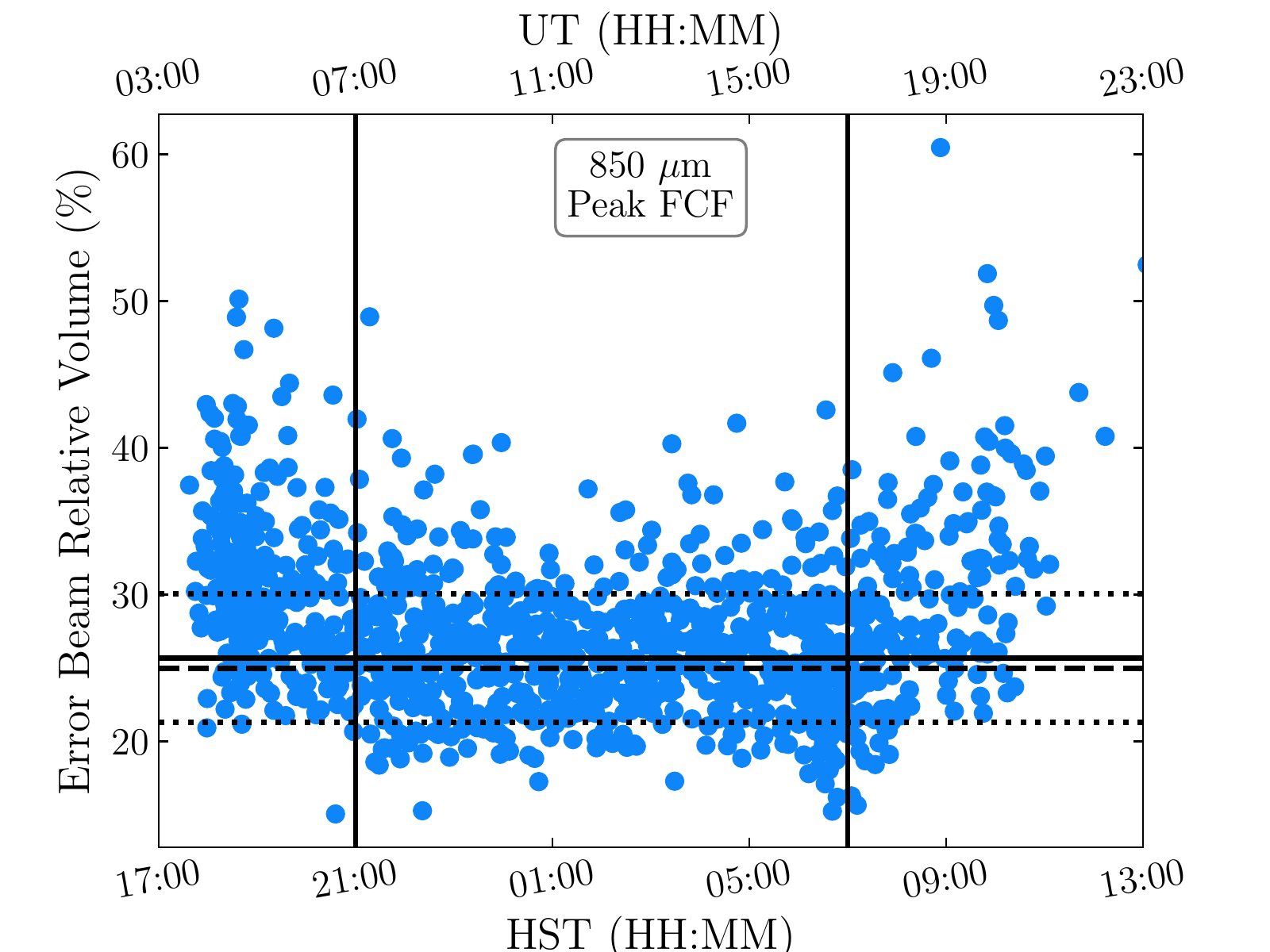}
\caption{The empirical measurements of the effective beam FWHM derived using Equation~\ref{eq:EffBeam} (\textit{top}) and the relative volume of the error beam derived by the the two-component fit (Equation~\ref{eq:TwoCompGauss}) (\textit{bottom}) at 450 (\textit{left}) and \mbox{850 $\mu$m} (\textit{right}) as a function of observation time. The primary calibrator Uranus was used to derive the FCFs in the top panels and the model fits in the bottom panels. The vertical lines mark the beginning and end of the stable observation period from 07:00--17:00 (UTC). The horizontal (dotted) lines show the standard deviation in the data for the stable observation period, centered on the median value over the stable observation period (horizontal, solid lines). The dashed, horizontal lines represent the original values derived by D13. The expansion and contraction of the dish during cooling and heating affects the beam size and, subsequently, the Peak FCFs.}
\label{fig:FWHMBeam}
\end{figure*}

% Flux in the wings
\begin{figure*}
\plottwo{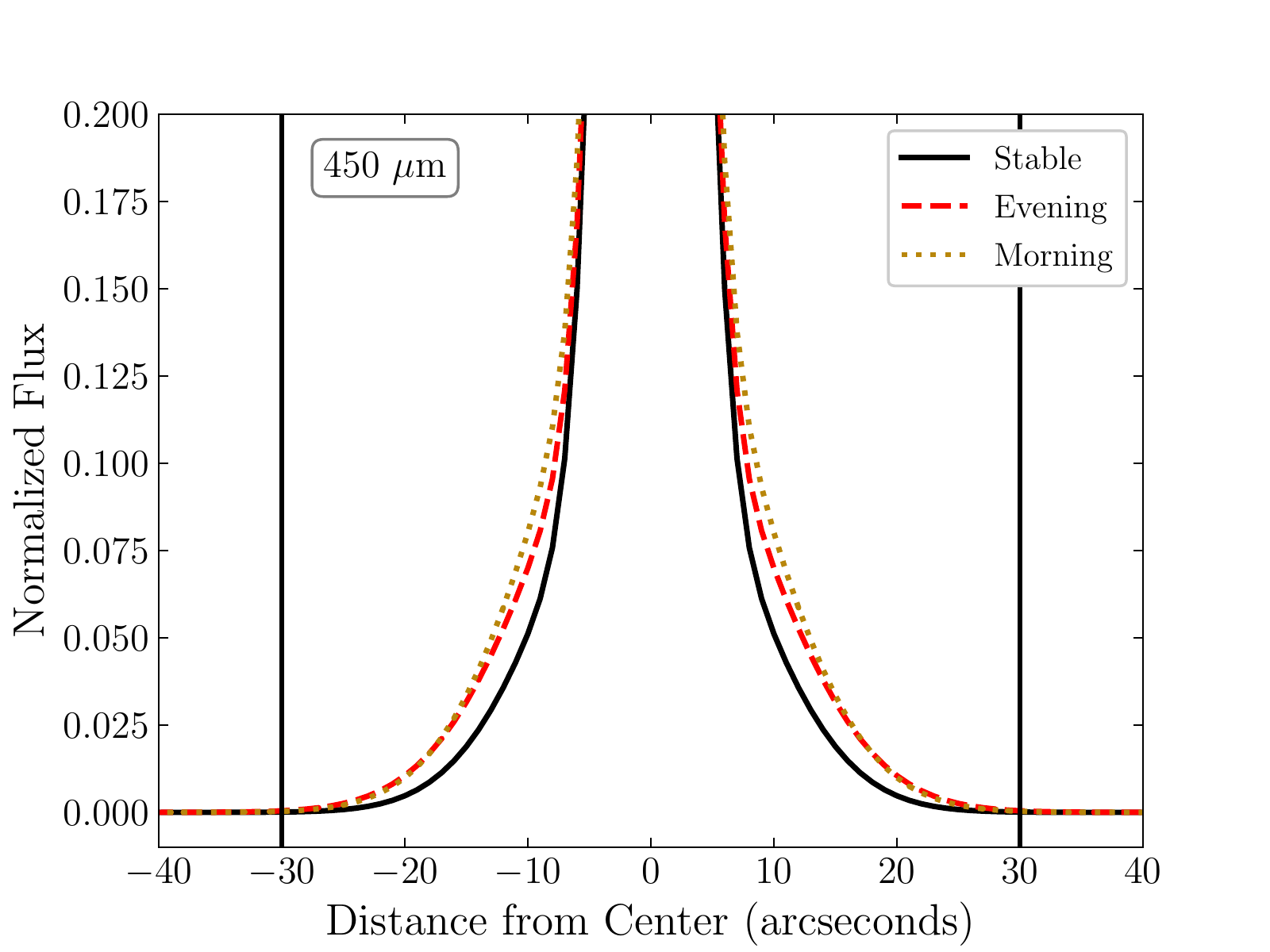}{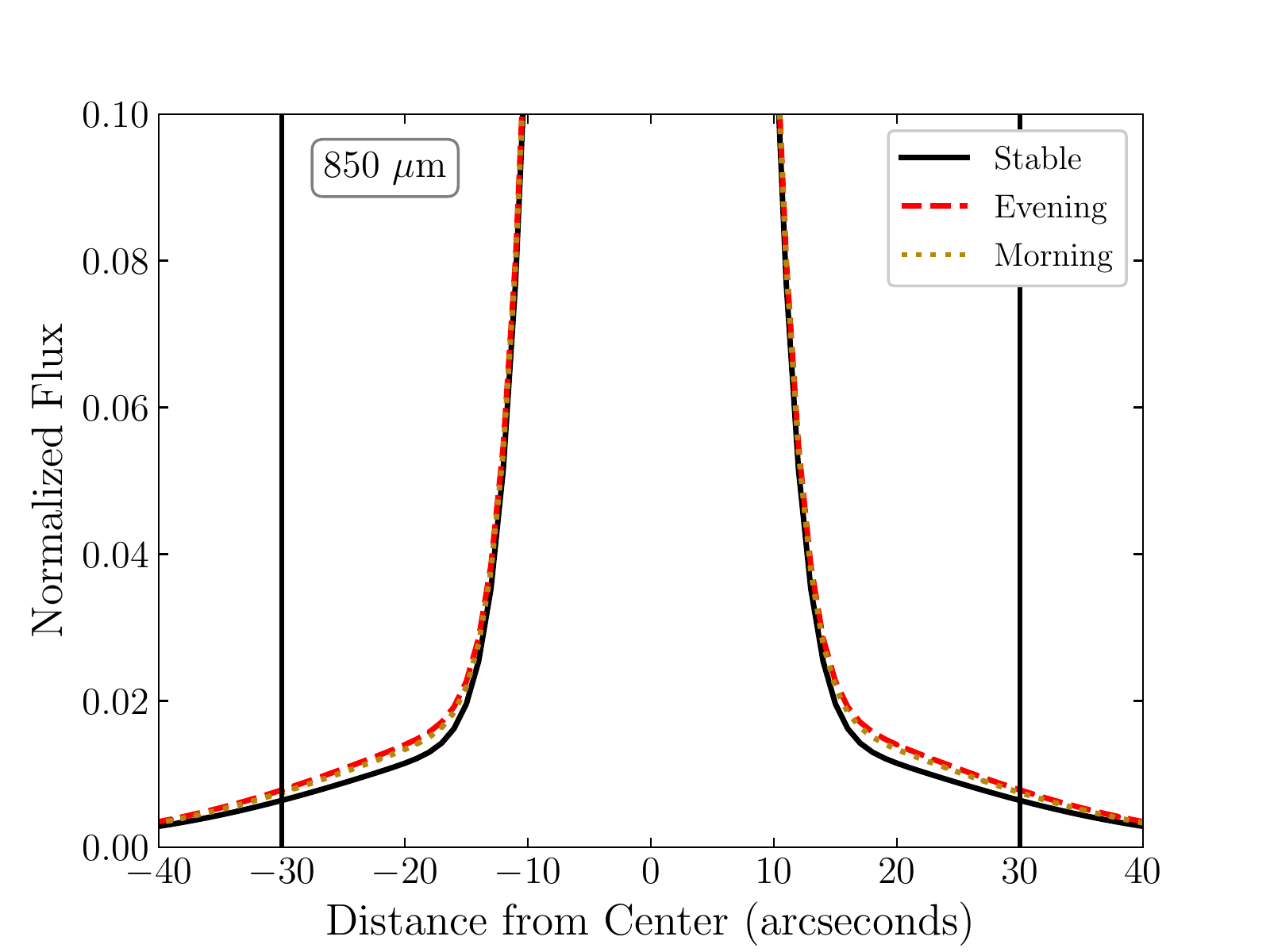}
\caption{The two-component Gaussian beam model described in Equation~\ref{eq:TwoCompGauss} plotted at three different times: Evening (before 04:00 UTC; red, dashed line), Stable period of the night (07:00-17:00, UTC; black, solid line), and Morning (after 17:00 UTC; dotted, gold line) at 450 (\textit{left}) and \mbox{850 $\mu$m} (\textit{right}). Throughout the course of the evening, stable, and morning intervals of a typical observing run, the main-beam component remains largely unchanged while the ``wings'' of the Gaussian flare during the evening and the morning as the secondary-beam (error) component is affected by changes in the dish shape as it heats and cools.}
\label{fig:Wings}
\end{figure*}

To illustrate the dish instability described in the previous section, an empirical measurement 
of the effective beam width (full width at half-maximum; $\mathrm{FWHM}_{\mathrm{eff}}$) can be made by comparing the 
derived Peak and Arcsecond FCFs (using the primary calibrator Uranus). In Equations~\ref{eq:FCFP} and
\ref{eq:FCFA}, the flux of the source 
is measured in \mbox{Jy beam$^{-1}$} and  \mbox{Jy arcsec$^{-2}$}, respectively. 
Therefore, dividing $\mathrm{FCF}_{\mathrm{peak}}$ by
$\mathrm{FCF}_{\mathrm{arcsec}}$ results in 
a measurement of the total beam area in arcsec$^{2}$. Writing
this in terms of the Gaussian $\mathrm{FWHM}$ introduces a factor of $\pi/[4\ln(2)]$,
\begin{equation}
\label{eq:EffBeam}
    \mathrm{FWHM_{\mathrm{eff}}} \approx \sqrt{\frac{\mathrm{FCF}_{\mathrm{peak}}/\mathrm{FCF}_{\mathrm{arcsec}}}{\pi/[4\ln(2)]}}
\end{equation}
We find no significant difference when comparing the effective beam widths derived using 
Equation~\ref{eq:EffBeam} before and after the SMU 
maintenance of 2018 June 30 (see Sections~\ref{subsec:FCFsLongTerm} and 
\ref{subsec:SMUtrouble}). Additionally, there is no significant trend with atmospheric 
transmission. In the top panels of Figure~\ref{fig:FWHMBeam}, we present the empirical 
$\mathrm{FWHM}_{\mathrm{eff}}$ of the beam at 450 and 
\mbox{850 $\mu$m} (left and right panels, respectively) as a 
function of time. The ``U-shape'' is directly related to the Peak FCF trends shown in the 
top panels of Figure~\ref{fig:NightlyFCF}. As discussed previously, the total-flux measurements 
are consistent over the course of the night while the peak-flux measurements 
undergo significant changes in the early evening and late morning. This suggests that flux 
from the primary beam component is diluted into the wings of the profile while staying 
within the 1$\arcmin$ diameter aperture that was used to measure the total flux. The 
\mbox{450 $\mu$m} effective beam FWHM during the stable part of the night (07:00--17:00 UTC) is $\sim10.0\pm0.6\arcsec$. At \mbox{850 microns}, the effective beam FWHM 
during the stable part of the night is $\sim14.4\pm0.3\arcsec$, in good agreement with the 
values of $9.6\arcsec$ and $14.1\arcsec$ derived by D13.

In order to better separate secondary effects such as dish 
imperfections and distortions from 
the main lobe response of the JCMT beam, we model both the 450 and \mbox{850 $\mu$m} 
beam patterns as a combination of two Gaussian functions (D13):
\begin{equation}
\label{eq:TwoCompGauss}
    G_{\mathrm{total}}=\alpha G_{\mathrm{MB}}+\beta G_{\mathrm{S}},
\end{equation}
where each Gaussian profile, $G$, is of the form $\exp[-4\ln(2)\times(r/\theta)^{2}]$,
where $r$ is the radial distance from the centre and $\theta$
is the FWHM of the profile, both measured in units of arcseconds.
The first component, $G_{\mathrm{MB}}$ represents the ``main beam'' 
response.  The second component, $G_{\mathrm{S}}$, is an
approximation of the ``error beam'', which describes the flux in the
shoulders of the profile. $\alpha$ and $\beta$ are 
coefficients describing the relative contribution of each 
component (the amplitudes). The broad error beam includes factors such as sidelobes due to diffraction, static dish deformations, and dish deformations induced by thermal gradients. Though this component is asymmetric, a Gaussian approximation allows for a simple measurement of its approximate volume. Combining the integrals of each 
profile provides the total beam volume, $V$,
\begin{equation}
\label{eq:GaussVol}
V = \frac{\pi}{4\ln(2)}[\alpha(\theta_{\mathrm{MB}})^{2}+\beta(\theta_{\mathrm{S}})^{2}]\mathrm{\:arcseconds}^{2},
\end{equation}
where $\theta_{\mathrm{MB}}$ and $\theta_{\mathrm{S}}$ represent the FWHM of the main beam and secondary (error) beam, respectively. The beam FWHMs were deconvolved by subtracting in quadrature the known size of Uranus at the time of the observation.

All observations of Uranus that were obtained when the atmospheric transmission was greater than 10\% at \mbox{450 $\mu$m} and 25\% at \mbox{850 $\mu$m} since 2011 May 1 were mapped in Az-El coordinates and modeled 
using Equation~\ref{eq:TwoCompGauss}. The peak position and elongation 
angle of the planet were derived using Starlink's 
\citep{cupid2007} implementation of the {\sc{GaussClumps}}
source-extraction algorithm \citep{stutzki1990}. A slice 
through the peak position was then defined 
along the long axis of the beam profile and the position-flux
information was extracted. The derived FWHM values for both the main- and 
error-beam components are statistically constant over both date and observing time.
In contrast, the relative integrated volume of the error beam at each wavelength  (Equation~\ref{eq:GaussVol}) varied as shown in the bottom panels of Figure~\ref{fig:FWHMBeam} (450 and \mbox{850 $\mu$m} on the left and right, respectively). This figure reveals a downward trend in the relative importance of the error beam ($\beta$) in the overall profile in the early evening and an upward trend in the late morning, mimicking the trend seen for the empirically measured effective FWHM (top panels). An error beam with a larger volume indicates that flux from the central region of the beam is being diluted into the broad shoulders of the profile. During the stable portion of the night, the relative volumes of the error beams are $53\% \pm 12\%$ and $26\% \pm 4\%$ at 450 and \mbox{850 $\mu$m}, respectively. The \mbox{850 $\mu$m} value compares favorably with the previously derived $25\%$ by D13, while the \mbox{450 $\mu$m} value suggests the relative volume of the error beam is larger than the previously estimated $40\%$.

To illustrate the flux dilution in the early evening and late morning, Figure~\ref{fig:Wings} shows the summed beam components, taking the median of the parameters $\alpha$, $\beta$, $\theta_{\mathrm{MB}}$, and $\theta_{\mathrm{S}}$ (Equation~\ref{eq:TwoCompGauss}) in the evening, the stable period of the night, and the morning. The profile shoulders increase significantly during the unstable periods when the dish is cooling and heating with the setting and rising of the Sun. According to this symmetric model, at \mbox{450 $\mu$m} there is approximately 10\% more flux in the shoulders in the evening and morning than during the stable period of the night (8\% in the evening and 12\% in the morning). At \mbox{850 $\mu$m}, the beam is much more stable; it has 2\% more flux in the shoulders in the evening and morning when compared with the stable period of the night. These results suggest a slower degradation of the peak flux than seen in Figure~\ref{fig:EveningStableMorning}, but that is a consequence of the assumed symmetry in the idealised model. As in the case of the empirically measured effective FWHM, there is no significant trend with atmospheric transmission.

% Two-component Beam Parameters
\begin{deluxetable}{ccc}
\tablecaption{Two-Component Gaussian Model Parameters (Equation \ref{eq:TwoCompGauss})}
\label{tab:BeamProps}
\tablecolumns{3}
%\tablenum{1}
\tablewidth{0pt}
\tablehead{
\colhead{} &
\colhead{450 $\mu$m}  &
\colhead{850 $\mu$m}
}
\startdata
$\theta_{\mathrm{MB}}^{a}$ (arcsec) & $6.2\pm1.0$  & $11.0\pm1.6$ \\
$\theta_{\mathrm{S}}^{b}$ (arcsec)  & $18.8\pm5.7$ & $49.1\pm8.4$ \\
$\alpha$ & $0.89\pm0.08$ & $0.98\pm0.01$ \\
$\beta$ & $0.11\pm0.08$ & $0.02\pm0.01$ \\
Rel. vol. Main Beam  & $0.47\pm0.12$ & $0.74\pm0.04$ \\
Rel. vol. Error Beam & $0.53\pm0.12$ & $0.26\pm0.04$ \\
\enddata
\tablecomments{$^{a}$The FWHM of the Main Beam component.\\$^{b}$The FWHM of the Error Beam component.} 
\end{deluxetable}

The \textit{expected} effective
 FWHM can be obtained by adding the two FWHM values calculated for each beam component in quadrature along with their 
 respective weights,
 \begin{equation}
     \mathrm{\theta_{\mathrm{eff,2-comp}} = \sqrt{\alpha\theta_{\mathrm{MB}}^{2}+\beta\theta_{\mathrm{S}}^{2}}}
 \end{equation}
Using the median values for $\alpha$, $\beta$, $\theta_{\mathrm{MB}}$ and 
$\theta_{\mathrm{S}}$ across the stable period of the night (see Table \ref{tab:BeamProps}), we calculate an 
effective FWHM derived by the two-component fit of
\mbox{$\theta_{\mathrm{eff,2-comp}} = 8.6\pm1.3\arcsec$}, for \mbox{450 $\mu$m} and \mbox{$12.6\pm1.9\arcsec$} for \mbox{850 $\mu$m}. These are consistent with, though lower than the values of 9.8 and 14.6$\arcsec$ derived by D13.

%%%%%%%%%%%%%%%%%%%%%%%%%%%%%%%%%%%%%%%%
%%%%%%%%%%%%%%%%%%%%%%%%%%%%%%%%%%%%%%%%
%%%%%%%%%%%%%%%%%%%%%%%%%%%%%%%%%%%%%%%%
\section{Secondary-Calibrator Fluxes}
\label{sec:SecCals}

% Secondary Calibrator Archival Comparison Figure
\begin{figure*}
\plotone{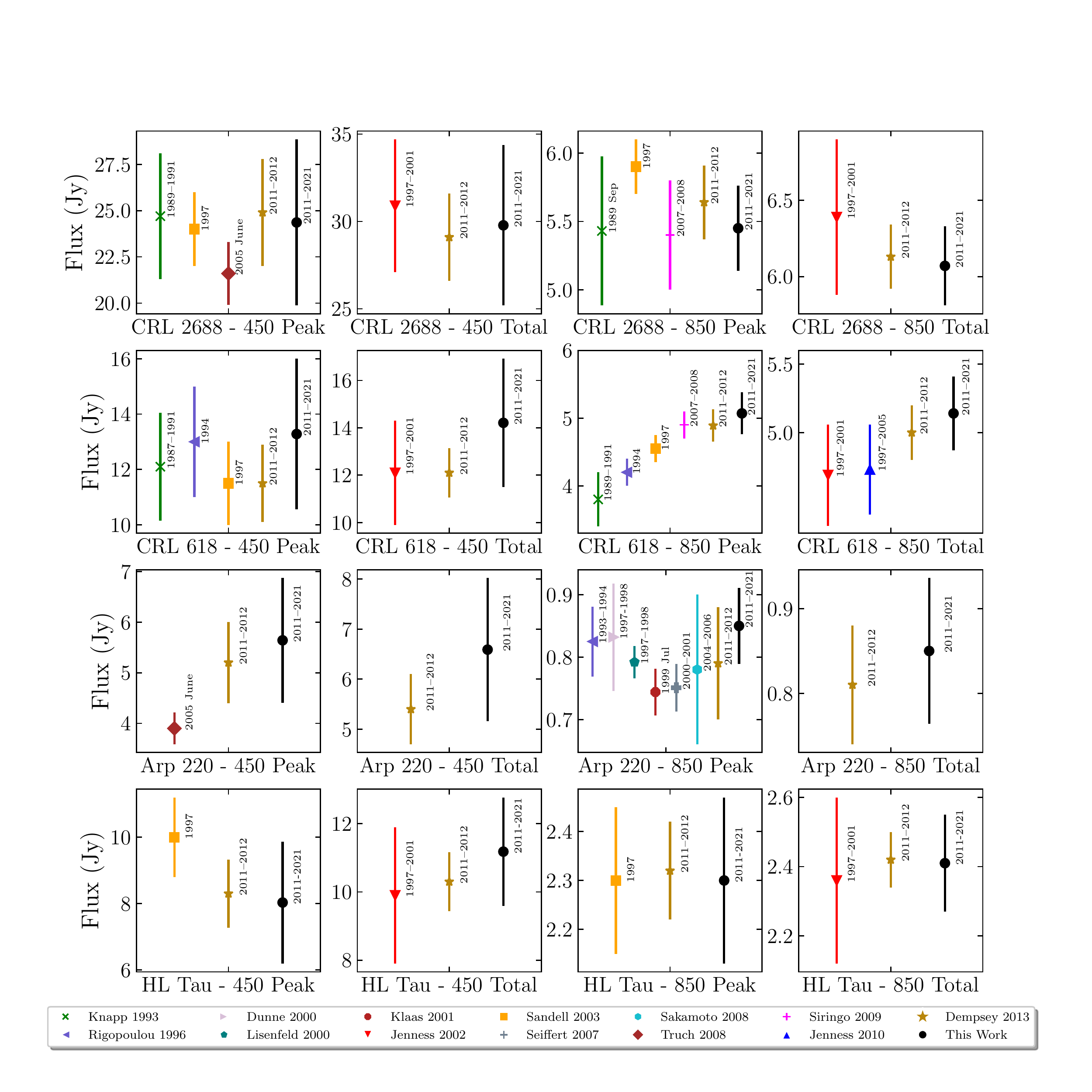}
\caption{Secondary-calibrator fluxes compared in the literature. From top to bottom: \mbox{CRL 2688}, \mbox{CRL 618}, \mbox{Arp 220}, and \mbox{HL Tau}. From left to right: \mbox{450--500 $\mu$m} peak flux, \mbox{450 $\mu$m} total flux, \mbox{800--870 $\mu$m} peak flux, and \mbox{850 $\mu$m} total flux. Reported peak flux measurements were derived by fitting the source, whereas total-flux measurements were derived using aperture photometry. Green exes are the median values of the fluxes presented in \mbox{Table 9} of \cite{knapp1993} (JCMT, UKT14). Left-facing triangles are from \cite{rigopoulou1996} (JCMT, UKT14, \mbox{800 $\mu$m}). Right-facing triangles are from \cite{dunne2000} (JCMT, SCUBA). Green pentagons are from \cite{lisenfeld2000} (JCMT, SCUBA). Red octagons are from \cite{klaas2001} (JCMT, SCUBA). Downward-facing triangles are from \cite{jenness2002} (JCMT, SCUBA). Orange squares are from \cite{sandell2003} (JCMT, SCUBA). Gray crosses are from \cite{seiffert2007} (JCMT, SCUPOL). Cyan hexagons are from \cite{sakamoto2008} (SMA, \mbox{860 $\mu$m}). Brown diamonds are from \cite{truch2008} (BLAST, \mbox{500 $\mu$m)}. Magenta crosses are from \cite{siringo2009} (APEX, LABOCA \mbox{870 $\mu$m}). Blue upward-facing triangles are from \cite{jenness2010} (JCMT, SCUBA). Light brown stars are from D13 (JCMT, SCUBA-2). Black circles are from this work (JCMT, SCUBA-2).}
\label{fig:SecCalLit}
\end{figure*}

%Secondary Calibrator Flux Table
\begin{deluxetable*}{ccccccccc}
\tablecaption{Secondary-calibrator Fluxes.}
\label{tab:SecCals}
\tablecolumns{8}
%\tablenum{1}
\tablewidth{0pt}
\tablehead{
\colhead{Source} &
\colhead{R.A.} &
\colhead{Dec.} &
\colhead{450 $\mu$m Peak} &
\colhead{850 $\mu$m Peak} &
\colhead{450 $\mu$m Total} &
\colhead{850 $\mu$m Total} &
\colhead{Num. 450} &
\colhead{Num. 850} \\
\colhead{} &
\colhead{(J2000)} &
\colhead{(J2000)} &
\colhead{(Jy beam$^{-1}$)} &
\colhead{(Jy beam$^{-1}$)} &
\colhead{(Jy)} &
\colhead{(Jy)}
}
\startdata
CRL 2688 & 21:02:18.27 & 36:41:37.00 & 24.36 $\pm$ 4.49 & 5.45 $\pm$ 0.31 & 29.78 $\pm$ 4.59 & 6.07 $\pm$ 0.26 & 691 & 1482\\
CRL 618 & 04:42:53.67 & 36:06:53.17 & 13.28 $\pm$ 2.72 & 5.07 $\pm$ 0.31 & 14.21 $\pm$ 2.71 & 5.14 $\pm$ 0.27 & 608 & 1237\\
Arp 220 & 15:34:57.27 & 23:30:10.48 & 5.64 $\pm$ 1.23 & 0.85 $\pm$ 0.06 & 6.59 $\pm$ 1.43 & 0.85 $\pm$ 0.09 & 322 & 501\\
HL Tau & 04:31:38.44 & 18:13:57.65 & 8.03 $\pm$ 1.84 & 2.30 $\pm$ 0.17 & 11.18 $\pm$ 1.59 & 2.41 $\pm$ 0.14 & 34 & 37 \\
\enddata
\tablecomments{Secondary-calibrator fluxes derived by applying the FCFs presented in Table~\ref{tab:3EpochFCFs} to the extinction-corrected raw flux measurements obtained during the stable part of the night. The number of \mbox{450 $\mu$m} observations used is less than its \mbox{850 $\mu$m} counterpart due to the exclusion of low-transmission data, which could not be used to fit a reliable peak flux. Secondary-calibrator light curves including all data from evening, stable, and morning observations are presented in Appendix~\ref{appsec:SecCalFluxes}.} 
\end{deluxetable*}

Since D13 presented the first on-sky calibration results, several of the secondary 
calibrators addressed in that work have been found to vary or be otherwise unreliable. 
In the present era, only four flux calibrator sources are routinely observed: 
\mbox{CRL 2688}, \mbox{CRL 618}, \mbox{Arp 220}, and \mbox{HL Tau} 
(in order of most to least frequent). 
Extinction-corrected peak and total-flux measurements were obtained of these 
most commonly used 
secondary calibrator sources (in units of pW) throughout the stable part of the night 
(07:00--17:00 UTC). Then, by applying the FCFs presented in Table~\ref{tab:3EpochFCFs} 
over the full 
10-year datasets, updated 450 and \mbox{850 $\mu$m} peak and total fluxes of 
these sources were derived in units of $\mathrm{Jy}$. The results are presented in 
Table~\ref{tab:SecCals} and full light curves (including evening and morning observations) are presented in Appendix~\ref{appsec:SecCalFluxes}. 
The same analysis was applied using the 
original extinction correction and FCF results derived by D13 on the full datasets and it
was found that the updated values presented in this work reduced the scatter by $1-4\%$
in the \mbox{450 $\mu$m} light curves and $0.5-2\%$ in the \mbox{850 $\mu$m} light curves. This is due to the improved opacity relation and the approximately order of 
magnitude more observations used
in each case. Figure~\ref{fig:SecCalLit} shows the four secondary calibrators
in the context of previous flux measurements drawn from the literature, separating
measurements that were made by fitting the peak of the source from those measurements that
were made by aperture photometry. Some fluxes derived at wavelengths near to the 450 and \mbox{850 $\mu$m} windows of interest are included such as: \mbox{500 $\mu$m} observed by the Balloon-borne Large Aperture Submillimeter Telescope (BLAST), \mbox{800 $\mu$m} observed by the now-retired UKT14 instrument of the JCMT, \mbox{860 $\mu$m} observed by The Submillimeter Array (SMA), and \mbox{870 $\mu$m} observed by the Atacama Pathfinder Experiment Telescope (APEX). See the figure caption for all literature references\footnote{None of the four secondary calibrators presented in this work are consistently observed with ALMA and when dust continuum flux measurements are performed, it is over a much smaller area of the source, focusing on substructures that are unresolved with single-dish telescopes.}. Below, each secondary calibrator is discussed in more detail.

\subsection{CRL 2688}

CRL 2688 (also known as the Egg Nebula; \citealt{CRL2688}) is a protoplanetary nebula located in the constellation of Cygnus at a distance of \mbox{920 pc}. Its significant, consistent thermal dust emission that is bright at submillimetre wavelengths qualify the source to be a robust standard calibrator. The new FCFs derived
in this work reduce the uncertainty in the \mbox{450 $\mu$m} peak and total 
flux by $2\%$ and $4\%$, respectively, when compared with data corrected by 
the original extinction correction and FCFs derived in D13. 
The \mbox{850 $\mu$m} peak and total-flux uncertainties are 
reduced by $0.5\%$ and $1\%$, respectively. 
For \mbox{CRL 2688}, we derive median peak-flux values of \mbox{24.36 $\pm$ 4.49 Jy beam$^{-1}$} and 
\mbox{5.45 $\pm$ 0.31 Jy beam$^{-1}$}
and total-flux values of \mbox{29.78 $\pm$ 4.59 Jy} and \mbox{6.07 $\pm$ 0.26 Jy} at 450 and 
\mbox{850 $\mu$m}, respectively.  The measured fluxes derived in this work are consistent with previous values 
noted in the literature (see Figure~\ref{fig:SecCalLit}). The \mbox{850 $\mu$m} peak flux obtained in 1997 \citep{sandell2003} appears to be anomalously high. While there is no direct $\sim850\mathrm{\:}\mu\mathrm{m}$ measurement during the 2005 BLAST mission \citep{truch2008}, the \mbox{850 $\mu$m} flux derived by the spectral energy distribution (SED) fit is also consistent with the most-recent SCUBA-2 measurements. Compared with D13, both the 450 and the \mbox{850 $\mu$m} peak and total-flux values agree to within $3\%$ with the original work. This source does not appear to vary significantly over time.

\subsection{CRL 618}

CRL 618 (also known as the Westbrook Nebula; \citealt{westbrook1975}) is another strong source of thermal dust emission: a bipolar
reflection nebula illuminated by a central B0 star and located in the constellation of Auriga at a distance of \mbox{920 pc}. The new FCFs derived
in this work reduce the uncertainty in the \mbox{450 $\mu$m} peak and total 
flux by $1\%$ and $2\%$, respectively, when compared with the original extinction 
correction and FCFs derived in D13. The \mbox{850 $\mu$m} peak and total-flux uncertainties are both reduced by $2\%$. 
For \mbox{CRL 618}. we derive median peak-flux values of \mbox{13.28 $\pm$ 2.72 Jy beam$^{-1}$} and \mbox{5.07 $\pm$ 0.31 Jy beam$^{-1}$} 
and total-flux values of \mbox{14.21 $\pm$ 2.71 Jy} and \mbox{5.14 $\pm$ 0.27 Jy} at 450 and 
\mbox{850 $\mu$m}, respectively. Comparing with D13, the \mbox{450 $\mu$m} peak and total-flux values are $16\%\mathrm{\:and\:}18\%$ higher than those 
derived in the previous work, while the \mbox{850 $\mu$m} peak and 
total-flux values are within $4\%$. While \cite{jenness2002,jenness2010} report no variability in the \mbox{850 $\mu$m} flux of \mbox{CRL 618} throughout 1997--2005 (using aperture photometry to measure the total flux), there is an indication that the source is now significantly brighter than it was between 1989--1994. \cite{knapp1993} report 1.3 and \mbox{1.1 mm} data that tentatively show a steady increase with time, while their 450 and \mbox{800 $\mu$m} data showed no increase to within the measured uncertainties. The authors note that the \mbox{450 $\mu$m} emission is likely to be dominated by warm
dust, while longer (near-millimeter wavelengths) are dominated by emission from the small central HII region.
We present 10-year light curves between 2011--2021 in Appendix~\ref{appsec:SecCalFluxes}.  These show no significant overall changes in either peak or total flux during this time period at \mbox{450 $\mu$m}; no significant changes in the total flux at \mbox{850 $\mu$m}; and a slight, but discernible, upward trend in the peak flux at \mbox{850 $\mu$m}. The linear increase in \mbox{850 $mu$m} over this 10-year period appears to be at the rate of \mbox{$0.036\pm0.003$ Jy/yr}. A simple comparison between the most-recent flux measurement of \mbox{5.2 Jy} derived in this work to the flux of
\mbox{3.8 Jy} derived in 1991 by \cite{knapp1993} suggests an average increase of \mbox{$\sim0.046$ Jy/yr}, implying that the brightening of 
\mbox{CRL 618} at \mbox{850 $\mu$m} may have been approximately steady over a 30-year period. It is recommended that \mbox{CRL 618} is used as a calibrator with caution, ensuring that flux measurements on a given night are compared with data obtained over at least the previous several months.

\subsection{Arp 220}

Arp 220 \citep{ARP220,IRASgeneral} is the closest ultra-luminous infrared galaxy (ULIRG)  and the 
most-luminous object in the local universe. The new FCFs derived
in this work reduce the uncertainty in the \mbox{450 $\mu$m} peak and total 
flux by $2\%$ and $3\%$, respectively, when compared to the original extinction 
correction and FCFs derived in D13. The \mbox{850 $\mu$m} uncertainties are the same. 
For \mbox{Arp 220}, we derive median peak-flux values of \mbox{5.64 $\pm$ 1.23 Jy} and \mbox{0.85 $\pm$ 0.06} 
and total-flux values of \mbox{6.59 $\pm$ 1.43 Jy} and \mbox{0.85 $\pm$ 0.09 Jy} at 450 and 
\mbox{850 $\mu$m}, respectively. At \mbox{850 $\mu$m}, the measured peak 
and total flux of the source are statistically indistinguishable. 
As discussed in Section~\ref{sec:FCFs}, a true point source has an equivalent 
value in units of \mbox{Jy beam$^{-1}$} and \mbox{Jy}. Comparing with D13, the \mbox{450 $\mu$m} peak and 
total-flux values are 9\% and 23\% higher than those 
derived in the previous work, while the \mbox{850 $\mu$m} peak and 
total-flux values are 5\% and 8\% higher. These new flux results are otherwise broadly 
consistent with previous measurements in the literature with some evidence of the source dimming, then brightening at \mbox{850 $\mu$m} over the past \mbox{$\sim$30 years}. The \mbox{500 $\mu$m} BLAST data \citep{truch2008} indicates a low flux in 2005 relative to the 2011--2021 measurements, while the predicted \mbox{850 $\mu$m} flux extrapolated from the SED fit is consistent with the flux measured by \cite{sakamoto2008} with the SMA during the same time period. The 10-year light curves presented in Appendix~\ref{appsec:SecCalFluxes} may indicate a slight downward trend in flux since 2012.

\subsection{HL Tau}

HL Tau \citep{HLTAU} is a young T Tauri star still associated with its nascent dust and gas located at a distance of \mbox{140 pc} towards the Taurus constellation. T Tauri stars are variable by nature, so this is the least frequently used flux calibrator source for SCUBA-2 observations. Only $\sim30$ observations have been performed during the stable part of the night over the course of 10 years ($\sim$60 in total). The new FCFs derived
in this work reduce the uncertainty in both the \mbox{450 $\mu$m} peak and total 
flux by $4\%$ when compared with the original extinction 
correction and FCFs derived in D13 while the \mbox{850 $\mu$m} peak and total-flux 
uncertainties are approximately the same. 
For \mbox{HL Tau}, we derive median peak-flux values of \mbox{8.03 $\pm$ 1.84 Jy beam$^{-1}$} and \mbox{2.30 $\pm$ 0.17 Jy beam$^{-1}$} 
and total-flux values of \mbox{11.18 $\pm$ 1.59 Jy} and \mbox{2.41 $\pm$ 0.14 Jy} at 450 and 
\mbox{850 $\mu$m}, respectively. Comparing with D13, the \mbox{450 $\mu$m} peak-flux value is within $3\%$, but the total-flux value is $10\%$ higher than those derived in the previous work 
(this is still consistent to within the uncertainty). The \mbox{850 $\mu$m} peak and 
total-flux values are within $1\%$. \mbox{450 $\mu$m} peak flux measured in 1997 \citep{sandell2003} appears significantly brighter than recent flux measurements. The total \mbox{450 $\mu$m} flux and both \mbox{850 $\mu$m} fluxes are consistent with the other measurements from the literature. Note that \mbox{HL Tau} was measured to have a significantly lower peak flux at both wavelengths in 2014 June/July (see Appendix~\ref{appsec:SecCalFluxes}).

%%%%%%%%%%%%%%%%%%%%%%%%%%%%%%%%%%%%%%%%%
%%%%%%%%%%%%%%%%%%%%%%%%%%%%%%%%%%%%%%%%%
%%%%%%%%%%%%%%%%%%%%%%%%%%%%%%%%%%%%%%%%%

\section{Case Study: Quasar 3C 84}
\label{sec:CaseStudy}

% 3C84 light curves
\begin{figure*}
\plottwo{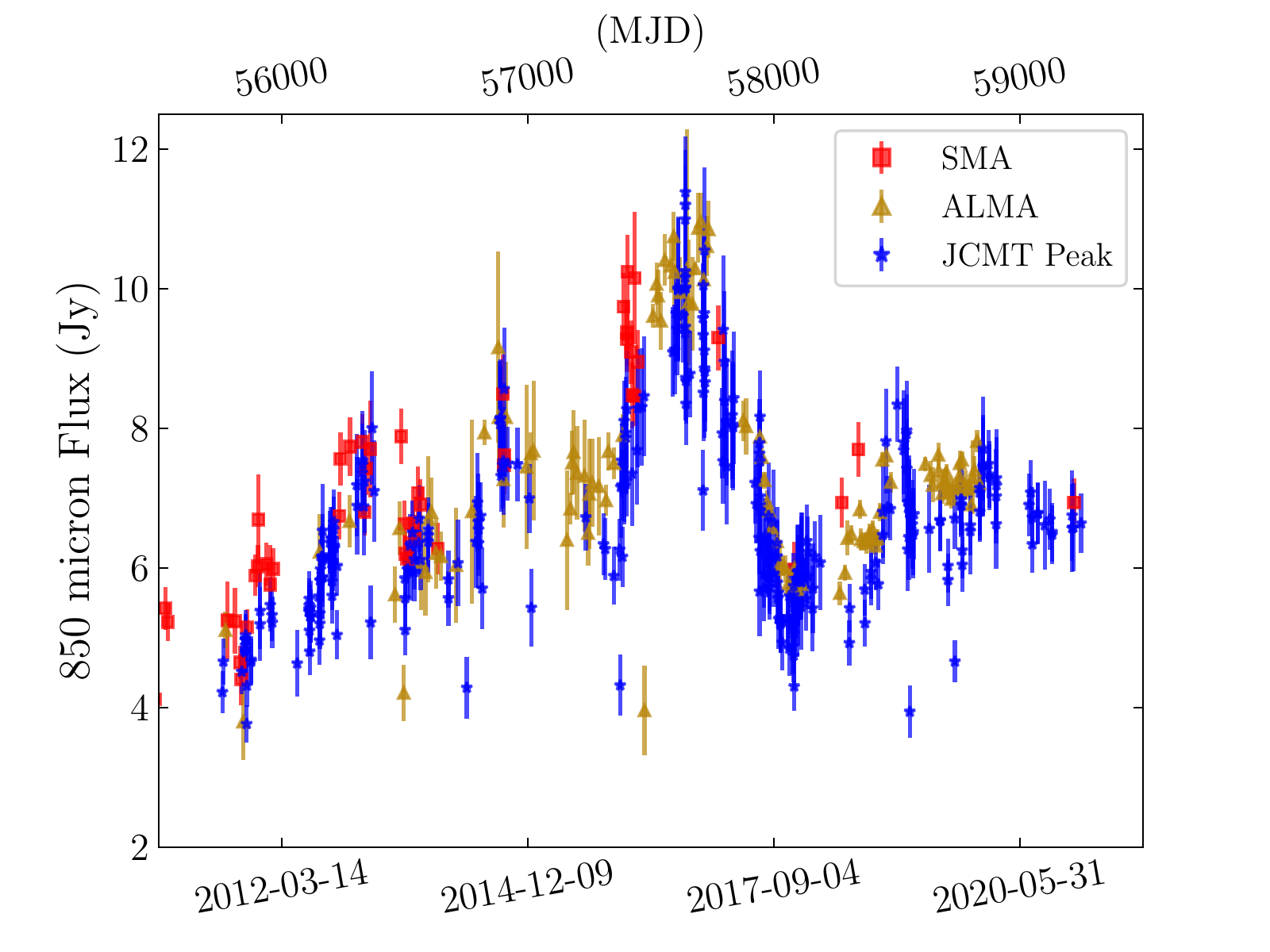}{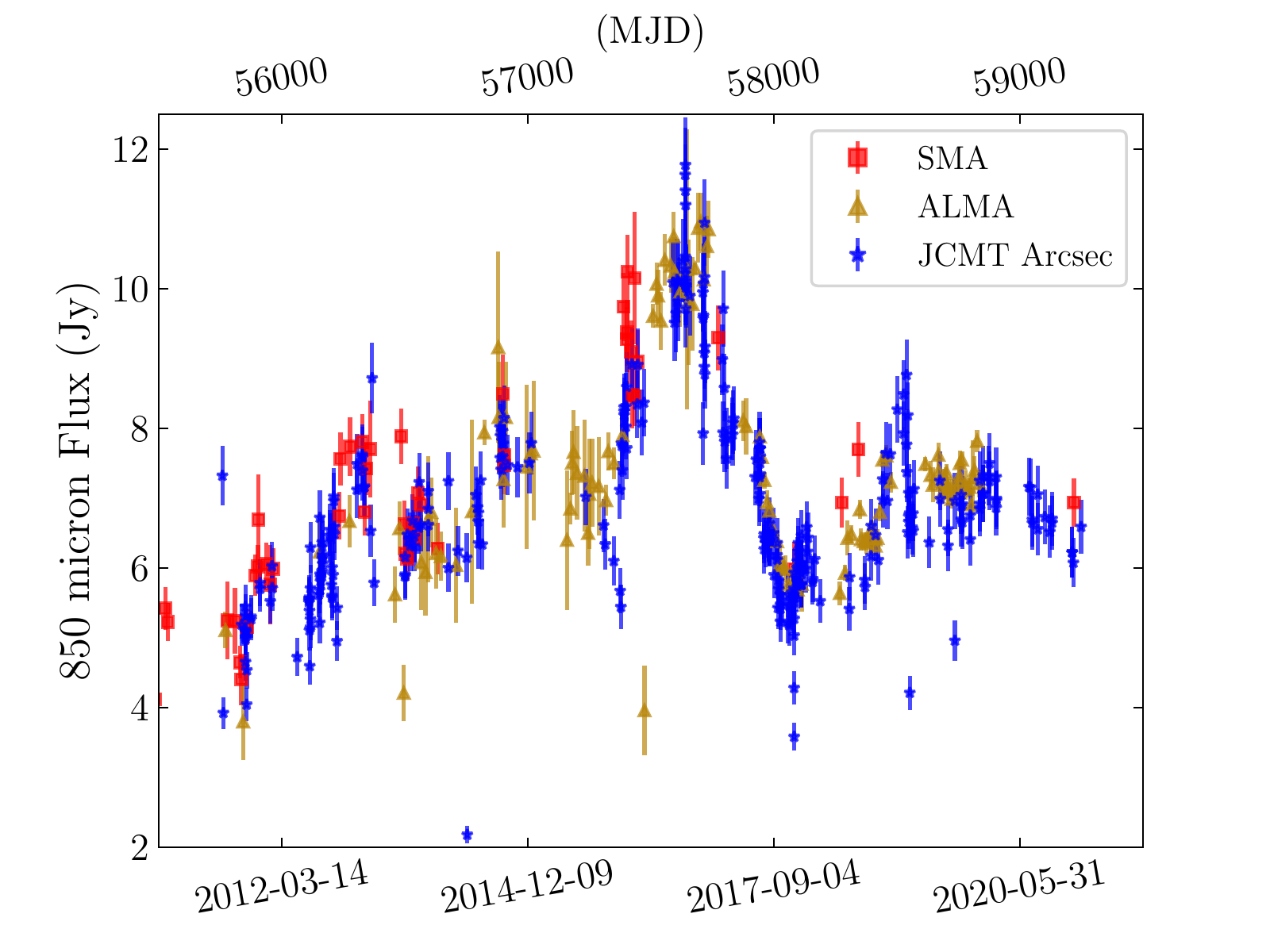}
\caption{The \mbox{850 $\mu$m} flux of Quasar 3C84 plotted over 10 years. Data obtained by the Submillimeter Array (SMA; red squares), the Atacama Large Millimeter/Submillimeter Array (ALMA; gold triangles), and the JCMT (blue stars) are included in each panel. The JCMT calibration employs the results presented in Table~\ref{tab:3EpochFCFs} with the modifications presented in Table \ref{tab:NightlyFCFMods}. \textit{Left:} JCMT flux measured using a Gaussian fit to the data. Peak FCFs are applied. \textit{Right:} JCMT flux is derived using aperture photometry with a \mbox{60 arcsecond} diameter aperture and background subtraction calculated over a \mbox{90--120 arcsecond} (inner, outer) diameter annulus. Arcsecond FCFs are applied. The JCMT flux measurements in each panel are in agreement with one another and largely in agreement with the SMA and ALMA observations.}
\label{fig:CaseStudy}
\end{figure*}

Quasars are common bandpass calibrator sources for interferometers due to their high brightness and small angular size. While quasars are not used at the JCMT as standard flux calibrators due to their variable nature, they are used throughout the night to calibrate the pointing model\footnote{Structural imperfections of the telescope result in small offsets from the commanded slew position. Bright sources with known positions are observed frequently to update the pointing model, yielding a reasonable accuracy of $\sim<3\arcsec$.}.  Figure~\ref{fig:CaseStudy} shows 10 years of pointing-correction observations toward Quasar 3C84 using the JCMT (blue stars), along with calibration observations for the SMA (red squares), and ALMA (gold triangles). The SMA data were obtained using the ``Submillimeter Array Calibrator List'' website\footnote{{\url{http://sma1.sma.hawaii.edu/callist/callist.html}}} and the ALMA data were obtained vis the ``ALMA Calibrator Catalogue''\footnote{{\url{https://almascience.eso.org/alma-data/calibrator-catalogue}}}. The raw JCMT data has been post-processed by applying the new extinction correction presented in Equation~\ref{eq:ORForm} (using the values given in Table~\ref{tab:OR}) and the recommended FCFs presented in Table~\ref{tab:3EpochFCFs} (along with the modifications in Table~\ref{tab:NightlyFCFMods}). In the left panel, the Peak FCFs were used and the resulting peak fluxes were measured with a 2D Gaussian function using {\sc{cupid}}'s {\sc{gaussclumps}} program \citep{CUPID2013,stutzki1990}. Background subtraction is treated as a free parameter in the fit. In the right panel, the Arcsecond FCFs were used and the total fluxes were measured using a 60$\arcsec$-diameter aperture employing a background subtraction where the background level was measured in an annulus with inner and outer diameters of 90 and 120$\arcsec$.

The JCMT data calibrated in both ways is self consistent (the peak flux and total flux should be equal for a true point source) and broadly consistent with the interferometric data. The SMA data appear to consistently return higher fluxes than the ALMA data while the JCMT data agrees better with the ALMA data and in some cases reports lower fluxes. Applying the previous recommended extinction correction and FCFs presented by D13 across the full light curve results in the JCMT data better agreeing with the SMA data, often reporting values higher than those measured at ALMA. The 2D Gaussian fit used to measure the flux for the left panel of Figure~\ref{fig:CaseStudy} is subject to higher uncertainty, especially in poor weather when the flux concentration is diluted from the main beam into the wings of the profile. In this case, the 2D Gaussian fit to the peak will underestimate the true flux. Anomalously low flux measurements are the result of focus issues. While the aperture photometry used to produce the measurements in the right panel of Figure~\ref{fig:CaseStudy} is more robust to variations in the beam, the background subtraction can add uncertainty if there are issues producing a smooth, uniform map background as a result of uncertainties in the data reduction. 

These results indicate that it is not only the applied extinction correction and FCFs for a given data set that fully determine the uncertainty in the measurement; the flux determination method also contributes to the scatter. Assuming a robust background subtraction can be performed, the aperture photometry method mitigates the affect of beam dilution due to poor weather or temperature gradients across the dish and is, overall, preferred over peak fitting.  

%%%%%%%%%%%%%%%%%%%%%%%%%%%%%%%%%%%%%%%%%
%%%%%%%%%%%%%%%%%%%%%%%%%%%%%%%%%%%%%%%%%
%%%%%%%%%%%%%%%%%%%%%%%%%%%%%%%%%%%%%%%%%

%%%%%%%%%%%%%%%%%%%%%%%%%%%%%%%%%%%%%%%%
%%%%%%%%%%%%%%%%%%%%%%%%%%%%%%%%%%%%%%%%
%%%%%%%%%%%%%%%%%%%%%%%%%%%%%%%%%%%%%%%%
%\section{Starlink Implementation}
%\label{sec:StarlinkUpdates}
%%%%%%%%%%%%%%%%%%%%%%%%%%%%%%%%%%%%%%%%
%%%%%%%%%%%%%%%%%%%%%%%%%%%%%%%%%%%%%%%%
%%%%%%%%%%%%%%%%%%%%%%%%%%%%%%%%%%%%%%%%

%-Highlight what the data reduction assumes when a user applies ORAC-DR to 
%their current project. ``As of Starlink 21a....the new opacity relations will be used and the  FCFs in this paper will be automatically applied depending on when your data were observed in both date and time?''

%%%%%%%%%%%%%%%%%%%%%%%%%%%%%%%%%%%%%%%%
%%%%%%%%%%%%%%%%%%%%%%%%%%%%%%%%%%%%%%%%
%%%%%%%%%%%%%%%%%%%%%%%%%%%%%%%%%%%%%%%%
\section{Summary}
\label{sec:Summary}
%%%%%%%%%%%%%%%%%%%%%%%%%%%%%%%%%%%%%%%%
%%%%%%%%%%%%%%%%%%%%%%%%%%%%%%%%%%%%%%%%
%%%%%%%%%%%%%%%%%%%%%%%%%%%%%%%%%%%%%%%%

Since SCUBA-2's commissioning phase ended a decade ago, thousands of observations of calibrator sources have been obtained in a wide range of weather conditions, spanning the evening through the late morning. D13 derived the initial atmospheric extinction correction and FCFs based on one year of data between 2011--2012. Since that time, the telescope has undergone hardware updates and changes that have impacted the calibration at both 450 and \mbox{850 $\mu$m}. Below is a summary of the updated recommendations for calibrating both archival and new SCUBA-2 data obtained by the JCMT:

\begin{enumerate}
    \item A new opacity relation to be used in the atmospheric extinction correction was derived by minimising the variance in the corrected flux of the primary calibrator, Uranus, and secondary calibrators, \mbox{CRL 2688} and \mbox{CRL 618} as a function of atmospheric transmission. The form of the non-linear relation is presented in Equation~\ref{eq:ORForm} with the best-fit parameters presented in Table~\ref{tab:OR}. The correction is performed by substituting the opacity relation into Equation~\ref{eq:EXTCOR} (see Section~\ref{sec:OR} for details).
    
    \item After the new extinction correction is applied to the raw data, the flux is calibrated by applying the (multiplicative) Peak and Arcsecond FCF values presented in Table~\ref{tab:3EpochFCFs} (see Section~\ref{sec:FCFs} for details). The FCFs vary as a function of date, accounting for: a) the upgrade of the thermal-filter stack in 2016 November, which improved transmission at \mbox{850 $\mu$m} and b) the SMU maintenance in 2018 May/June, which improved the flux concentration and lowered the FCF values at both wavelengths (see Section~\ref{subsec:SMUtrouble}).
    
    \item For data observed outside the hours of \mbox{07:00--17:00 UTC}, temperature gradients across the dish due to the setting and rising of the sun necessitate corrections to the Peak FCFs to account for beam degradation (wherein flux from the main-beam component is diluted into the wings of the approximately Gaussian beam profiles). These linear corrections as a function of time of night are presented in Table~\ref{tab:NightlyFCFMods} (see Section~\ref{sec:FCFsnightly} for details). The effective beam widths and relative volumes of the main and error (secondary) beam are presented in detail as a function of time in Section~\ref{subsec:Beam}. Results of the two-component beam models can be found in Table \ref{tab:BeamProps}.
\end{enumerate}

The updated opacity relations and stable time-of-night FCFs presented in this work (first two points) are now the default used in the {\sc{starlink}} data reduction and calibration software as of the software's 21A release\footnote{\url{https://starlink.eao.hawaii.edu}}. If desired, the FCF corrections for data observed before 07:00 UTC or after 17:00 UTC (during the evening and morning) must be applied manually\footnote{Detailed information can be found on the East Asian Observatory tutorial page: \url{https://www.eaobservatory.org/jcmt/science/reductionanalysis-tutorials/}}. Between 07:00 and 17:00 UTC, the portion of the night that is most stable to temperature gradients that cause dish deformation, the total and peak calibrator flux uncertainties measured at \mbox{450 $\mu$m} are found to be 14\% and 17\%, respectively. Measured at \mbox{850 $\mu$m}, the uncertainties are 6\%, and 7\%. During the evening (pre 07:00 UTC) and morning (post 17:00 UTC), the increased scatter and uncertainty in the linear-fit corrections of the Peak FCF values result in a few percent increase in the relative calibration errors presented in Table~\ref{tab:3EpochFCFs}.

In Section~\ref{sec:SecCals}, we applied the new extinction correction and FCFs to the four most commonly used secondary calibrators: \mbox{CRL 2688}, \mbox{CRL 618}, \mbox{Arp 220}, and \mbox{HL Tau} to derive updated flux estimates and to investigate long-term changes in calibrator brightnesses. While \mbox{CRL 2688} and \mbox{HL Tau} are consistent with previous measurements derived in the literature and show no discernible variability over time, \mbox{CRL 618} and \mbox{Arp 220} present less reliable light curves and require significant updates to the results presented by D13 (see Figure~\ref{fig:SecCalLit}). 10-year light curves for all calibrator sources are presented in Appendix~\ref{appsec:SecCalFluxes}, Table \ref{tab:SecCalFlux} (machine-readable tables are available in the online version).

The updated calibrations are also applied to \mbox{850 $\mu$m} observations of Quasar 3C84 (a JCMT pointing offset calibrator) and compared with fluxes obtained by SMA and ALMA. The measured JCMT fluxes corrected by the Peak and Arcsecond FCFs are consistent with one another and are broadly consistent with the interferometric data. The method of flux determination contributes to the overall uncertainty of the measurement (see Section~\ref{sec:CaseStudy} for details).

Calibrator maps can be downloaded from the Canadian Astronomy Data Centre (CADC)\footnote{{\url{https://www.cadc-ccda.hia-iha.nrc-cnrc.gc.ca/}}} by searching ``Proposal ID: JCMTCAL''. Up-to-date information is posted on the JCMT SCUBA-2 calibration website: {\url{https://www.eaobservatory.org/jcmt/instrumentation/continuum/scuba-2/calibration/}}. 

\acknowledgments

The authors wish to recognise and acknowledge the very significant 
cultural role and reverence that the summit of
Maunakea has always had within the indigenous Hawaiian community. 
We are most fortunate to have the opportunity
to conduct observations from this mountain.
The comments provided by the anonymous referee have
significantly strengthened this work.
The James Clerk Maxwell Telescope is operated by the East Asian 
Observatory on behalf of The National Astronomical Observatory of 
Japan; Academia Sinica Institute of Astronomy and Astrophysics; the 
Korea Astronomy and Space Science Institute; the Operation, 
Maintenance and Upgrading Fund for Astronomical Telescopes and 
Facility Instruments, budgeted from the Ministry of Finance (MOF) 
of China and administrated by the Chinese Academy of Sciences 
(CAS), as well as the National Key R\&D Program of China (No. 
2017YFA0402700). Additional funding support is provided by the 
Science and Technology Facilities Council of the United Kingdom and 
participating universities in the United Kingdom and Canada. 
Additional funds for the construction of SCUBA-2 were provided by 
the Canada Foundation for Innovation. This research used the 
facilities of the Canadian Astronomy
Data Centre operated by the National Research Council of
Canada with the support of the Canadian Space Agency.
The James Clerk Maxwell Telescope has historically been operated by the Joint Astronomy Centre on behalf of the Science and Technology Facilities Council of 
the United Kingdom, the National Research Council of Canada and the Netherlands Organisation for Scientific Research. The Starlink software \citep{currie2014}
is currently supported by the East Asian Observatory.

\vspace{5mm}

%%%%%%%%%%%%%%%%%%%%%%%%%%%%%%%%%%%%%%%%
%%%%%%%%%%%%%%%%%%%%%%%%%%%%%%%%%%%%%%%%
%%%%%%%%%%%%%%%%%%%%%%%%%%%%%%%%%%%%%%%%
% Facility and Software Used

\facilities{JCMT}

\software{astropy \citep{astropy},  
          matplotlib \citep{matplotlib},
          aplpy \citep{aplpy}, 
          Starlink \citep{currie2014}, 
          }
%%%%%%%%%%%%%%%%%%%%%%%%%%%%%%%%%%%%%%%%
%%%%%%%%%%%%%%%%%%%%%%%%%%%%%%%%%%%%%%%%
%%%%%%%%%%%%%%%%%%%%%%%%%%%%%%%%%%%%%%%%

\appendix

%%%%%%%%%%%%%%%%%%%%%%%%%%%%%%%%%%%%%%%%
%%%%%%%%%%%%%%%%%%%%%%%%%%%%%%%%%%%%%%%%
%%%%%%%%%%%%%%%%%%%%%%%%%%%%%%%%%%%%%%%%
\section{Flux Uncertainty}
\label{appsec:SingleObs}

%%%%%%%%%%%%%%%%%%%%%%%%%%%%%%%%%%%%%%%%
%%%%%%%%%%%%%%%%%%%%%%%%%%%%%%%%%%%%%%%%
%%%%%%%%%%%%%%%%%%%%%%%%%%%%%%%%%%%%%%%%

% Intrinsic Scatter in the Peak and Arcsec FCF
\begin{figure*}
\plotthree{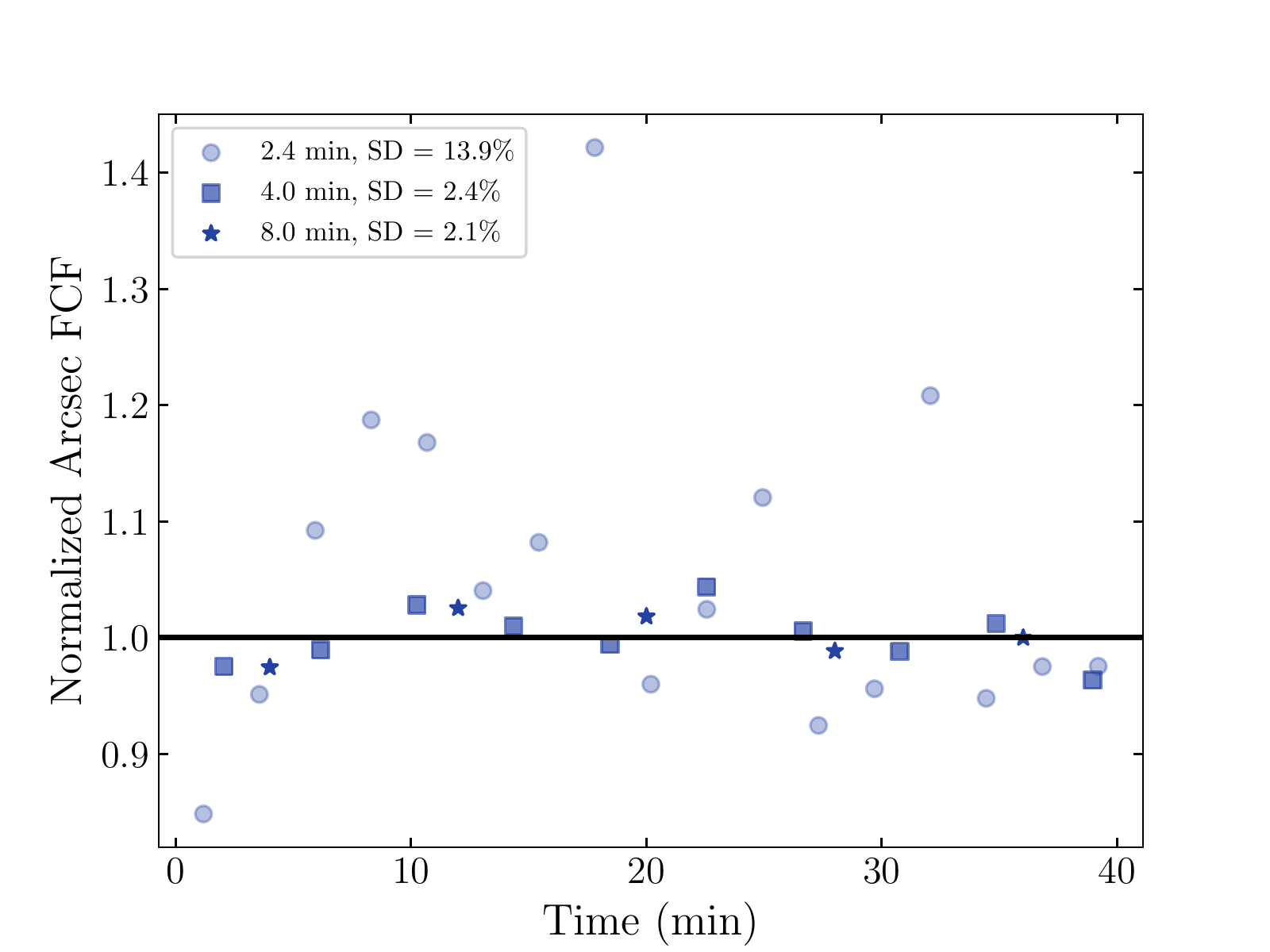}{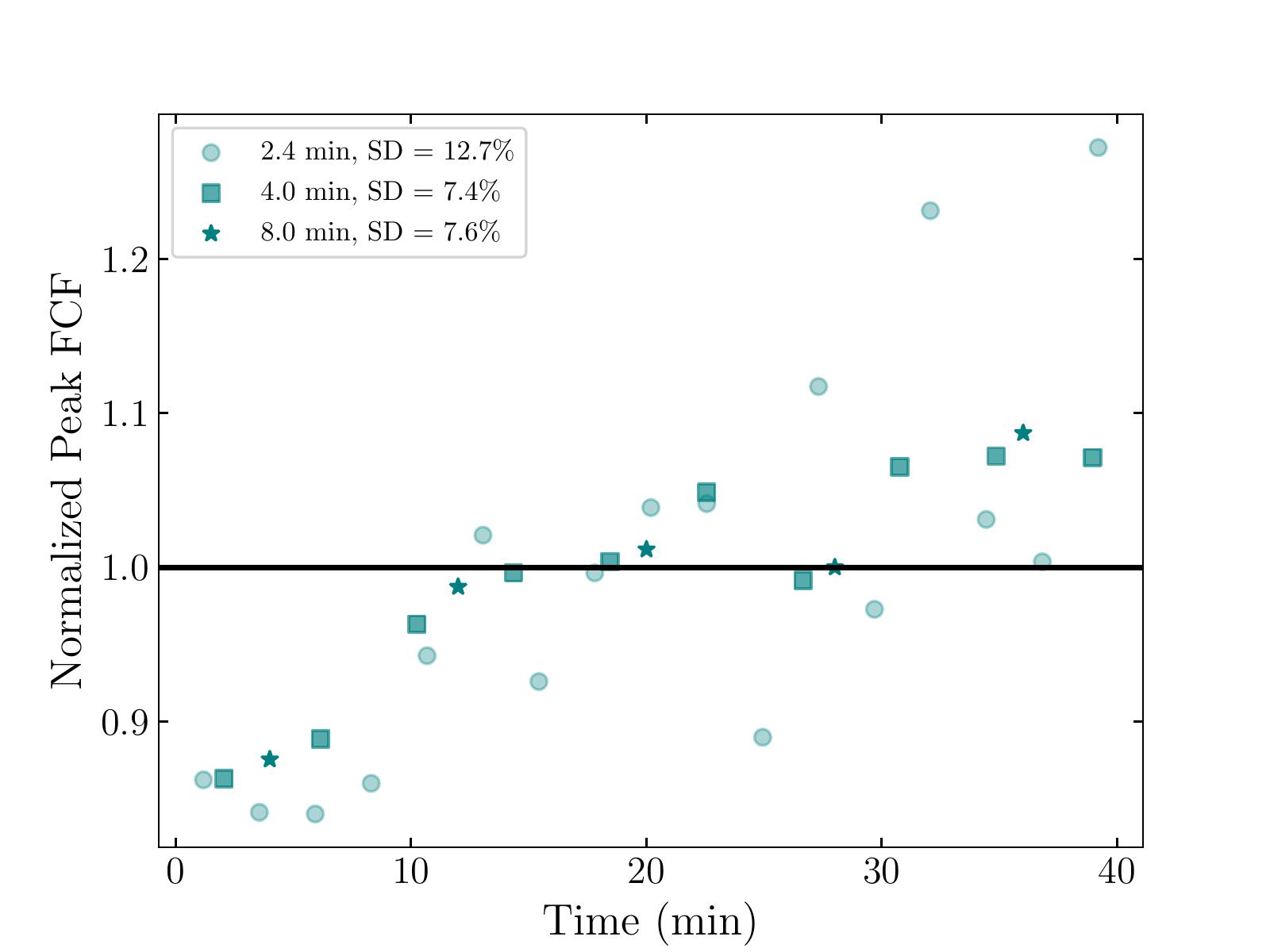}{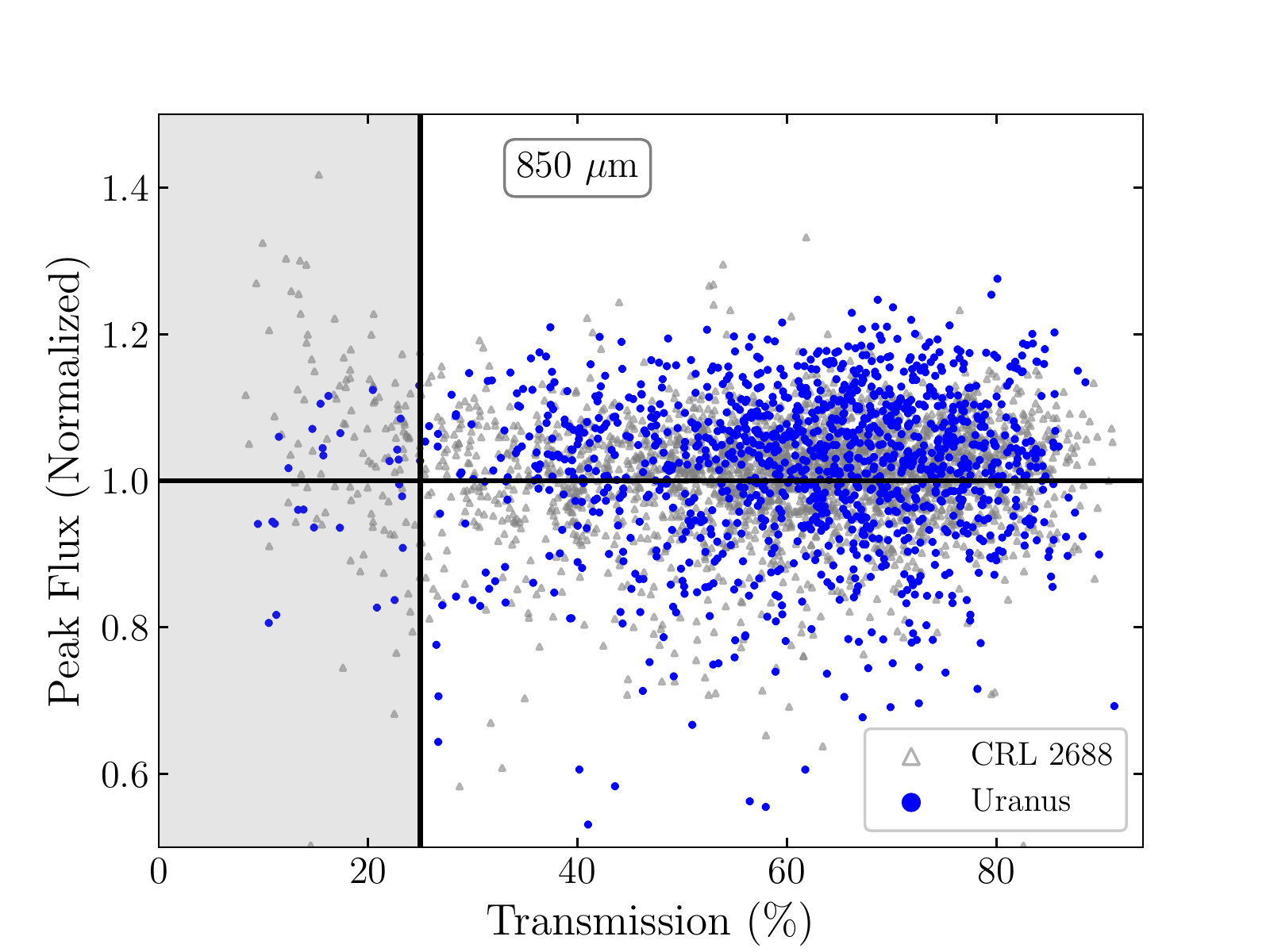}
\caption{\textit{Top:} Normalized Arcsecond (\textit{left}) and Peak (\textit{right}) FCFs derived for $\sim2.4\mathrm{\:minute}$ (circles), $4\mathrm{\:minute}$ (squares), and $8\mathrm{\:minute}$ (stars) subsections of the raw data for a single, 40 minute, \mbox{850 $\mu$m} observation of \mbox{CRL 2688}. The data were taken with $\tau_{225}\sim0.3$ and $\mathrm{Airmass}$ ranging between $1.18-1.31$. Solid, horizontal lines are shown at a value of 1.0 to represent the normalized median FCF values. \textit{Bottom:} The normalised \mbox{850 $\mu$m} peak fluxes of Uranus (blue circles) and \mbox{CRL 2688} (gray triangles) as a function of atmospheric transmission. The peak-flux response of the primary calibrator Uranus is flat across the full transmission range whereas the peak flux of CRL 2688 is over-corrected below a transmission of 25\%.}
\label{fig:40min850}
\end{figure*}

While a full analysis of the myriad contributions to flux uncertainty is beyond the scope of this paper, here we provide a broad discussion of the more notable issues. There is seemingly no obvious parameter that is the singular dominant cause of the inherent scatter in the calibrator light curves. Before the light is received by the detector, the effect of the bright and variable atmosphere at submillimeter wavelengths is removed by an approximation that assumes an atmospheric model that is plane-parallel. The true atmosphere may not be well described by this model and low-lying variations in weather conditions with scale heights comparable to several times the height of the JCMT facility complicate the matter further. In addition, the calibration of the water-vapor monitor brightness temperatures and the parabolic fit used to estimate the properties of the \mbox{183 GHz} water line introduce uncertainties in the PWV measured for use in the extinction correction. Subtle atmospheric effects such as the submillimeter seeing, wind speed, ambient-temperature changes, and humidity will all contribute lower-level contributions to beam distortions and, subsequently, flux measurements. Measuring fluxes by fitting the peak of a compact source will be affected more than measurements performed using aperture photometry as each of the more subtle atmospheric effects works to dilute the main-beam flux. 

The GORE-TEX\texttrademark$\:$ membrane that protects the dish from the high-altitude elements is not 100\% transmissive at submillimeter wavelengths. During the commissioning of the POL-2 instrument, the membrane was removed in order to increase the sensitivity at \mbox{450 $\mu$m} and to better understand its affect on the instrumental polarisation (IP). The weather conditions that allow for the JCMT to observe while the membrane is removed are pristine and limited. A provisional analysis of calibrator observations observed during the membrane removal suggests a decrease in \mbox{850 $\mu$m} Peak FCF values by 6--7\% and a decrease in \mbox{850 $\mu$m} Arcsecond FCF values by 4--5\%. The uncertainty, however, is increased as the dish is exposed and unstable due to wind\footnote{While the membrane is removed, the maximum wind speed during which the JCMT can operate is \mbox{18 mph}.}. At \mbox{450 $\mu$m}, the increase in flux uncertainty masked any improvement in sensitivity due to the removal of the membrane. 

Finally, instrumental noise, non-cooled optical components, and uncertainties in the {\sc{makemap}} routine's modeling and removal of noise components from the time stream \citep[or introduction of map artifacts, see][for details]{chapin2013} can all introduce uncertain contributions to the final flux. 

None of these individual factors alone dominate the scatter. Therefore, quantifying each 
noise contribution is a difficult task that requires a significant amount of observations taken under controlled conditions. Even then, there are atmospheric conditions that cannot be measured. To assess the changes in FCFs taken throughout approximately stable conditions during a time of poor atmospheric transmission (where the FCFs have a larger scatter and tend to over-correct, see Figure~\ref{fig:RawCorFluxOR}), a single 40 minute long observation of \mbox{CRL 2688} was performed. The observation took place on 2018 October 10 at 08:15 (UTC) (stable period of the night) in 
\mbox{Weather Band 5}, ($\tau_{225} = 0.3$). The setting source had an airmass that ranged from 1.18--1.31 over the course of the observation, yielding an atmospheric transmission range at \mbox{850 $\mu$m} from 21--17\%\footnote{Note that these observations were excluded from the analysis in the main paper as the atmospheric transmission did not pass the 25\% threshold.}.

This \mbox{40 minute} observation was subdivided into 
17 $\sim2.4\mathrm{\:minute}$ maps using {\sc{makemap}}'s {\sc{shortmaps}} command and FCF values were derived for each sub-observation. A \mbox{2.4 minute} integration time was chosen to match the typical individual calibrator observations taken in the wettest weather. The {\sc{shortmaps}} command first produces a ``total'' map using all data. Then, the map for each \mbox{2.4 minute} period (each \textit{shortmap}) is constructed by adding the total map onto a 
map made from the time-series residuals for each shortmap.  In this way, each shortmap is made using the same final sky model (produced for the total map after noise subtraction; see \cite{chapin2013} for details) and the increased noise in each shortmap is due only to the shorter amount of input time-series data. Using the same sky model for each observation reduces the scatter in the FCFs when compared to performing a series of individual observations (which will each produce a slightly different sky model based on the available data). The normalized \mbox{850 $\mu$m} results are shown in the top panels of Figure~\ref{fig:40min850} (circles). 

The Arcsecond FCFs are consistent with a constant value over the 40 minute observation despite the increasing transmission as the source was setting. The Peak FCFs, however, show an obvious increasing trend as the transmission decreases. In the bottom panel of Figure~\ref{fig:40min850}, the peak-flux response as a function of atmospheric transmission is shown for Uranus and  \mbox{CRL 2688} over all calibrator observations (compare with the right panel of Figure~\ref{fig:RawCorFluxOR}). Uranus's peak flux is constant across all atmospheric transmissions whereas the peak flux of \mbox{CRL 2688} is clearly over-corrected below a transmission of 25\%. \mbox{CRL 2688} is not a point source even at the JCMT's resolution of $\sim14.5\arcsec$ at \mbox{850 $\mu$m}, and it is the recovery of this extended structure in poor transmission, short integration time (i.e. high uncertainty) observations that causes an artificial depression in the peak flux while the total flux across the source remains conserved. Therefore, it is especially important to calibrate poor transmission maps of compact sources with the Arcsecond FCFs and perform aperture photometry to measure robust fluxes. An investigation of each shortmap, individually, revealed that the $2.4\mathrm{\:minute}$ typical integration time resulted in SNR values ranging from $2.8-5.1$ (decreasing with time as the source was setting). Therefore, additional sets of shortmaps were produced that divided the $\sim$40 minute observation into 4 and 8 minute sub-observations and the analysis was repeated (see squares and stars plotted in Figure~\ref{fig:40min850}). The increased integration times improved each flux measurement resulting in minimum SNR values of $\sim$4.4. There is no significant difference between the FCF distributions derived by the 4 and 8 minute shortmap datasets, so for efficiency, calibrator integration times are now increased to $4\mathrm{\:minutes}$ in the rare times that SCUBA-2 is used during very wet weather conditions in order to improve the flux estimates. The poor transmission during the observation resulted in an SNR that was insufficient to conduct the same analysis at \mbox{450 $\mu$m}.

%%%%%%%%%%%%%%%%%%%%%%%%%%%%%%%%%%%%%%%%
%%%%%%%%%%%%%%%%%%%%%%%%%%%%%%%%%%%%%%%%
%%%%%%%%%%%%%%%%%%%%%%%%%%%%%%%%%%%%%%%%
\section{Secondary-Calibrator Light Curves}

\label{appsec:SecCalFluxes}
%%%%%%%%%%%%%%%%%%%%%%%%%%%%%%%%%%%%%%%%
%%%%%%%%%%%%%%%%%%%%%%%%%%%%%%%%%%%%%%%%
%%%%%%%%%%%%%%%%%%%%%%%%%%%%%%%%%%%%%%%%

% Ten year light curves
\begin{figure*}
\plotfour{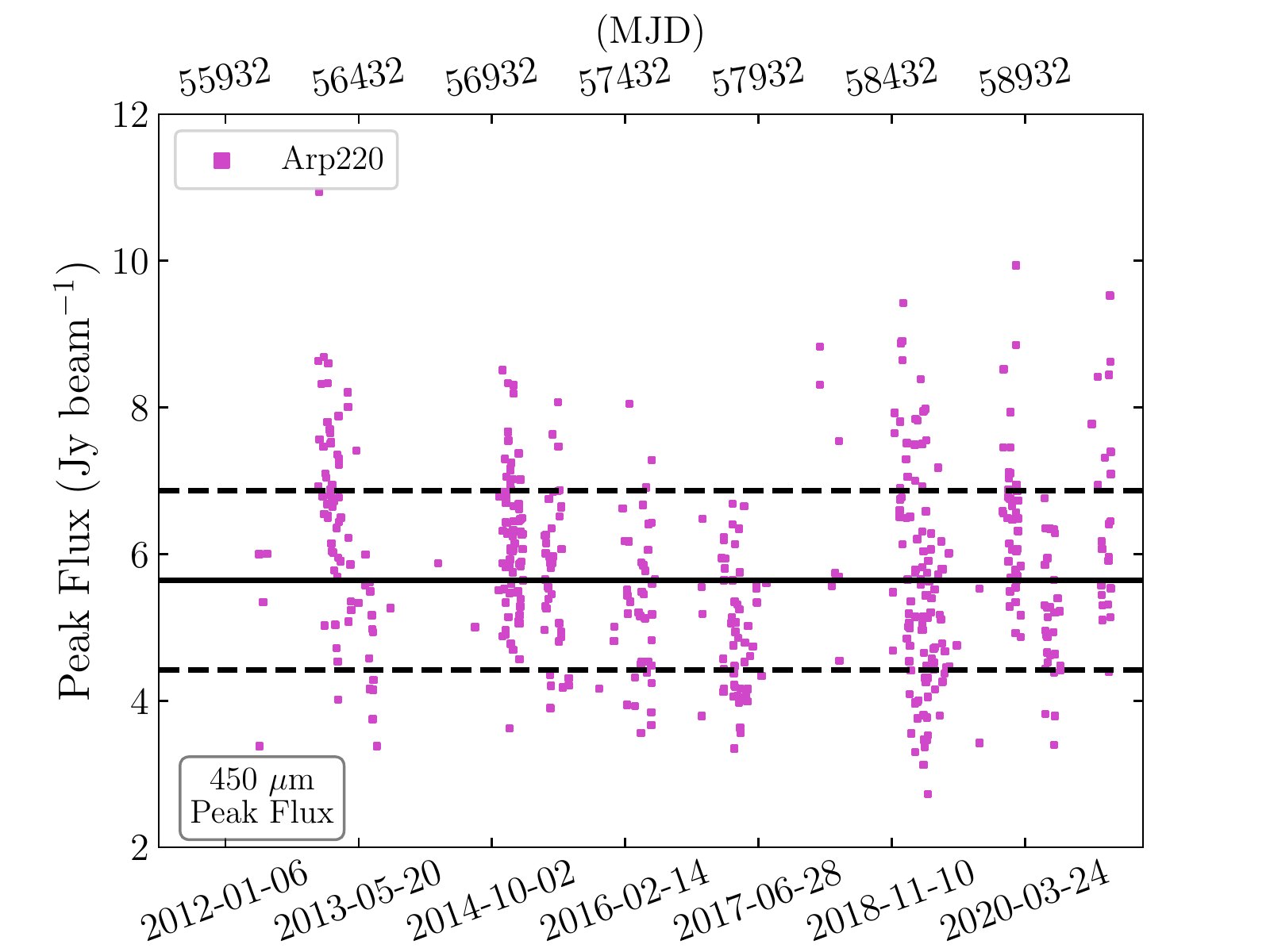}{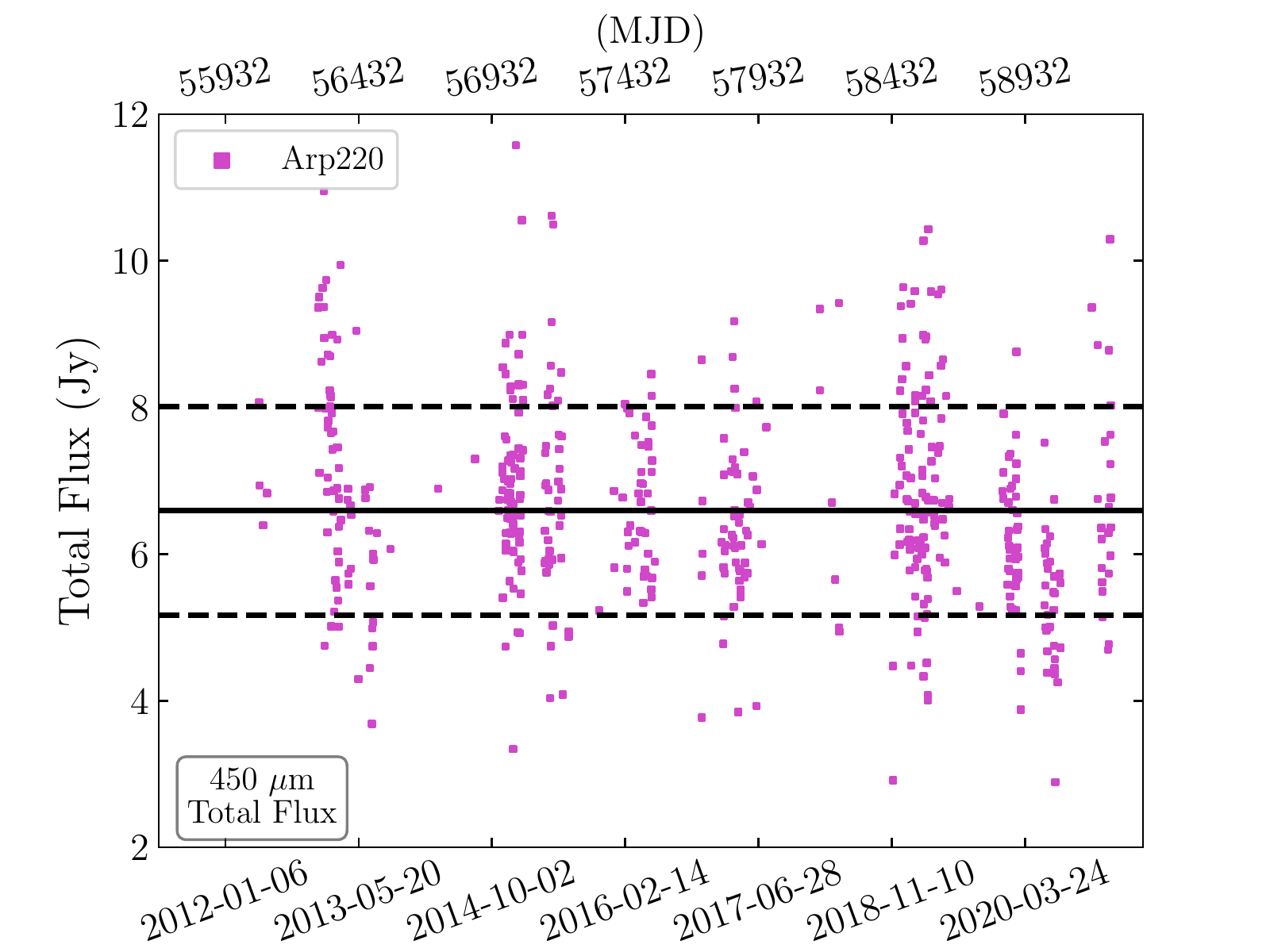}{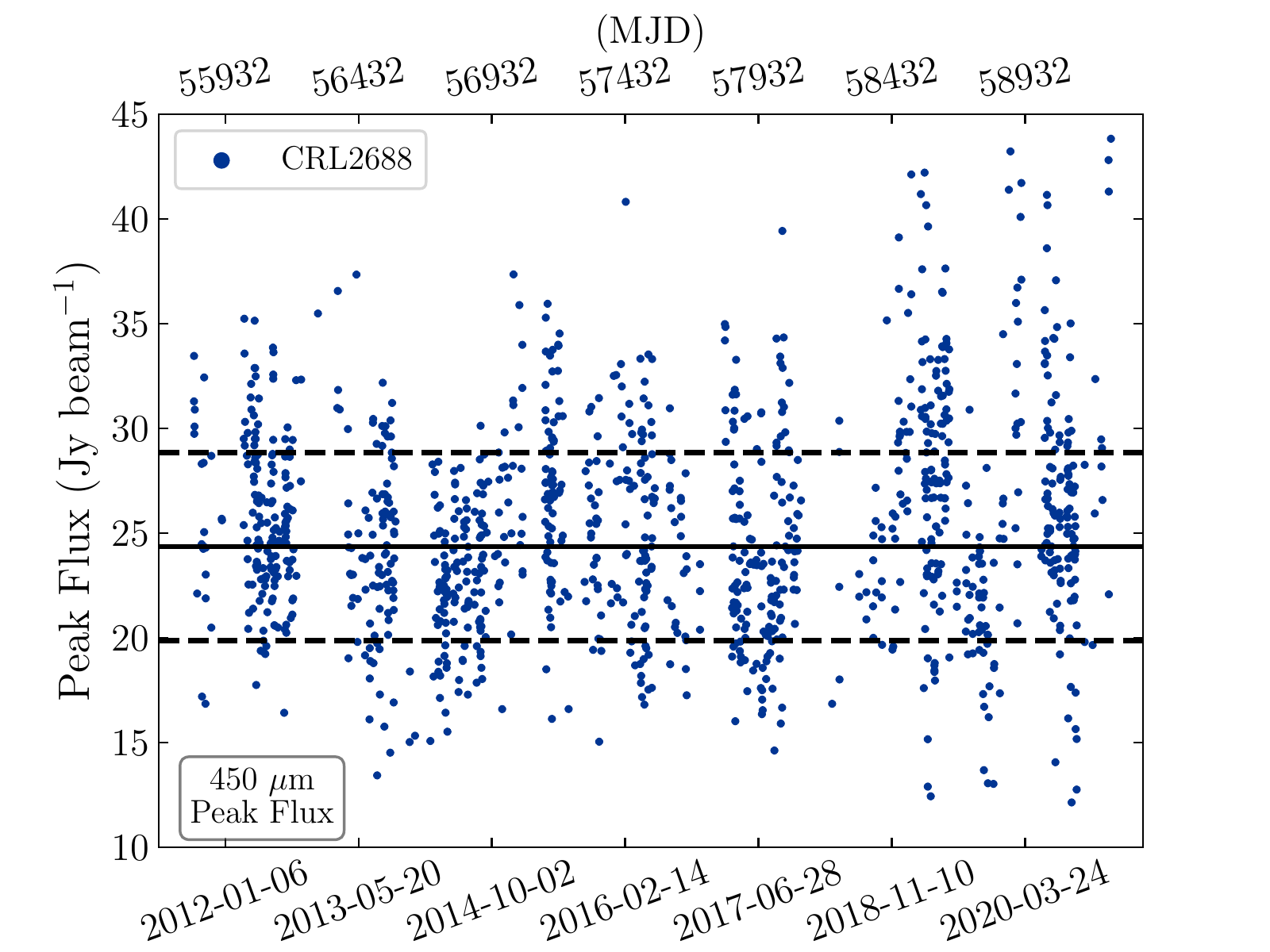}{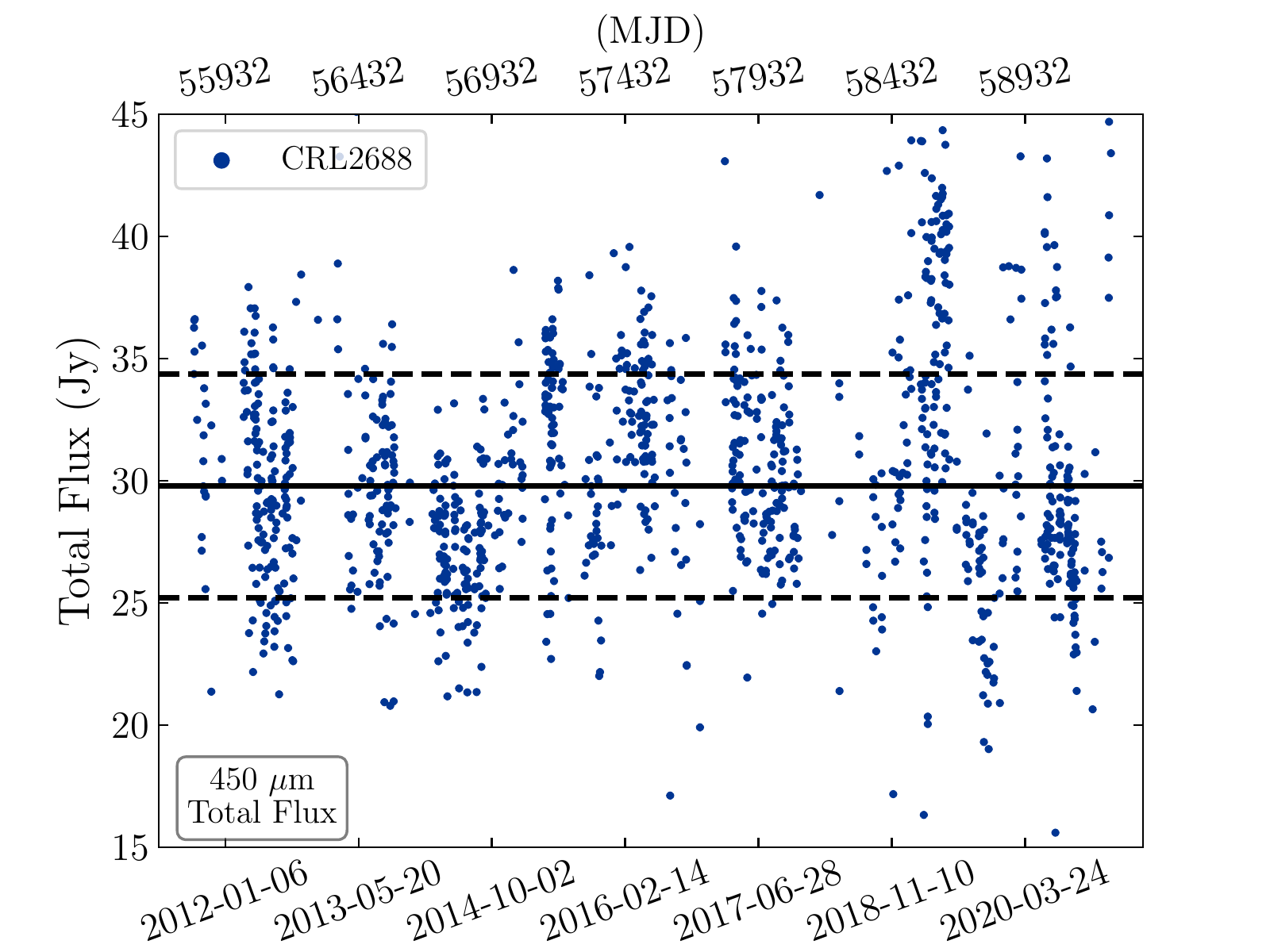}
\caption{The peak (\textit{left}) and total (\textit{right}) fluxes derived for secondary-calibrator sources \mbox{Arp 220} (\textit{top}) and \mbox{CRL 2688} (\textit{bottom}) at \mbox{450 $\mu$m} using the FCFs presented in Table~\ref{tab:3EpochFCFs}, along with the modifications presented in Table~\ref{tab:NightlyFCFMods}. The solid line represents the median value of the distribution and the dashed line shows the standard deviation of the light curve.}
\label{fig:SecCalFlux1}
\end{figure*}
\begin{figure*}
\plotfour{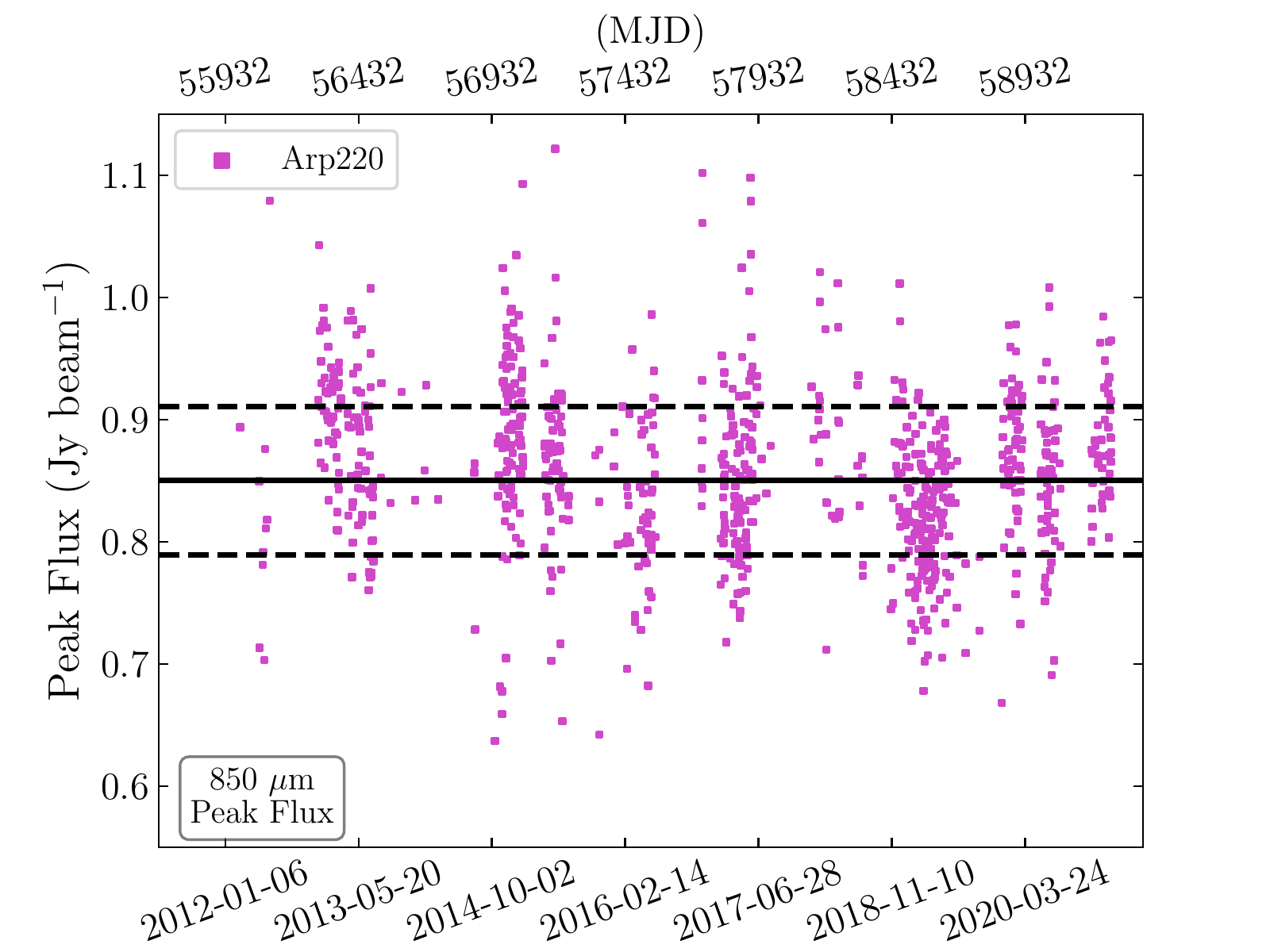}{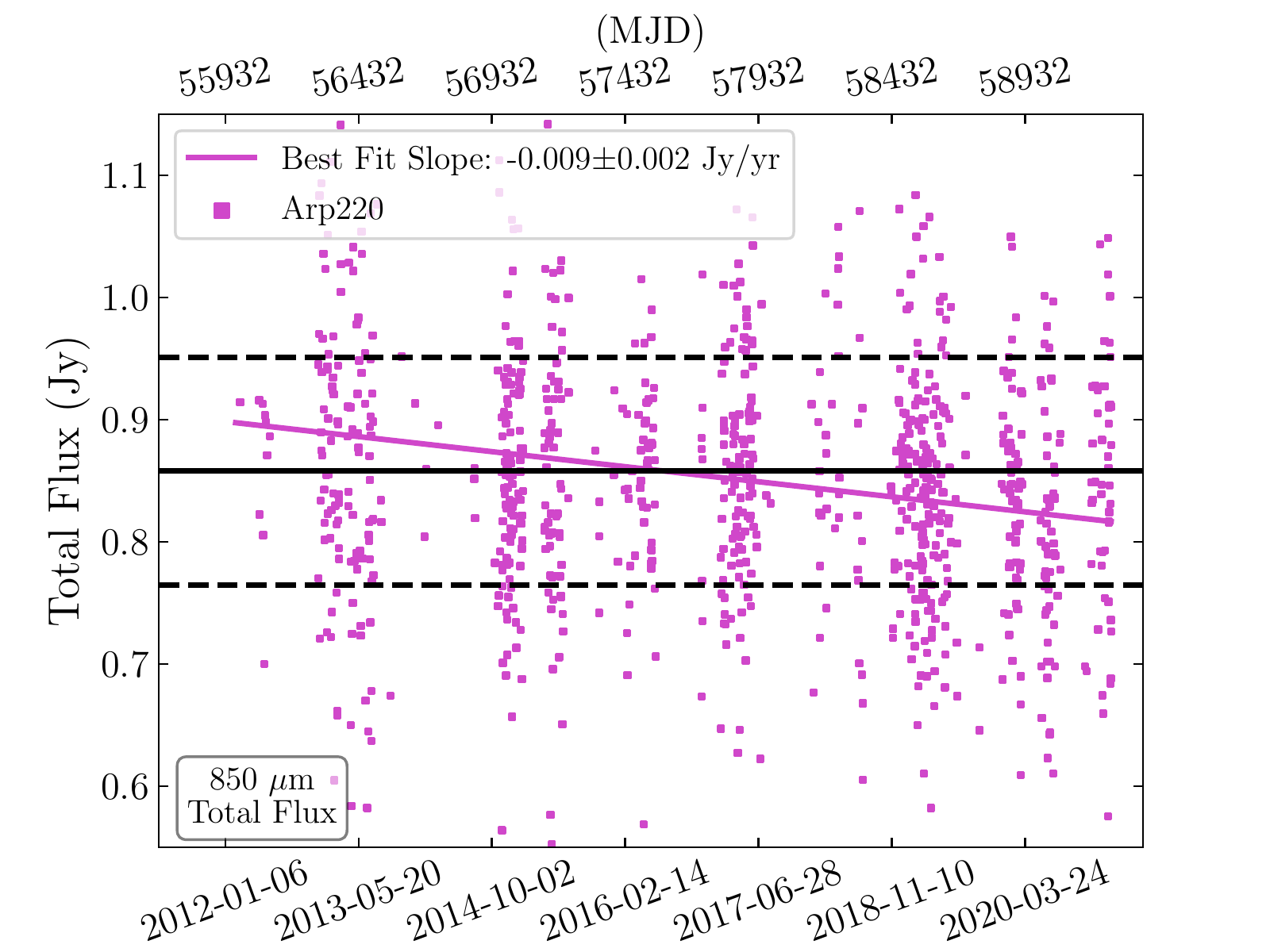}{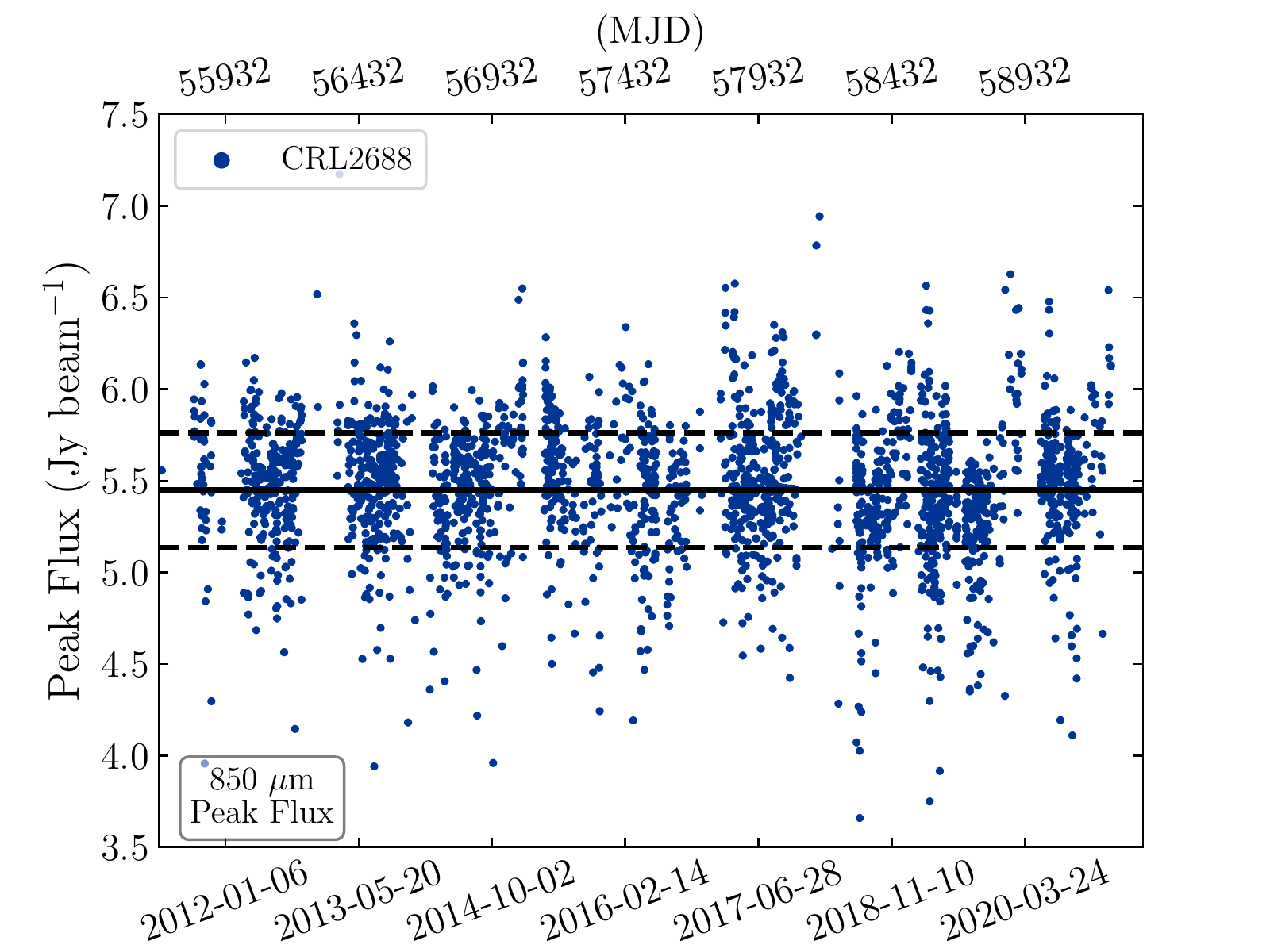}{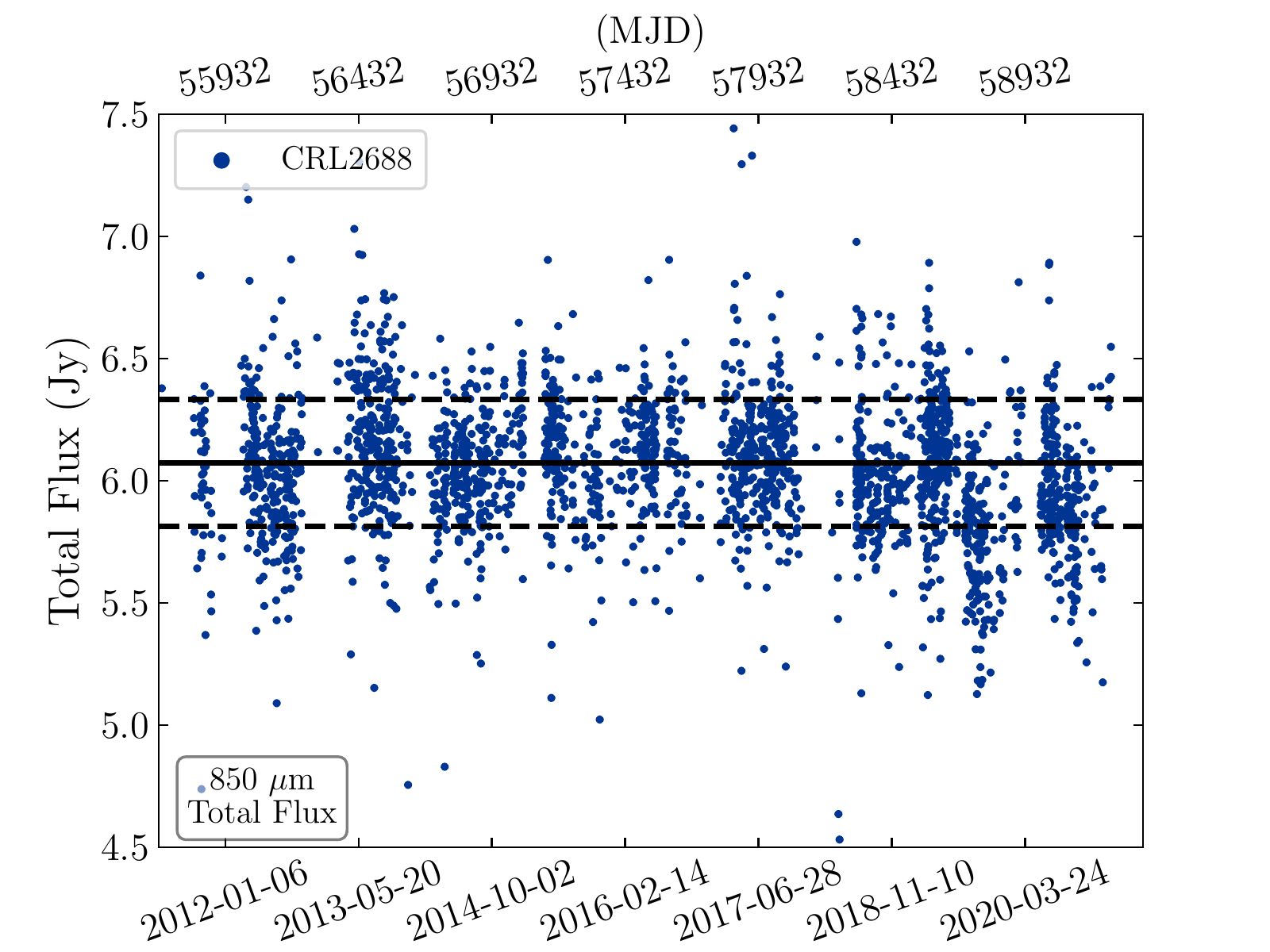}
\caption{The peak (\textit{left}) and total (\textit{right}) fluxes derived for secondary-calibrator sources \mbox{Arp 220} (\textit{top}) and \mbox{CRL 2688} (\textit{bottom}) at \mbox{850 $\mu$m} using the FCFs presented in Table~\ref{tab:3EpochFCFs}, along with the modifications presented in Table~\ref{tab:NightlyFCFMods}. The solid line represents the median value of the distribution and the dashed line shows the standard deviation of the light curve.}
\label{fig:SecCalFlux2}
\end{figure*}
\begin{figure*}
\plotfour{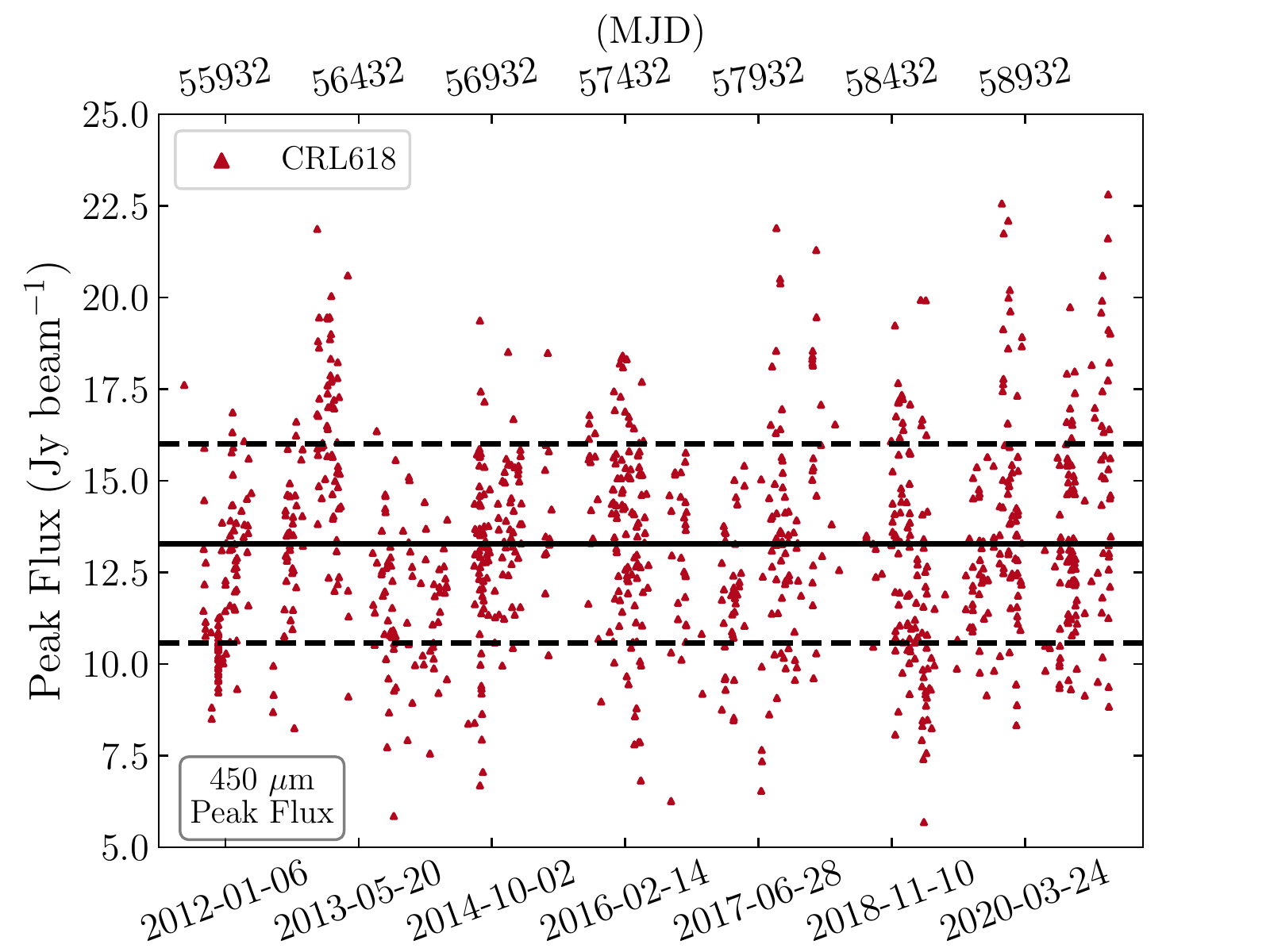}{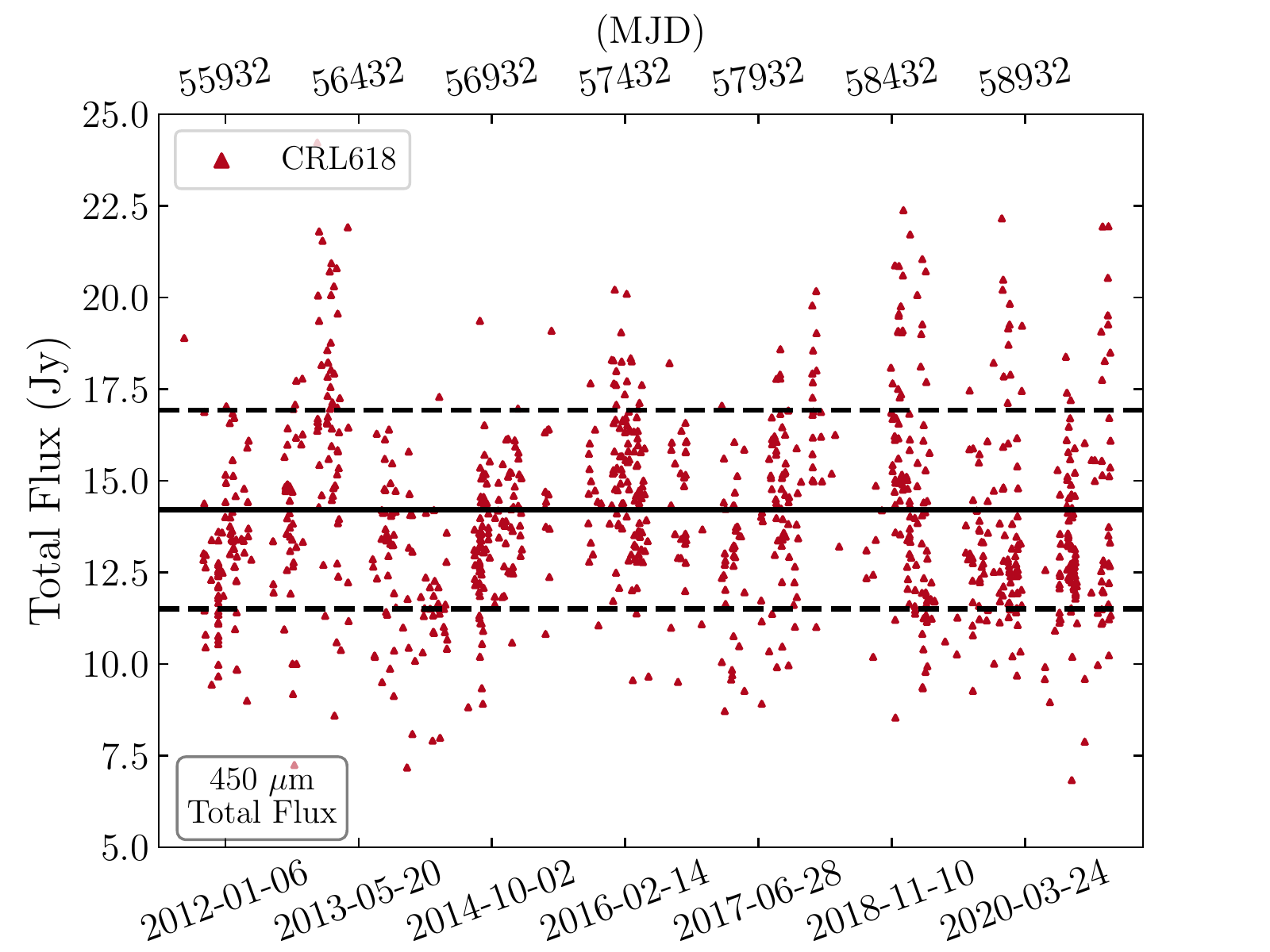}{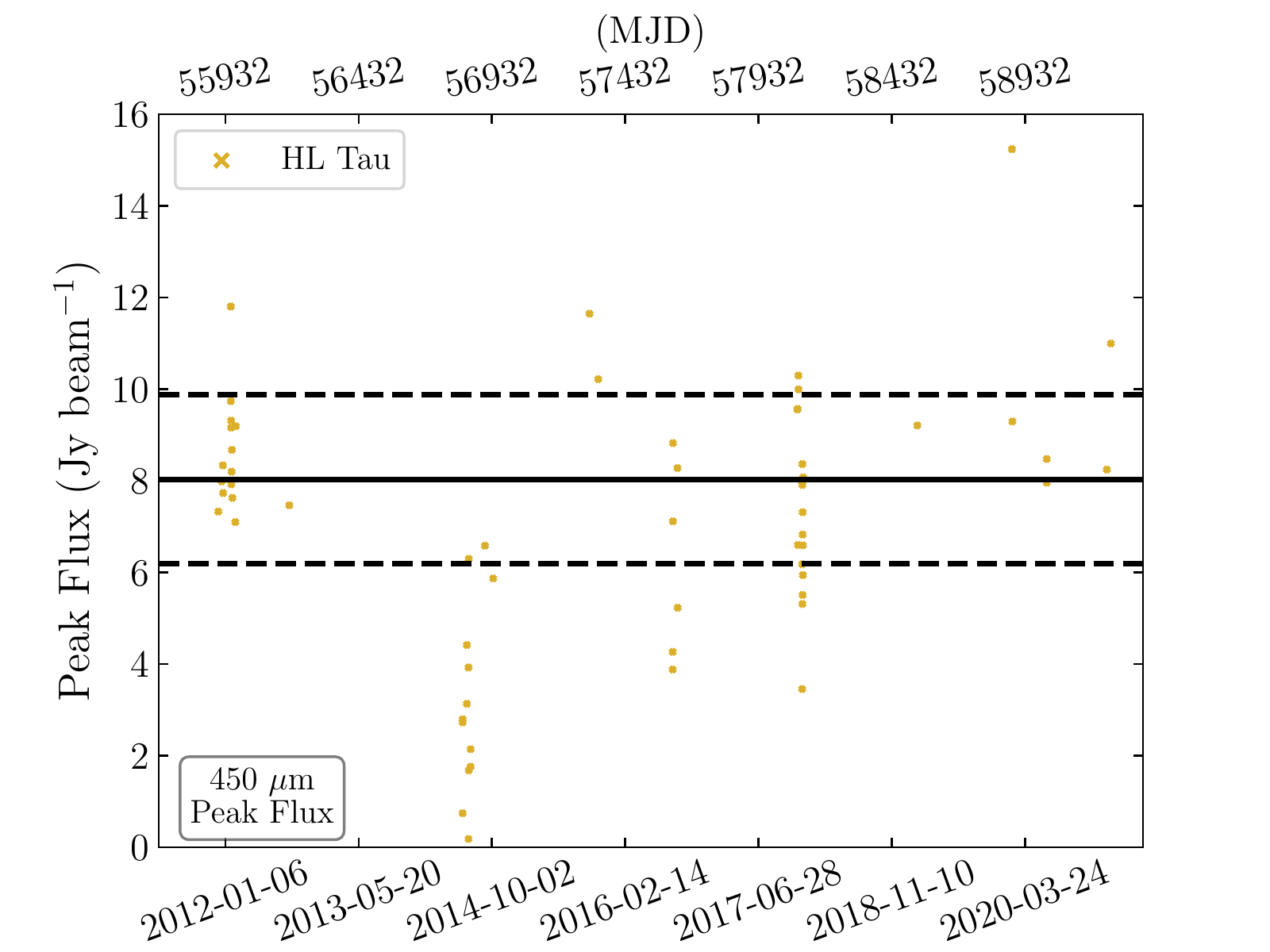}{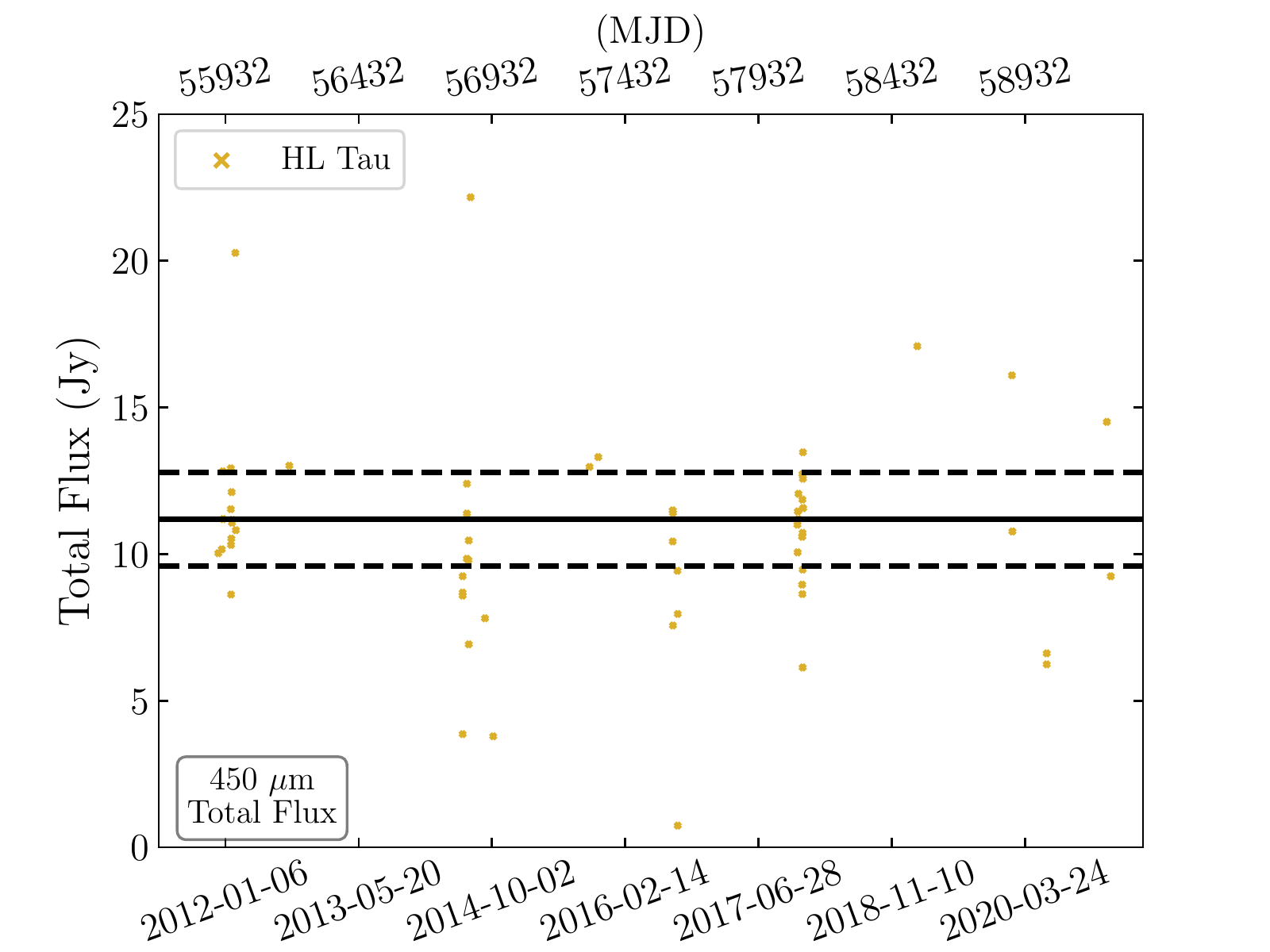}
\caption{The peak (\textit{left}) and total (\textit{right}) fluxes derived for secondary-calibrator sources \mbox{CRL 618} (\textit{top}) and \mbox{HL TAU} (\textit{bottom}) at \mbox{450 $\mu$m} using the FCFs presented in Table~\ref{tab:3EpochFCFs}, along with the modifications presented in Table~\ref{tab:NightlyFCFMods}. The solid line represents the median value of the distribution and the dashed line shows the standard deviation of the light curve.}
\label{fig:SecCalFlux3}
\end{figure*}
\begin{figure*}
\plotfour{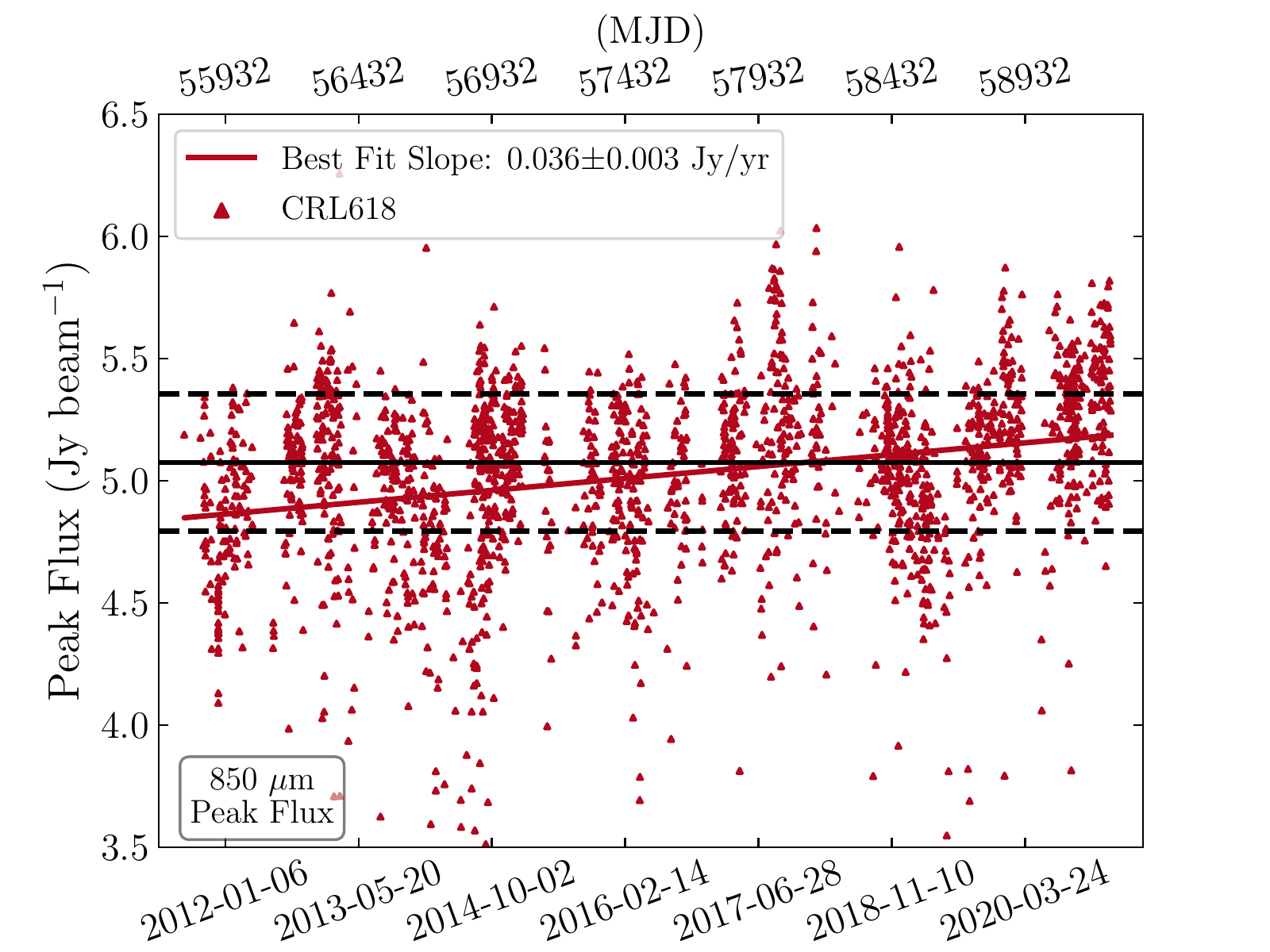}{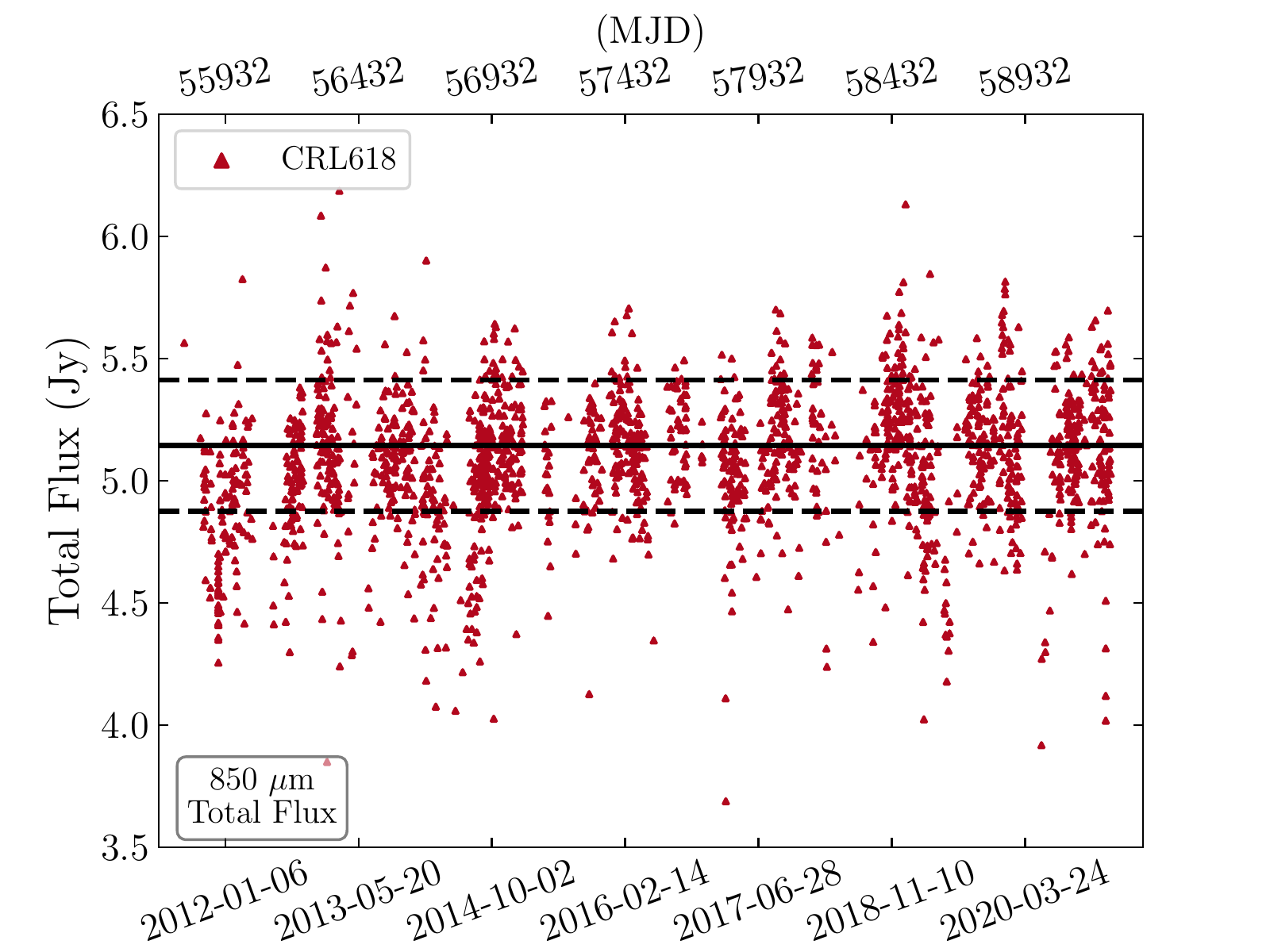}{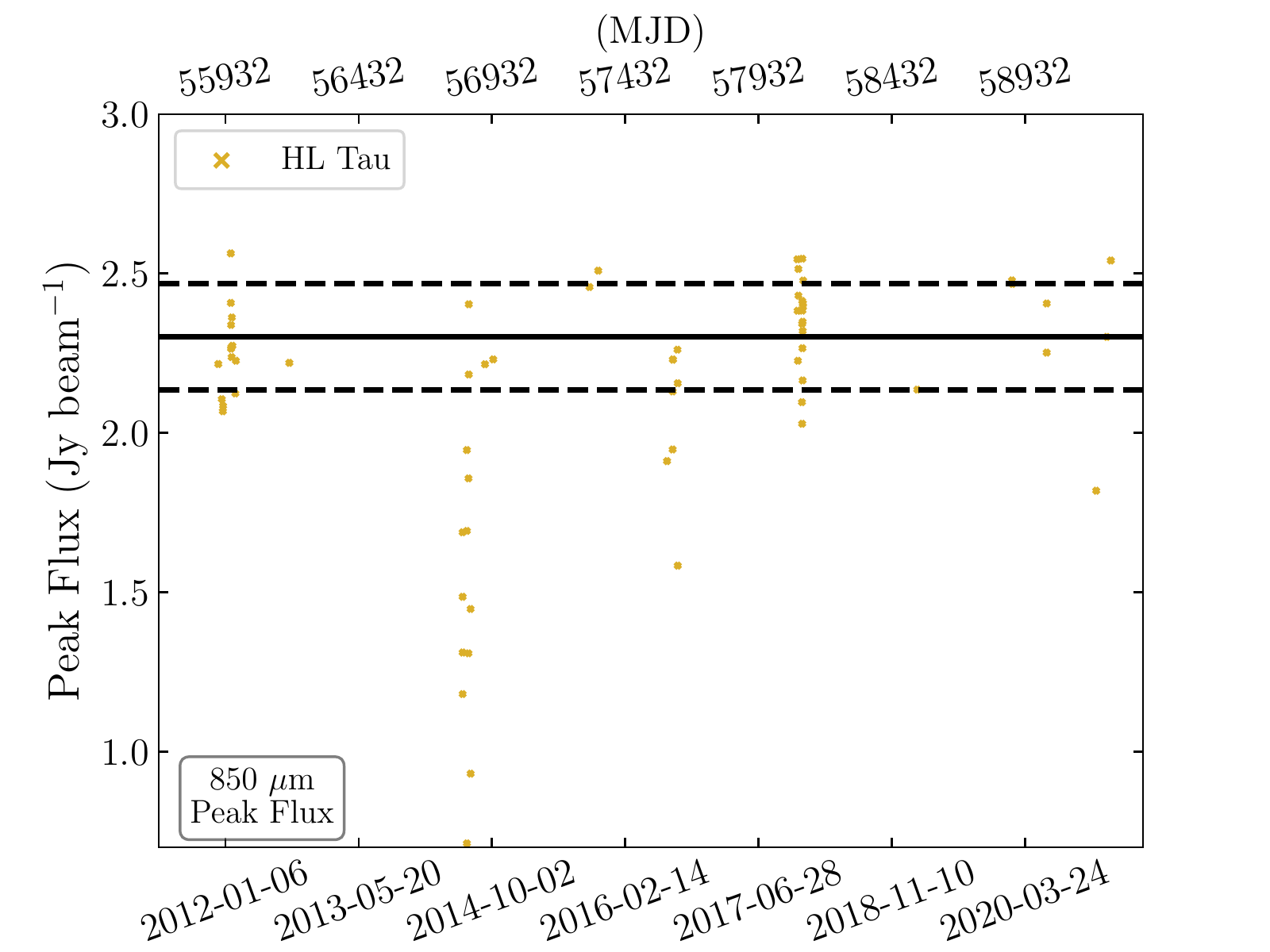}{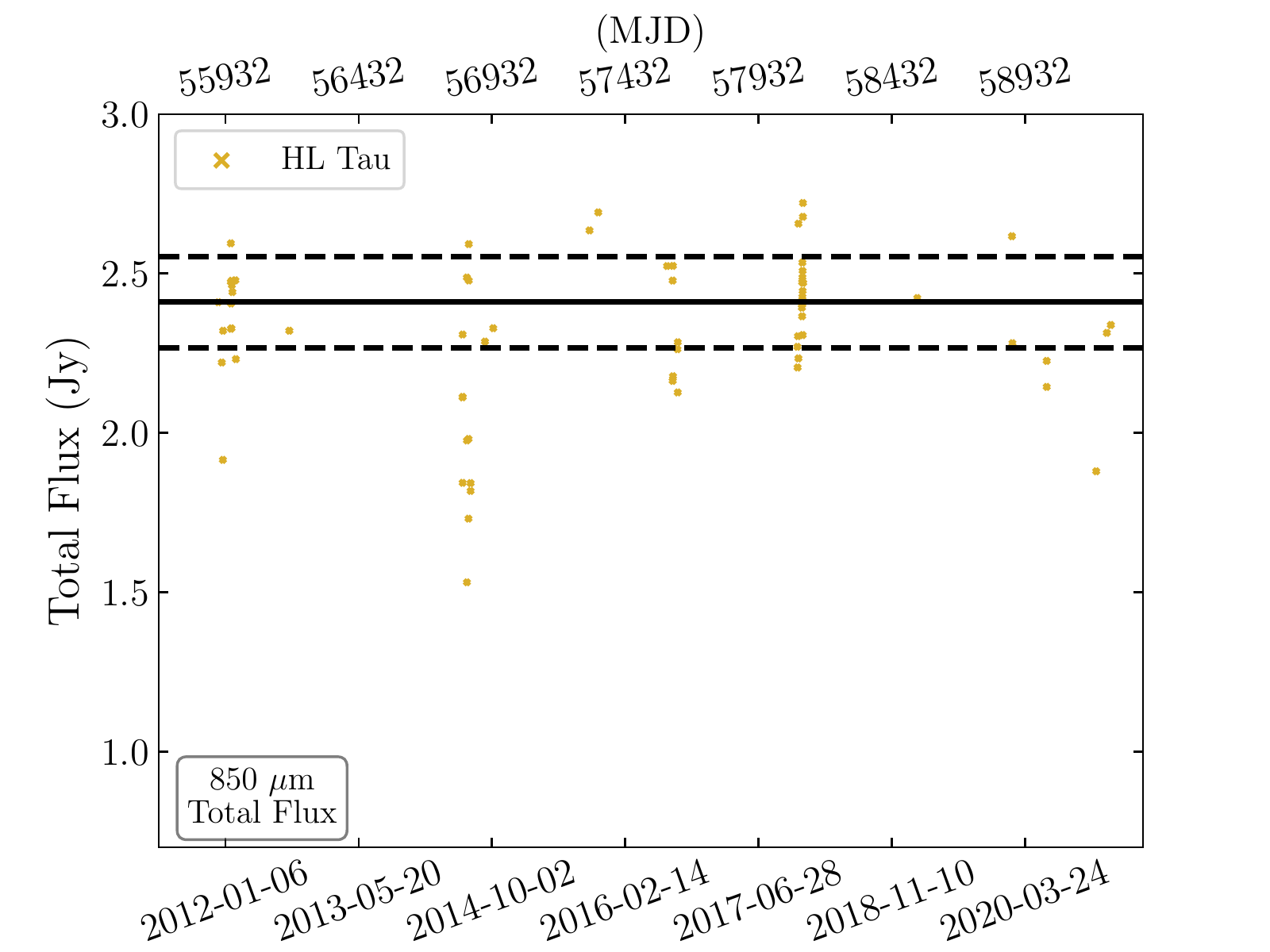}
\caption{The peak (\textit{left}) and total (\textit{right}) fluxes derived for secondary-calibrator sources \mbox{CRL 618} (\textit{top}) and \mbox{HL TAU} (\textit{bottom}) at \mbox{850 $\mu$m} using the FCFs presented in Table~\ref{tab:3EpochFCFs}, along with the modifications presented in Table~\ref{tab:NightlyFCFMods}. The solid line represents the median value of the distribution and the dashed line shows the standard deviation of the light curve.}
\label{fig:SecCalFlux4}
\end{figure*}

% Ten year light curve Table (Sec Cals)
\begin{deluxetable*}{cccccccc}
\tablecaption{Secondary-calibrator Flux Information}
\label{tab:SecCalFlux}
\tablecolumns{8}
\tablewidth{0pt}
\tablehead{
\colhead{Source} &
\colhead{MJD} &
\colhead{$\tau_{225}^{a}$} &
\colhead{$A^{b}$} &
\colhead{450 $\mu$m Peak$^{c}$} &
\colhead{450 $\mu$m Total$^{d}$} &
\colhead{850 $\mu$m Peak$^{c}$} &
\colhead{850 $\mu$m Total$^{d}$}
\\
\colhead{} &
\colhead{} &
\colhead{} &
\colhead{} &
\colhead{(Jy beam$^{-1}$)} &
\colhead{(Jy)} &
\colhead{(Jy beam$^{-1}$)} &
\colhead{(Jy)}
}
\startdata
... & ... & ... & ... & ... & ... & ... & ...  \\
 Arp220 & 56997.8166  &     0.064 &  1.053 &  5.53 &   6.55 &   0.87 &   0.83\\
 Arp220 & 56998.7641  &     0.057 &  1.216 &  5.88 &   5.68 &   0.84 &   0.77\\
 Arp220 & 56999.7709  &     0.057 &  1.173 &  5.98 &   6.35 &   0.86 &   0.85\\
 Arp220 & 57000.7346  &     0.060 &  1.360 &  7.06 &   6.90 &   0.93 &   0.81\\
 Arp220 & 57000.7840  &     0.058 &  1.116 &  6.35 &   7.40 &   0.89 &   0.87\\
 Arp220 & 57001.7033  &     0.049 &  1.625 &  7.19 &   8.26 &   0.93 &   0.96\\
 Arp220 & 57001.7754  &     0.049 &  1.135 &  7.16 &   6.84 &   0.95 &   0.89\\
 Arp220 & 57001.8224  &     0.053 &  1.025 &  6.11 &   6.99 &   0.99 &   0.80\\
 Arp220 & 57002.8384  &     0.051 &  1.008 &  4.79 &   6.74 &   0.91 &   1.66\\
 Arp220 & 57002.7228  &     0.045 &  1.408 &  6.88 &   8.28 &   0.88 &   0.93\\
 ... & ... & ... & ... & ... & ... & ... & ... \\
\enddata
\tablecomments{$^{a}$The opacity of the atmosphere at 225 GHz at the time of the observation.\\$^{b}$The airmass at the time of the observation.\\$^{c}$Peak fluxes were measured by performing a Guassian fit to the source while the data was still in units of picowatts, then applying the FCFs and FCF corrections in Tables \ref{tab:3EpochFCFs} and \ref{tab:NightlyFCFMods}.\\$^{d}$Total fluxes were measured using aperture photometry while the data was still in units of picowatts. The total flux was calculated within a 1 arcminute diameter aperture centered on the source. The background level was determined using an annulus with inner diamter 1.5 arcminutes and outer diameter 2 arcminutes. The FCFs and FCF corrections in Tables \ref{tab:3EpochFCFs} and \ref{tab:NightlyFCFMods} were applied.}
\end{deluxetable*}

10-year light curves of the 450 and \mbox{850 $\mu$m} peak and total fluxes of the four most commonly used secondary flux calibrators are presented in Figures~\ref{fig:SecCalFlux1} (\mbox{Arp 220} and \mbox{CRL 2688} at \mbox{450 $\mu$m}), \ref{fig:SecCalFlux2} (\mbox{Arp 220} and \mbox{CRL 2688} at \mbox{850 $\mu$m}), \ref{fig:SecCalFlux3} (\mbox{CRL 618}, and \mbox{HL Tau} at \mbox{450 $\mu$m}), and \ref{fig:SecCalFlux4} (\mbox{CRL 618}, and \mbox{HL Tau} at \mbox{850 $\mu$m}). The solid lines represent the median value of the distribution and the dashed lines show the median absolute deviation added in quadrature to the 5\% uncertainty in Uranus' (the primary calibrator) flux model.  Table~\ref{tab:SecCalFlux} gives an example of the information that is provided for all these secondary-calibrator sources in machine-readable tables in the online version of the paper. 

The 450 and \mbox{850 $\mu$m} \mbox{CRL 2688} light curves are consistent with constant flux values over time. The \mbox{450 $\mu$m} Peak flux of \mbox{Arp 220} perhaps shows an indication of a parabolic shape with lower flux values reported throughout 2016 and early 2017, but the \mbox{450 $\mu$m} total-flux light curve appears constant over time within the uncertainties. At \mbox{850 $\mu$m}, however, the \mbox{Arp 220} peak-flux values appear constant, whereas the total-flux values appear to decrease linearly at the rate of $0.009\pm0.002\mathrm{\:Jy/year}$. The \mbox{HL Tau} peak and total fluxes at each wavelength also appear broadly constant over time with the exception of 2014, where fluxes appear anomalously low. The majority of these data were obtained in the months of June and July.

\mbox{CRL 618} shows the most indications of variability when compared with the other secondary-calibrator sources. At \mbox{850 $\mu$m} the peak flux clearly increases over time. A simple linear fit to the data suggests a brightening at a rate of $0.036\pm0.003\mathrm{\:Jy/yr}$. Therefore, over the course of 10 years, the peak flux has increased by $0.36\mathrm{\:Jy}$, or, $\sim7\%$. There also appears to be a quasi-periodic trend of brightening and dimming on much shorter ($\sim$yearly) timescales than the overall secular increase. The total flux does not show the same linear, upward trend, reporting a roughly constant value equivalent to the brightest peak flux (measured in early 2021). In both the peak and total flux light curves, the data obtained between 2011--2012 appears to be at a lower value than the subsequent data obtained, highlighting the difference between the measured values in this paper and those in D13. At \mbox{450 $\mu$m}, the larger scatter in the distribution masks any upward trend that may exist in the peak flux, though there are still indications of a quasi-periodic brightening and dimming, especially before 2017. 

%%%%%%%%%%%%%%%%%%%%%%%%%%%%%%%%%%%%%%%%
%%%%%%%%%%%%%%%%%%%%%%%%%%%%%%%%%%%%%%%%
%%%%%%%%%%%%%%%%%%%%%%%%%%%%%%%%%%%%%%%%
\bibliography{SCUBA2Calbib}
%%%%%%%%%%%%%%%%%%%%%%%%%%%%%%%%%%%%%%%%
%%%%%%%%%%%%%%%%%%%%%%%%%%%%%%%%%%%%%%%%
%%%%%%%%%%%%%%%%%%%%%%%%%%%%%%%%%%%%%%%%

\end{document}